\documentclass{aa}
\usepackage{graphicx}
\usepackage{longtable}
\usepackage{colordvi}
\usepackage[varg]{txfonts}
\usepackage{color}
\usepackage{natbib}
\bibpunct{(}{)}{;}{a}{}{,} 
\definecolor{darkgreen}{rgb}{0.0,0.5,0.0}
\definecolor{darkred}{rgb}{0.5,0.0,0.0}
\definecolor{brown}{rgb}{0.65,.16,0.16}
\definecolor{grey}{rgb}{0.4,0.5,0.6}

\newcommand{\greenitalics}[1]{\textcolor{darkgreen}{\it #1}}
\newcommand{\blueitalics}[1]{\textcolor{blue}{\it #1}}
\newcommand{\reditalics}[1]{\textcolor{red}{\it #1}}
\newcommand{\rpi}{\mathrm{\pi}}
\renewcommand{\d}{\mathrm{d}}

\newcommand{\gNFWgTTAL}{1}
\newcommand{\BCGgT}{2}
\newcommand{\BCGgTTAL}{3}
\newcommand{\gNFWTLSgTTAL}{4}
\newcommand{\BCGggOMTAL}{5}
\newcommand{\gNFWggOMTAL}{6}
\newcommand{\gNFWgT}{7}
\newcommand{\BCGPSfourgTTAL}{8}
\newcommand{\NFWgTTAL}{9}
\newcommand{\NFWTLSgTTAL}{10}
\newcommand{\gNFWTTAL}{11}
\newcommand{\NFWgT}{12}
\newcommand{\NFWTTAL}{13}
\newcommand{\BCGPSfourgT}{14}
\newcommand{\NFWggOMTAL}{15}
\newcommand{\NFWisoETTAL}{16}
\newcommand{\EisoELTTAL}{17}
\newcommand{\gNFWisoELT}{18}
\newcommand{\gNFWisoELgTTAL}{19}
\newcommand{\NFWisoELT}{20}
\newcommand{\gNFWisoELTTAL}{21}
\newcommand{\NFWisoELgTTAL}{22}
\newcommand{\EsixisoELTTAL}{23}
\newcommand{\NFWisoELTTAL}{24}
\newcommand{\cNFWisoELTTAL}{25}
\newcommand{\HisoELTTAL}{26}
\newcommand{\NFWisoELgOMTAL}{27}
\newcommand{\NFWiso}{28}
\newcommand{\NFWisoESTTAL}{29}
\newcommand{\NFWisoLSTTAL}{30}
\newcommand{\numruns}{30}

\begin{document}
   \title{Structural and dynamical modeling of WINGS clusters. II. The
     orbital anisotropies of elliptical, spiral and lenticular galaxies.}
\titlerunning{Orbital anisotropies of ellipticals, spirals \& S0s in WINGS clusters}
\author{ G. A. Mamon \inst{1} 
\and A. Cava\inst{2} 
\and A. Biviano\inst{3,1,4}
\and A. Moretti\inst{5}
\and B. Poggianti\inst{5}
\and D. Bettoni\inst{5}
}
\offprints{G. A. Mamon, \email{gam@iap.fr}} 
\institute{Institut d'Astrophysique de Paris (UMR 7095: CNRS \&
     Sorbonne Universit\'e), 98 bis Bd Arago, F-75014 Paris, France
\and Department of Astronomy, University of Geneva, 51 Ch. des Maillettes, 1290 Versoix, Switzerland
\and INAF-Osservatorio Astronomico di Trieste, Via Tiepolo 11, 34143 Trieste,
Italy
\and
IFPU - Institute for Fundamental Physics of the Universe, via Beirut 2, I-34014, Trieste, Italy
\and INAF-Osservatorio Astronomico di Padova, Vicolo Osservatorio 5, 35122 Padova, Italy
}
\authorrunning{Mamon et al.}

  \date{Received; accepted}
  
  \abstract
      {
        The orbital shapes of galaxies of different classes is a probe of
  their formation and evolution.
  The Bayesian MAMPOSSt mass/orbit modeling algorithm is used to jointly
        fit the distribution of elliptical, spiral (and irregular), and
        lenticular galaxies in projected phase space, on 3 pseudo-clusters
        (built by stacking the clusters after re-normalizing their
       positions and velocities) of 54
        regular clusters from the Wide-field Nearby Galaxy-clusters
  Survey (WINGS),
  with at least 30 member velocities. Our pseudo-clusters (stacks) contain nearly 5000
  galaxies with available velocities and morphological types. \numruns\ runs of
  MAMPOSSt with different priors are presented.
         The highest MAMPOSSt likelihoods are obtained for generalized
        NFW models with steeper inner slope, free-index Einasto models, and
        double NFW models for  the cluster
        and the brightest cluster
        galaxy. However, there is no strong Bayesian evidence for a steeper
        profile than the NFW model.
        The mass concentration
        matches the predictions from cosmological simulations.
        Ellipticals usually trace best the mass distribution, while S0s are close.
        Spiral galaxies show increasingly radial orbits at increasing radius,
        as do S0s on two stacks, and ellipticals on one stack.
        The inner orbits of all three types in the 3 stacks are consistent with isotropy.
           
        Spiral galaxies should transform rapidly into early-types given their
        much larger extent in clusters.
        Outer radial orbits are expected for the spirals, a consequence of
        their recent radial infall into the cluster. The less radial orbits
        we find for early-types could be related to the longer time spent by these
        galaxies in the cluster.
        We demonstrate that two-body relaxation is too slow to explain
        the inner isotropy of the early types, which suggests that inner
        isotropy is the
        consequence of violent relaxation during major cluster mergers or
        dynamical friction and tidal braking acting on subclusters.
        We propose that
the inner isotropy of the short-lived spirals is a
        selection effect of spirals passing only once through pericenter
        before being transformed into early-type morphologies. 
      } 
  
      
      \maketitle
      
      \section{Introduction}
\label{sec:intro}      
      
The orbits of galaxies in galaxy clusters are a useful tool to understand the
evolution of clusters. Galaxies detaching themselves from their initial
Hubble expansion should enter clusters on fairly radial orbits.
In the inner regions of clusters, most galaxies have
arrived at early times, and the two-body relaxation time is often thought to
be shorter than the age of the Universe
deep inside the cluster where the crossing times are very short. Hence, the
galaxies in the inner regions should have forgotten their initial trajectories and
the inner population should have isotropic velocities.
But clusters grow by mergers, from very minor to major. In the limit of  very minor
cluster mergers, clusters are relatively isolated systems  accreting individual
galaxies, whose orbits should be fairly radial on their first infall.
In the opposite limit of major cluster mergers, galaxies suffer violent
relaxation that isotropizes their orbits. Moreover, the angular momentum
of the secondary cluster will be transferred into individual galaxies, which
may lead to an excess of more circular orbits.

The measure of the elongations of galaxy orbits is therefore a fundamental tool to
understand the formation of clusters, and how
galaxy  orbits vary with cluster mass, elongation,
and large-scale environment. One may go further and understand how
orbital elongations depend on galaxy
stellar mass or luminosity, as well as galaxy specific star formation rate or color or
even morphological type, as in the present study.

Galaxy orbits in clusters are best studied through \emph{mass-orbit
  modeling}.
Galaxies can be  considered tracers of the gravitational potential.
The prime method to extract orbital shapes
from observational data is through the use of the \emph{Jeans equation} of local
dynamical equilibrium (JE), which states that the  divergence of
the  dynamical  pressure tensor is the opposite of the product of
the tracer density
times the potential gradient. 
The pressure tensor is the tracer density times the tensor of mean squared
velocities,
which may be anisotropic (i.e. a \emph{velocity ellipsoid} with
unequal eigenvalues).  In 
spherical symmetry,
the stationary JE is
\begin{equation}
  {\d \left (\nu \left\langle v_r^2\right\rangle\right) \over \d r} + 2\,{\beta \over r}\,
  \nu \left\langle v_r^2\right\rangle  = -\nu {G\,M \over r^2} \ ,
\label{jeans}
\end{equation}
where
$\nu(r)$ and $M(r)$ are the radial profiles of respectively the tracer
density and the total mass,
$\left\langle v_r^2\right\rangle(r)$ is the mean squared radial velocity profile, and
\begin{equation}
\beta = 1 - {\left\langle v_\theta^2 \right\rangle + \left\langle v_\phi^2
  \right\rangle\over 2\,\left\langle v_r^2\right\rangle} \ ,
\label{beta}
\end{equation}
represents
the (spherical, scalar) anisotropy of the mean squared velocities, called
the \emph{velocity anisotropy} parameter (or simply
\emph{anisotropy}), and is expressed in terms of the root-mean-squared (rms)
velocities ($\beta$ usually varies with radius).
For equilibrium systems, there are no net meridional streaming
motions, hence $\left\langle v_\theta^2\right\rangle = \sigma_\theta^2$. For
galaxy clusters, one neglects rotation leading to
$\left\langle v_\phi^2\right\rangle = \sigma_\phi^2$.
By symmetry,
 $\sigma_\phi=\sigma_\theta$. 
On the other hand, the JE contains radial streaming motions, e.g. from infall.
Radial, isotropic and circular orbits
 respectively have
 $\beta=1$, $\beta=0$, and $\beta\to -\infty$.

In additional to such a \emph{Jeans analysis}, another class of analysis uses
the collisionless Boltzmann (Vlasov) equation (CBE), which states that the
six-dimensional fluid is incompressible in the absence of galaxy collisions
(see, e.g., chap. 5 of \citealp{Courteau+14} for a quick review of these stellar dynamics).
In fact, the JE is a direct consequence (1st velocity moment) of the CBE.
One can attempt to find a well-behaved and realistic 6D distribution
function (DF),
expressed in terms of energy and angular momentum or in action angle space,
whose moments match the data. In particular, the distribution of tracers in
\emph{projected phase space} (projected radii and relative line-of-sight velocities, PPS) can
be expressed as a triple integral of the DF \citep{Dejonghe&Merritt92}.

There are, however,  many hurdles to extract orbital shapes from either Jeans
or DF analyses.
1) Clusters are observed in projection, with only 2 positional (sky)
coordinates and one (\emph{line-of-sight}, LOS) velocity.
2) The lack of information on the depth coordinate causes observers to mix
clusters along the LOS.
3) Clusters tend to be prolate systems \citep{Skielboe+12} (although not far from
spherical symmetry), as well as prolate in phase space \citep{Wojtak13}.
4) The spherical JE contains a single (radial) equation linking the unknown
mass profile ($M(r)=(r^2/G)\,\d\Phi/\d r$) and anisotropy (linked to the
orbital shapes), an issue called the \emph{mass / anisotropy degeneracy}
(MAD, \citealp{Binney&Mamon82}),
while DF analysis suffers from an analogous degeneracy between potential and DF. 
5) Streaming motions complicate the analysis.
6) The JE includes partial time derivatives, which are very difficult to
estimate (see \citealp{Falco+13}).

There are many variants of Jeans and DF analyses (see chap. 5 of
\citealp{Courteau+14} for a partial review, focused on galaxy masses).
A first class of methods attempts to fit models to the data.
The simplest method is to fit models of the LOS velocity dispersion profile to
the observed one. 
While the LOS velocity dispersion profile is a double integral of the tracer
density and total mass profiles, it can be simplified to single integral
formulations for simple anisotropy profiles
\citep{Mamon&Lokas05b}. 
In a non-rotating system, the 
velocity dispersion anisotropy
affects the shapes of
the distribution of LOS velocities \citep{Merritt87}.
The MAD can thus be (partially) lifted by folding in the
LOS velocity kurtosis profile
\citep{Lokas02,Lokas&Mamon03,Richardson&Fairbairn13,Read&Steger17}.
Another class of methods inverts the data to recover models. 
For example, assuming a mass profile, which may be deduced from other types
of observations (e.g. X-rays, strong and weak gravitational lensing), one can
derive a non-parametric anisotropy profile
\citep{Binney&Mamon82,Tonry83,Bicknell+89,Solanes&Salvador-Sole90,Dejonghe&Merritt92},
a process 
called \emph{anisotropy inversion}.\footnote{Conversely,
  one can assume an anisotropy profile and deduce a non-parametric
mass profile \citep{Mamon&Boue10,Wolf+10}, although this \emph{mass inversion} is less relevant for
the study of orbits of the present work.}

A disadvantage of Jeans analysis methods is that they usually require the radial
binning of the data.\footnote{One can instead bin the model, i.e. the radial
  profile of 
  velocity variance \citep{Diakogiannis+17} or mass density
  \citep{Read&Steger17}, but the 
  choice of bins and regularization are complex issues.} One way around this is to specify the form of the LOS
velocity distribution function, and starting with \cite{Walker+09}, it is
popular to assume a Gaussian LOS velocity distribution function (in studies
of dwarf spheroidal galaxies, but this has not yet been done for
clusters). But since the LOS velocity distribution function depends on the
velocity anisotropy \citep{Merritt87}, it is not desirable to measure the
anisotropy in this manner.
In DF methods, one specifies a form for the DF written as $f = f(E,L)$, where
$E$ is energy and $L$ is angular momentum. In particular, models with
constant anisotropy $\beta$ have $f \propto L^{-2\beta}$.
One can then compute not only moments in radial bins, but also at specific
positions of the tracers in PPS.
Early studies used fairly arbitrary choices for the DF. \cite{Wojtak+08}
assumed a separable form for $f(E,L)$  and found that it matched well the
halos in cosmological $N$-body simulations. 
This method, adapted to observational data by \cite{Wojtak+09}, is powerful,
but slow as it involves computing triple integrals 
\citep{Dejonghe&Merritt92}.
Nevertheless, it has been successfully applied to clusters
\citep{Wojtak&Lokas10,Wojtak+11} and galaxies \citep{Wojtak&Mamon13}.
A promising method is to express the DF in terms of action-angle variables \citep{Vasiliev19}.

In a hybrid method called MAMPOSSt (\citealp*{Mamon+13}, hereafter MBB13), the
DF is no longer expressed in terms of $E$ and $L$, but in terms of the
three-dimensional velocity distribution function, the simplest form being a Gaussian.
This greatly accelerates the method as it only involves single
integrals to predict the observed distribution of tracers in PPS. It is a
hybrid model, because while MAMPOSSt does not involve radial binning and
assumes a (velocity) distribution function, it uses parametric forms for the total mass and
velocity anisotropy profiles and 
solves the JE for $\left\langle v_r^2(r)\right\rangle$ to compute the likelihood of the  distribution
of tracers in PPS. Using mock clusters from cosmological simulations,
MBB13 found that
MAMPOSSt 
lifts the MAD, with slightly comparable accuracy on the mass normalization and
scale as the dispersion-kurtosis method of \cite{Lokas&Mamon03} (according to
the tests of \citealp*{Sanchis+04_DK}) and the DF
method of \cite{Wojtak+09}.\footnote{MAMPOSSt was the 2nd most efficient 
   among 23 algorithms in a recent cluster mass challenge \citep{Old+15}
   to determine the mass normalizations of 1000 realistic mock clusters
   The {\tt Num} richness-based
  method (Mamon et al., in prep.) that
  was the most efficient does not compute mass and anisotropy profiles.}
In both comparisons, MAMPOSSt did much better on
the velocity anisotropy, reaching double
the accuracy on
$\log \left(\left\langle v_r^2\right\rangle^{1/2}/\sigma_\theta\right)$.

There have been many attempts to measure the anisotropy of galaxy
orbits in clusters.
In a pioneering study, \cite{Merritt87} attempted anisotropy inversion on the
Coma cluster with 300 tracers, but was not able to settle whether the orbits
were circular, radial or isotropic, given his uncertainty on the mass profile.
\cite{Lokas&Mamon03} considered both LOS dispersion and  kurtosis profiles of
the same Coma cluster and
determined a slightly tangential (assumed constant) anisotropy (with large
uncertainty).

Another way to lift the MAD is to adopt the mass profile from other methods.
Applying anisotropy inversion (from \citealp{Bicknell+89} to the mass profile
of Abell~1689 ($z=0.18$) 
derived from weak lensing, \cite{Natarajan&Kneib96} found that
the velocities are isotropic in the core, and become radial ($\beta=0.5$) at
800 kpc, which corresponds to roughly $r_{200}/3$ given the published virial masses
\citep{Sereno+13,Lemze+08}.\footnote{The radius $r_\Delta$ is where the mean
  density of a system is $\Delta$ times the \emph{critical} density of the Universe at the
  cluster redshift. We will call $r_{200}$ the `virial' radius (in quotes),
  but will also refer to the theoretical virial radius, which is close to
  $r_{100} \simeq 1.35\,r_{200}$ with the analogous definition. We also
  define $M_\Delta = M(r_\Delta)$ and $v_\Delta = \sqrt{G\,M_\Delta/r_\Delta}$.\label{fn:rdelta}}
\cite{Benatov+06} applied the same anisotropy inversion to 5 clusters from the CAIRNS
survey ($z=0.03$ to 0.3) whose mass profiles were obtained from X-rays and/or
lensing. They 
derived anisotropy profiles that were radial at the very center, and showed a
diversity of profiles in their bodies, with 2 clusters showing slightly
tangential or isotropic velocities at $r_{200}$, 1 mildly radial
($\beta=0.3$) and 2 fully radial ($\beta \simeq 0.95$ at $r_{200}$).
Analyzing a mere 64 dwarf galaxies in Coma, \cite{Adami+09} adopted the mass
profile derived by \cite*{Geller+99} with the caustic
method \citep{Diaferio99}, then fit a constant
$\beta$ to the LOS dispersion profile, to obtain $\beta=0.4\pm0.2$ or
$0.7\pm0.1$ depending on their fit of the number density profile.
So for the Coma cluster,
 the orbital anisotropy may be a function of galaxy mass (since
the sample analyzed by \citeauthor{Lokas&Mamon03} was composed of more
luminous galaxies than that analyzed by \citeauthor{Adami+09}).

Different galaxy types are often thought to have different anisotropies. For
example, early-type galaxies
prefer dense regions of clusters,  and are expected to have fallen into the
cluster at early times and relaxed to isotropic velocities. In contrast,
spiral galaxies are thought to be falling in clusters on fairly radial orbits, and perhaps bouncing
out on similarly radial orbits. It is thus important to separate the two
populations, which are expected to have very different kinematics.

\cite{Biviano&Katgert04} used anisotropy inversion (following the method of
\citealp{Solanes&Salvador-Sole90}, hereafter SS90) on a joint analysis of 59 
ENACS clusters with over 20 member velocities per cluster, and were the
first 
to measure the radial variations of
the orbits of early versus late spiral morphological types. They assumed that
early-type galaxies had isotropic velocities (as inferred from the 
roughly Gaussian LOS velocity distribution that \citealp*{Katgert+04} found
for non-central early types, following the predictions of 
\citealp{Merritt87}), which enabled \cite{Katgert+04} to first 
determine the mass profile.
\citeauthor{Biviano&Katgert04} found that, for early type spirals
(Sa, Sb),  $\beta$ rises to 0.7
at half the cluster `virial' radius, $r_{200}$, then falls to 0.35 at $r_{200}$.
In late-type spirals, they found that $\beta=0$ (isotropic velocities) out to
$0.6\,r_{200}$ and then rises to $\beta=0.3$ at $r_{200}$.
They did not consider lenticular (S0) galaxies, nor has anybody else until now.
The results of \citeauthor{Biviano&Katgert04} were confirmed by
\cite*{Munari+14}, who studied Abell~2142 ($z=0.09$),  first determining the
 mass profile from a  combination of X-ray, lensing and 
 dynamical studies, and using  anisotropy inversion to deduce that
 red galaxies have isotropic orbits at all radii, while the orbits of blue
 galaxies are isotropic in the inner regions and more radial outside. 
\cite{Biviano&Poggianti09} analyzed two stacks of cluster galaxies
finding that the orbits of non-emission-line galaxies and
emission-line galaxies are similar in the $z \sim 0.56$ stack, while
non-emission-line galaxies move on more isotropic orbits in the $z \sim 0.07$
stack.
But the statistical evidence for this evolution is
very weak.

Other studies of Biviano and collaborators point to somewhat radial outer
orbits for passive galaxies.
Analyzing a $z=0.4$ cluster with 600 member velocities, \cite{Biviano+13}
concluded (using MAMPOSSt and performing SS90 anisotropy inversion from the
MAMPOSSt mass profile)  that both star forming and passive galaxies have isotropic orbits
inside and radial orbits outside.
Working on stacked $z\sim 1$ clusters from the Gemini Cluster Astrophysics
Spectroscopic Survey (GCLASS), \cite{Biviano+16} determined the mass profile
with MAMPOSSt and performed SS90 anisotropy inversion and
found that both passive and star forming galaxies show radial
outer anisotropy ($\beta=0.4$, with large error bars).
\cite{Annunziatella+16} performed SS90 anisotropy inversion
on a parametric
Navarro-Frenk-White (NFW, \citealp{Navarro+96}) model fit to
lensing data for
Abell~209 ($z=0.21$), and found that passive galaxies display radial outer anisotropy,
while their inner anisotropy depends
on their stellar mass (slightly radial at high mass and tangential at low
mass).
 \cite{Capasso+19} analyzed non-emission-line galaxies in 
 110 SZ-selected clusters at  $0.26 < z < 1.32$, using both MAMPOSSt and anisotropy inversion. They
 concluded
 that passive galaxies
have isotropic inner orbits and more radial outer orbits (their outer anisotropy varies with
increasing redshift in an oscillatory manner).

Other anisotropy inversions of clusters indicate different conclusions.
\cite{Hwang&Lee08} studied
Abell~1795 ($z=0.06$) with several mass profiles determined from X-ray analyses,
and performed anisotropy inversion (using the technique of
\citealp{Bicknell+89}) to deduce  that early- and
late-type galaxies had similar radial profiles of anisotropy starting radial
in the core, dropping to very tangential $\beta \approx  -3$ at $r_{200}/2$
and roughly flat beyond; but their study suffered from having only 160 galaxies. 
\cite{Aguerri+17} analyzed Abell~85
($z=0.06$)
using a parametric NFW model for the mass profiles
obtained by the caustic method \citep{Diaferio&Geller97} and from X-ray data, and applying
their own
anisotropy inversion equations (which turn out to be equivalent to those of SS90,
as shown in Appendix~\ref{sec:anisSS90A17}). They
found  isotropic outer orbits ($\beta=0.0\pm0.3$) for blue dwarf
galaxies, but very radial outer orbits for
red galaxies ($\beta=0.7\pm0.2$).
This is the first study to point to red (or passive or elliptical) galaxies
having more radial outer orbits than blue (or star forming or spiral)
galaxies. 
Could the hierarchy of outer radial anisotropy versus the morphological type
or specific star formation class depend on the cluster?

One can alternatively blame the high sensitivity of anisotropy inversion to
the  required  extrapolation of both the data and the model tracer density
and mass profiles both outwards to $r\to\infty$ and inwards to $r=0$.
Moreover, all anisotropy inversion algorithms
involve differentiating
the observational data --- they require the knowledge of
$\d \left[\Sigma(R)\,\sigma_{\rm los}^2 (R)\right]/\d R$, where $\Sigma$ is
the surface density.\footnote{The algorithms
of \cite{Bicknell+89} and \cite{Dejonghe&Merritt92} also involve
differentiating the dynamical pressure, which is the sum of integrals of the
model and of the data, see Appendix~\ref{sec:anisB89DM92}, which shows that
the two algorithms are equivalent
for the specific case of virial equilibrium.}
This is where the DF methods have an advantage.
\cite{Wojtak&Lokas10} studied nearby clusters with their state-of-the-art
DF method, which they analyzed individually and then jointly assuming common
anisotropy profiles. In 8 clusters out of 10, they found that the
distribution of galaxies in PPS implied isotropic inner orbits and radial
outer orbits ($\beta=0.7$ at 1 to $1.5\, r_{200}$).
Using the same technique, \cite{Wojtak&Mamon13} studied  the kinematics of satellite galaxies around galaxies themselves (i.e. in
small groups), and deduced that satellites around red galaxies  lied on orbits with radial
outer anisotropy, but they did not separate the satellites according to color
or morphological type.

In this article, we study the dependence of the velocity anisotropy profiles
of galaxies in clusters on their
morphological type, distinguishing between elliptical, spiral, and (for the
first time) lenticular galaxies.
We thus seek to settle the debate on the different orbital shapes of
elliptical and spiral galaxies in clusters, but also wish to understand
better S0 galaxies through their orbital shapes. Should the  S0 galaxies resemble
spiral galaxies in their orbital shapes, because they both have disks, or
should they resemble more elliptical galaxies, because they have large bulges
and old stellar populations?
This
may help understand whether
S0s originate from  spiral galaxies 
that saw their disks fade, grew their bulges by mergers, or possibly
even originate from ellipticals that accreted disks.

We use the WIde field Nearby Galaxy clusters Survey (WINGS), which contains
X-ray-selected (median luminosity
$L_X^{0.1-2.4\,\rm keV} = 10^{43.75}\,\rm erg\,s^{-1}$) clusters at redshifts
$0.04 < z < 0.07$ and that are located
on the sky at least
$20^\circ$ from the Galactic Plane
\citep{Fasano+06}.
The WINGS spectroscopic dataset \citep{Cava+09,Moretti+14} has been complemented with redshifts from
literature data collected through the SDSS-DR7\footnote{http://www.sdss.org/}
and NED\footnote{http://ned.ipac.caltech.edu/} databases.
Thanks to the
stacking procedure, the size of the sample we analyze here is roughly
an order of magnitude larger than those used in previous studies of
individual clusters (which were typically limited to a few hundred
members).
In the first article (\citealp{Cava+17}, hereafter Paper~I) of the present series of
articles on WINGS clusters, the sample of WINGS clusters was split between
regular and irregular clusters.
Here, we focus on the regular clusters, and consider the sub-populations of elliptical, S0 and
spiral+irregular galaxies as different tracers of the same gravitational potential.

We adopted MAMPOSSt as our primary tool to extract
simultaneously the mass profile and the anisotropy profiles of the different
classes of galaxies.
This will be the first application of MAMPOSSt to a large sample of stacked
nearby clusters.
Stacking clusters  reduces the intrinsic
effects of triaxiality that create unavoidable biases in the derived mass
profile parameters of individual halos (MBB13).

The outline of this article is the following. We present the
MAMPOSSt mass-orbit modeling algorithm in Sect.~\ref{sec:method}.
In Sect.~\ref{sec:data}, we explain the
data sample, the
stacking method, while we explain in Sect.~\ref{sec:mpo_practical} the
practical implementation of MAMPOSSt, in particular the radial profiles
adopted for number density, surface number density, mass and anisotropy.
In Sect.~\ref{sec:results}, we determine the mass and anisotropy profiles of the
stacked samples. We discuss our results in Sect.~\ref{sec:discuss}
and provide our conclusions in
Sect.~\ref{sec:concl}.  We assume a $\Lambda$CDM
cosmological model with $\Omega_{\rm m,0}=0.3$, $\Omega_{\Lambda,0}=0.7$, and
$H_0 = 70 \,\rm km \,s^{-1} \, Mpc^{-1}$.

\section{The MAMPOSSt algorithm}
\label{sec:method}

In its standard implementation, MAMPOSSt (MBB13) performs a maximum likelihood fit of the
distribution of galaxies in PPS, using parameterized models for the radial
profiles of total mass and velocity anisotropy, as well as for the radial
number profile and its corresponding surface number density and projected
number profiles.
MAMPOSSt assumes spherical symmetry, negligible streaming motions,
and a form for the 3D velocity distribution of the tracers
(taken to be a Gaussian in its current implementation).

In Paper~I, we determined the number density profiles of the three
morphological types by fits of NFW plus constant background models on the
photometric data.
Given the known distribution of projected radii $R$,
the MAMPOSSt likelihood of the
distribution of galaxies in PPS
(projected radii and LOS velocities $v_z$) is (MBB13)
\begin{eqnarray}
  - \ln {\cal L} &=& - \sum \ln p(v_z|R,r_\nu,\vec{\eta}) \ ,
\label{lnL}
\\
p(v_z|R,\vec{\eta}) &=& { \sqrt{2/\rpi}\over \Sigma(R)} \,
    \int_R^\infty {\nu\over \sigma_z}\,\exp\left[-{v_z^2\over
        2\sigma_z^2}\right]\,{r\over \sqrt{r^2-R^2}}\,\d r \ ,
\label{pofvz}
\\
\sigma_z^2(r,R) &=& \left [1-\beta(r) {R^2\over r^2}\right]\,\left\langle v_r^2(r)\right\rangle\ ,
\label{sigmaz}
\end{eqnarray}
where $\vec{\eta}$ is the vector of parameters describing the radial profiles
of mass and velocity anisotropy,
while $\nu$ and $\Sigma$ are the model number density and corresponding surface
density profiles, respectively.
The mean squared radial velocity, $\left\langle v_r^2\right\rangle$,
in the right-hand-side of equation~(\ref{sigmaz}) is previously determined
for a given $\vec{\eta}$, by solving the spherical stationary JE of
equation~(\ref{jeans}), 
which can be inverted by solving for $\d \ln K_\beta / \d \ln r = 2\,\beta(r)$,
yielding \citep{vanderMarel94,Mamon&Lokas05b}
\begin{equation}
  \nu(r)\,\left\langle v_r^2(r)\right\rangle = {1\over K_\beta(r)}\int_r^\infty
  K_\beta(s)\,\nu(s)\,{G\,M(s)\over s^2}\,\d s \ ,
  \label{sigmar}
\end{equation}
where $K_\beta(r)/K_\beta(s) = \exp\left[2\int_r^s \beta(t) \d t/t\right]$ depends
  on the anisotropy model, and its values 
  are given in Appendix~A of MBB13 for simple anisotropy models.

LOS velocity  uncertainties $\epsilon_v$ are accounted for by MAMPOSSt by
replacing
$\sigma_z$ in equation~(\ref{pofvz}) by $\sqrt{\sigma_z^2+\epsilon_v^2}$.
In our dataset, the typical velocity uncertainties are $\epsilon_v \approx 53
\, \rm km \, s^{-1}$, which turn out to be negligible relative to the cluster
velocity dispersions, hence have little effect on the MAMPOSSt results. 

MBB13 have tested that splitting the PPS into the separate determinations of the
number density profile from the distribution of projected radii on one hand,
and the mass and anisotropy profiles from the distribution of LOS velocities
at given projected radius on the other hand, leads to virtually the same
parameters as the standard joint fit of PPS.

\begin{figure}[ht]
  \centering
  \includegraphics[width=\hsize]{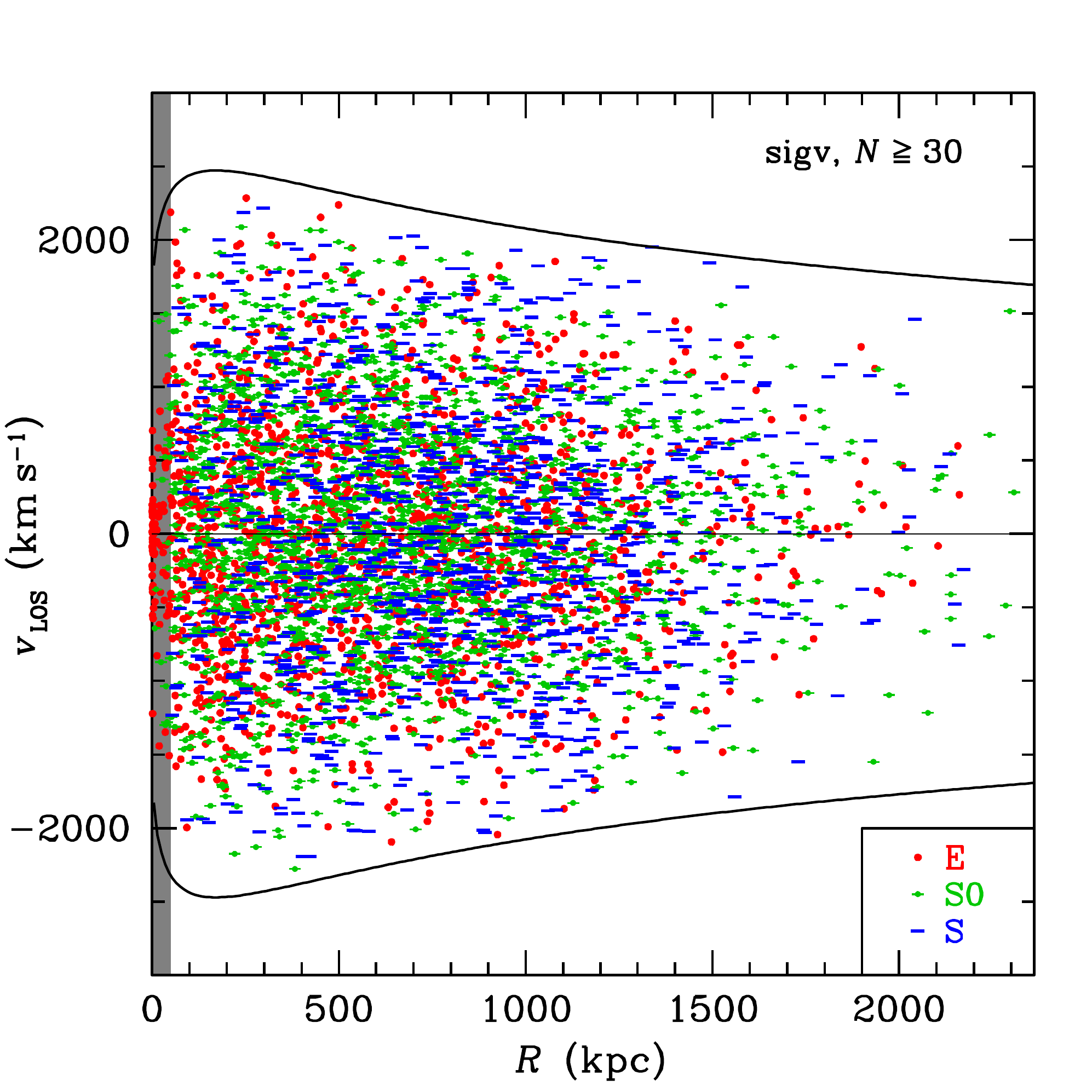}
  \caption{Projected phase space diagram of the {\tt sigv} stacked
    cluster. Each symbol is a galaxy, with shapes and colors provided in the legend. The
    maximum projected radius corresponds to our maximum allowed value of
    $1.35\,r_{200} \simeq r_{100}$.     The typical velocity errors are
    $53 \, \rm km \, s^{-1}$. 
The \emph{grey shaded region} denotes the
inner projected radii that are not considered in the MAMPOSSt analysis.
The \emph{curves} indicate the $\pm2.7\,\sigma_{\rm LOS}(R)$ conditions obtained
from the Clean algorithm (Sect.~\ref{sec:sample}).
  \label{fig:PPS}}
\end{figure}

Finally, MAMPOSSt can jointly analyze the positions in PPS of several independent tracers, such as
elliptical (E), lenticular (S) and spiral+irregular (S) galaxies.
Since the three populations of E, S0 and S galaxies move in the same
gravitational potential, but have different spatial and velocity
distributions (see Fig.~\ref{fig:PPS}), making a joint analysis of the three
populations, by allowing a different $\beta(r)$ for each of them, results in
a more stringent constraint on the remaining parameters for $M(r)$.
The joint likelihood is the product of those from each of the tracers, i.e. 
$\ln{\cal L} = \ln{\cal L}_{\rm E} + \ln{\cal L}_{\rm S0} + \ln{\cal L}_{\rm
  S}$.
In their comparison of mass modeling methods on mock clusters from a
semi-analytical model,
\cite{Old+15} found MAMPOSSt to
perform slightly better on the measure of the virial mass when jointly
analyzing red vs. blue tracers instead of grouping them together.
\citet{Biviano&Poggianti09} adopted this approach in their analysis of two sets of clusters.  

\begin{table*}[ht]
\caption{Stacked clusters}
\label{tab:numbers}
\begin{center}
\tabcolsep 3pt
\begin{tabular}{lccccccccrrrrrrrr}
\hline\hline
Method & $N_{\rm min}$ & $r_{200}$ & $\log M_{200}$
& $\log r_{\rm E}$ &$\log r_{\rm S0}$ &$\log r_{\rm S}$ &
\multicolumn{4}{c}{number (all)} & &  \multicolumn{4}{c}{number ($R_{\rm
    min}\leftrightarrow R_{\rm max}$)} \\
\cline{8-11}
\cline{13-16}
& & (kpc) & ($\rm M_\odot$) & (kpc) & (kpc) & (kpc)  &
total & \multicolumn{1}{c}{E} & S0 & \multicolumn{1}{c}{S} & & total &
\multicolumn{1}{c}{E} & S0 & \multicolumn{1}{c}{S}\\
(1) & (2) & (3) & (4) & (5) & (6) & (7) & (8) & (9) & (10) & (11) && (12) &
(13) & (14) & (15) \\
\hline
{\tt sigv} & 30 & 1749 & 14.79 & 2.690$\pm$0.097 & 2.813$\pm$0.107 & 3.314$\pm$ 0.105 & 4816 & 1701 & 1871 & 1244 & &  4682 & 1607 & 1845 & 1230 \\
{\tt sigv} & 81 & 1846 & 14.86 & 2.708$\pm$0.128 & 2.813$\pm$0.087 &
3.322$\pm$0.122 & 3361 & 1164 & 1334 &  863 & &  3289 & 1109 & 1321 &  859 \\
{\tt Num } & 30 & 1629 & 14.69 & 2.690$\pm$0.106 & 2.799$\pm$0.103 &
3.270$\pm$0.103 & 4730 & 1676 & 1838 & 1216 & &  4593 & 1583 & 1807 & 1203 \\
{\tt tempX  } & 30 & 1710 & 14.76 & 2.602$\pm$0.152 & 2.699$\pm$0.078 &
3.369$\pm$0.126 & 3607 & 1264 & 1441 &  902 & &  3523 & 1198 & 1426 &  899 \\
\hline
\end{tabular}
\end{center}

\noindent Notes: the columns are as follows.
(1): stacking method;
(2): minimum number of galaxies per cluster;
(3): $r_{200}$ obtained from Clean;
(4): $\log M_{200}$ corresponding to $r_{200}$;
(5--7): scale radii of the spatial
distributions of elliptical (5), lenticular (6) and spiral (7)
galaxies, determined from fits of the photometric data;
(8--11): total number of galaxies in sample;
(12--15): number of galaxies restricted to the projected radii analyzed by MAMPOSSt.
\end{table*}

\section{Data preparation}
\label{sec:data}
\subsection{Sample and interloper removal}
\label{sec:sample}

In Paper~I, we applied the substructure test of \cite{Dressler&Shectman88} to
our initial sample of 73 WINGS clusters containing over 10\,000 galaxies.
This test led to 15 irregular clusters (at 95 per cent confidence), leaving 58 regular clusters.
Irregular clusters  violate the condition of smooth gravitational potential
upon which  mass-orbit
modeling techniques are based.
\cite*{Foex+17} found that discarding  clusters identified as irregular with
the \cite{Dressler&Shectman88} statistic  leads to a much
better match between MAMPOSSt and X-ray based masses.
We thus discard the irregular clusters and
only consider
the 58 regular clusters.

The median apparent magnitude of galaxies in our spectroscopic sample is
$V=17.7$ (16.8--18.9 quartiles), 
which translates to an absolute magnitude $M_V = -19.2$ at
the median
cluster redshift $z=0.054$, i.e. a luminosity satisfying
$\log (L/{\rm L_\odot}) = 9.6$.
Assuming mass-to-light ratios of
$M/L_V = 3$ (ellipticals, from fig. 6 of \citealp{Auger+10}, who adopted a
\citealp{Chabrier03} initial mass function) and 2 (spirals), our median stellar mass is
$\log (M_{\rm stars}/{\rm M_\odot}) = 10.0$ for spirals and 10.4 for ellipticals.

The galaxy morphologies were determined with the MORPHOT automatic tool
\citep{Fasano+12}, and following Paper~I, we assigned galaxies to classes E
(ellipticals, $T_{\rm M} \leq -4$),
S0 (lenticulars, $-4 < T_{\rm M} \leq 0$),
and S (spirals, $T_{\rm M}>0$).

Following Paper~I, we assume that the clusters are centered on their
\emph{Brightest Cluster Galaxy} (BCG), defined within $0.5 \,r_{200}$.
We run the Clean algorithm (MBB13) to remove interlopers in LOS velocity
$v_{\rm LOS} = c_{\rm light}\,(z-\overline z)/(1+\overline z)$, where $\overline z$ is the
\emph{median} cluster redshift (not that of the BCG) and $c_{\rm light}$ is
the speed of light.

Clean
begins by searching for significant gaps
in the distribution  of LOS velocities using the gapper technique of
\cite{Wainer&Thissen76} with gapper parameter (not a concentration) $C=4$ \citep{Girardi+96}.
Clean then iterates on the membership defined by the criterion
$|v-{\rm median}(v)| < 2.7\,\sigma_{\rm LOS}^{\rm NFW}(R)$, where
the factor 2.7 was found to optimally recover the LOS velocity dispersion
profile of pure NFW models \citep*{Mamon+10}.
The term $\sigma_{\rm LOS}^{\rm NFW}(R)$ requires the knowledge of the scale
radius $r_{-2}$ and the mass within, $M(r_{-2})$, or equivalently the virial
radius and the \emph{concentration} $c_{200} = r_{200}/r_{-2}$.
Clean estimates the virial radius from 
the aperture velocity dispersion, assuming 
an NFW model
and the \cite{Mamon&Lokas05b} 
velocity anisotropy profile that goes from isotropic in the inner regions to
somewhat radial in the outer regions \citep{Mamon+10} with a transition
radius equal to the NFW scale radius.
On first pass, the aperture velocity dispersion is measured by the robust
median absolute deviation technique (e.g., \citealp*{Beers+90}) and the
concentration of the NFW model is taken as $c=4$, typical of rich clusters. 
On subsequent passes, the aperture velocity dispersion is measured using the
biweight estimator (e.g., \citeauthor{Beers+90})
and the concentration is taken from the concentration-mass
relation that \cite*{Maccio+08} fit to the haloes of dissipationless
cosmological $N$-body simulations.

We then restrict our cluster sample to the 54 clusters with at least $N_{\rm m}=30$
members within $R_{200}$ (median velocity dispersion $763 \, \rm km \, s^{-1}$).

\subsection{Stacking}

We stack the clusters into a pseudo-cluster
by rescaling the projected radii  $R_{i,j}$ of galaxies in cluster $j$ to the
mass-weighted average cluster with
$R_{i,j}' = R_{i,j} \left\langle r_{200}\right\rangle/r_{200,j}$ and the
LOS velocities 
$v_{i,j}$ to
$v_{i,j}'= v_{i,j} \left\langle v_{200}\right\rangle/v_{200,j}$
(see footnote~\ref{fn:rdelta}).
For each cluster, we estimate $r_{200}$ in three different manners:
\begin{enumerate}
\item A velocity dispersion based estimator, {\tt sigv}, obtained from the
  Clean algorithm (MBB13).
\item A richness based estimator called {\tt Num} (Mamon et al. in prep., see
  \citealp{Old+14}), which performs a linear fit between log richness and log
  virial radius. {\tt Num} 
  performed the best
  among over 20 algorithms in recovering the value of $M_{200}$, hence that
  of $r_{200}$ \citep{Old+14,Old+15}.
\item An X-ray temperature based estimator derived from the mass-temperature
  relation of \cite*{Arnaud+05}, which we call {\tt tempX}.
  But this is limited to the 40 regular clusters with
  observed X-ray temperatures.
\end{enumerate}

We then apply the Clean procedure one last time to each of these 3 stacks to
remove yet undetected interlopers.
We finally discard galaxies with projected radii beyond
$r_{100} \simeq 1.35\,r_{200}$
of our
stacked clusters, where we adopted the Clean values of $r_{200}$ found in Paper~I.
Indeed, $r_{100}$ is the theoretical virial radius, which is thought to
be the maximum physical radius where dynamical equilibrium is achieved
(i.e. no net radial velocities). Moreover, the Jeans equation is not valid
beyond $r = 2\,r_{100}$ \citep{Falco+13},
so the limiting \emph{projected} radius
must satisfy $R < 2\,r_{100}$, hence our conservative choice of
$r_{100}$.

This leaves us with up to nearly 5000 galaxies for our 3 stacks.
We also exclude the very inner region since it is dominated by the internal
dynamics of the BCG, rather than by the overall cluster
(see, e.g., \citealp{Biviano&Salucci06}).
Using a minimum projected radius of 50 kpc (roughly $0.03\,r_{200}$),
leaves us now with a total of up to 4682 galaxies
(for the {\tt sigv} stack), as displayed in Table~\ref{tab:numbers}.

While, according to the mass challenge of
\cite{Old+15}, {\tt Num} recovers $M_{200}$ (hence $r_{200}$) with much less scatter
than Clean ({\tt sigv}), it has the drawback that the recovered log mass
varies as roughly $0.5\,\log M_{\rm true}$, thus leading to positive
(negative) bias for clusters of low (high) mass.
It thus seems preferable to avoid this non-unity
recovered vs. true log mass slope of {\tt Num}, which may bias our results.
We
thus adopt {\tt sigv} as our main stacking method, but will compare in
Sect.~\ref{sec:discuss} the 
{\tt sigv} results with those from the {\tt Num} and {\tt tempX} stacking methods. 

\subsection{Quick look at the stacked data}
Figure~\ref{fig:PPS} displays the projected phase space diagram, highlighting
the distributions of galaxies of different morphological types.
\begin{figure}[ht]
  \centering
  \includegraphics[width=\hsize]{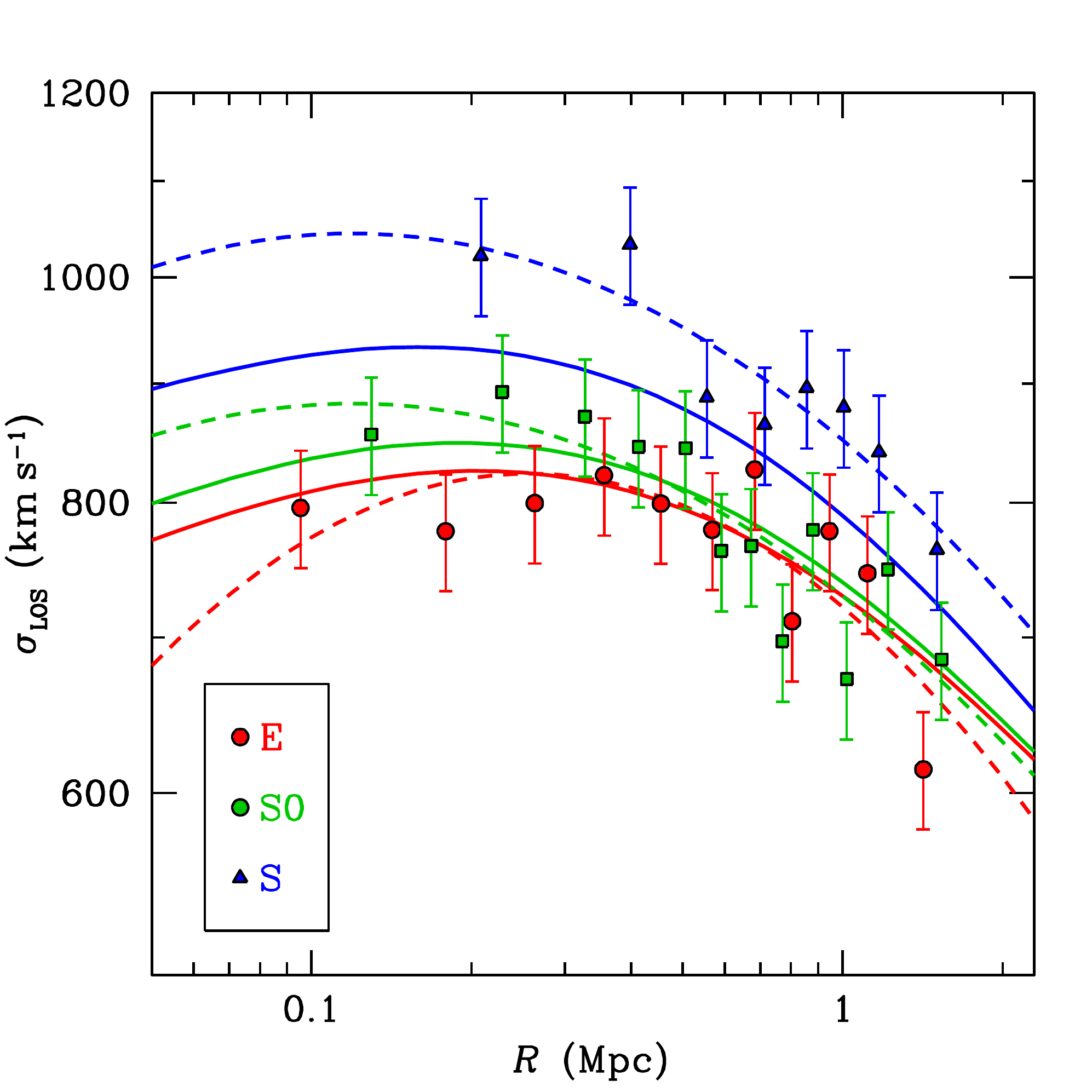}
  \caption{Line-of-sight velocity dispersion profiles for the elliptical
    (\emph{red}), lenticular (\emph{green}), and spiral (\emph{blue})
    galaxies in the {\tt sigv} stacked cluster. Radial bins of
    $\simeq$ 150
    galaxies were used. Error bars are $\sigma_{\rm LOS}/\sqrt{2(N-1)}$ (expected for
    Gaussian LOS velocity distributions). 
    The \emph{solid curves} are predictions assuming the tracer number density profiles
    given in Paper~I, isotropic velocities, for an NFW total mass profile of
    concentration $c_{200} = 4$, while the \emph{dashed curves} are 2nd-degree
    polynomial fits in log-log space.}
\label{fig:sigmalos}
\end{figure}
Figure~\ref{fig:sigmalos} shows the LOS velocity dispersion
profiles for the 3 morphological types of galaxies. A major part of the
differences in these velocity dispersion profiles arises from the different
scale radii of the 3 number density profiles. The solid curves in
Fig.~\ref{fig:sigmalos} show the predicted LOS velocity dispersion profiles
for an NFW model, with `virial' radius and tracer radii adopted from Table~2 of
Paper~I, also assuming isotropic orbits and a mass concentration of $c_{200} = 4$.
The curves match well the observed velocity dispersions for elliptical and S0
galaxies, suggesting that these early types do not depart much from velocity
isotropy. The sharp reader may notice that at projected radii $R < 330\,\rm
kpc$, the 3 S0 galaxy LOS velocity dispersions are all above the isotropic
prediction, while the 3 elliptical LOS velocity dispersions are all below the
isotropic prediction. This suggests that the inner anisotropies of S0s may be
somewhat radial while those of ellipticals may be somewhat tangential.

On the other hand, the LOS velocity dispersions of spiral galaxies are
significantly higher
than expected from the isotropic model. This may be the signature of an
anisotropic velocities of the spiral population. Alternatively, this may
signify that the spiral population is not in dynamical equilibrium. We will
discuss this possibility in Sect.~\ref{sec:betainS}.

\section{Practical implementation of MAMPOSSt}
\label{sec:mpo_practical}

We now present the practical implementation of the version of MAMPOSSt that
we used here,\footnote{https://gitlab.com/gmamon/MAMPOSSt} starting with our
adopted radial profiles for number density, mass and anisotropy.

\subsection{Tracer number density profiles}
MAMPOSSt allows for the density profiles of the observed tracers to have up to 2 parameters (scale and shape
--- no normalization is required for tracers with negligible
mass). 
In this work, we assume 1-parameter NFW number density profiles, following the fits to the photometric
data performed in Paper~I.
We express the NFW number density profile as
\begin{eqnarray} 
  \nu(r) &=& {N(r_\nu) \over 4\pi r_\nu^3}\,\widetilde \nu
  \left({r\over r_\nu} \right)
  \label{nuofr}\\
  \widetilde \nu(x) &=& {x^{-1}\,(x+1)^{-2}\over \ln 2 -
    1/2} \ ,
  \label{nutildeNFW}
\end{eqnarray} 
where
$r_\nu$ is the \emph{tracer scale radius}, defined as
the radius of logarithmic slope --2, which matches the
usual scale radius for the NFW model (but not for the other models presented below).
The NFW surface density profile is expressed as (see \citealp{Bartelmann96}
and \citealp{Lokas&Mamon01} for similar formulations)
\begin{eqnarray} 
  \Sigma(R) &=& {N_{\rm p}(r_\nu)\over \pi r_\nu^2}\,
  \widetilde \Sigma(X) \nonumber \\
  &=& {1/(x^2-1)-C^{-1}(1/X) / |X^2-1|^{3/2}\over 2\ln 2
    - 1} \ ,
  \label{SigmaofR}
\end{eqnarray} 
where
\begin{equation}
  C^{-1}(x) = \left\{ \begin{array}{ll}
    \cos^{-1} x & (x<1) \ , \\
    \cosh^{-1} x & (x>1) \ ,
  \end{array}
  \right.
  \label{Cinv}
\end{equation}
and $\widetilde \Sigma(1) = (1/3)/(2\ln2-1)$.

We adopt the values  of the decimal logarithm of
the NFW scale radii of the stacked clusters, obtained from fits
of a (projected) NFW
cluster model plus constant background to the 
photometric data of the stacked clusters. These values (and their
uncertainties) are provided in
Table~\ref{tab:numbers} for all three stacks, 
for galaxy morphological types E
(ellipticals), S0 (lenticulars) and S (spirals).
We adapted MAMPOSSt to take priors on these log scale radii assuming a
Gaussian distribution centered on the value and with a dispersion equal to
the uncertainty on log scale radius, and truncated at $3\,\sigma$.

\subsection{Mass profile}
In MAMPOSSt, the total mass profile is generally specified by a dark
component,
possibly massive tracer components, and a possible central black hole.
In the present work, the tracer components are generally massless and the
black hole mass is assumed negligible (fixed to zero), hence the `dark'
component generally refers to the total mass, and we call it the `cluster'
component, which refers to the total mass unless we include a massive tracer
for the BCG.
The cluster mass model can have up to 3 parameters (normalization, scale, and
possible shape).

We express the mass profile in terms of the mass at the scale radius times a
dimensionless function of radius over scale radius:
\begin{equation}
  M(r) = M(r_{\rho})\,\widetilde M \left ({r\over r_{\rho}}\right) \ ,
\end{equation}
where $r_{\rho}$ is the \emph{mass density radius}, defined as that where the
logarithmic slope of the mass density
profile is --2.
In virial units, one would then have $M(r)/M_{\rm vir}=\widetilde M(c
r/r_{\rm vir})/\widetilde M(c)$, where $c=r_{\rm vir}/r_{\rho}$ is the \emph{concentration parameter}.
We consider the following dimensionless mass profiles (noting that
$\widetilde M(1) = 1$ by definition):
\begin{description}
  \itemsep 0.5\baselineskip
\item [{\bf NFW:}] the usual NFW model, whose density profile is given in
  equations~(\ref{nuofr}) and (\ref{nutildeNFW}),
 for which the inner and outer logarithmic slopes are respectively --1 and 
  --3:
  \begin{equation}
    \widetilde M(x) = {\ln(x+1)-x/(x+1) \over \ln 2-1/2} \ .
\label{MNFWtilde}
  \end{equation}

\item [{\bf cNFW:}] the cored NFW model (generalized NFW with zero inner slope),
  \begin{equation}
    \widetilde M(x) = {\ln(2x+1)\,(3x+1)/(2x+1)^2\over \ln 3 - 8/9} \ .
  \end{equation}
In the cored-NFW model, $r_{\rho}$ is equal to twice the scale radius.

\item [{\bf gNFW:}] the  generalized NFW model, whose density profile is
$\rho(r) \propto r^\gamma\,(r+r_{\rm s})^{-3-\gamma}$, for which the inner and outer
  logarithmic slopes are respectively $\gamma$ (e.g. --1 for NFW) and --3 again:
\begin{eqnarray} 
  \widetilde M(x)&=&
             {I_{-(\gamma+2)x}\left[\gamma+3,-(\gamma+2)\right]\over
                            I_{-(\gamma+2)}\left[\gamma+3,-(\gamma+2)\right]}
             \ ,
  \nonumber \\
 &=&
        x^{\gamma+3}\,{^2F_1\left[\gamma+3,\gamma+3,\gamma+4,-(\gamma+2)x\right]\over
          ^2F_1\left[\gamma+3,\gamma+3,\gamma+4,-(\gamma+2)\right]} \ ,
\label{MgNFW}
\end{eqnarray}
where $I_x(a,b)$ is the regularized incomplete  beta function, while
$^2F_1(a,b,c,x)$ is the ordinary (Gaussian) hypergeometric function.
In the generalized NFW model, $r_{\rho} = (\gamma+2)\,r_{\rm s}$.

\item [{\bf Hernquist:}] the \cite{Hernquist90}  model with density profile
  $\rho(r) \propto r^{-1}\,(r+r_{\rm s})^{-3}$, for which the inner and outer
  logarithmic slopes are respectively  --1 and --4:
  \begin{equation}
    \widetilde M(x) = 9\,\left({x\over x+2}\right)^2 \ .
  \end{equation}
  In the Hernquist model, $r_{\rho} = r_{\rm s}/2$.

\item [{\bf Einasto:}] the \cite{Einasto65} model was introduced for stellar
  distributions in the Milky Way, but was found by \cite{Navarro+04} and
  confirmed by many to fit even better the density profiles of haloes than
  the NFW model:
  \begin{equation}
       \widetilde M(x) = {\gamma(3n,2n\,x^{1/n})\over \gamma(3n,2n)} \ .
  \end{equation}
  
  \end{description} 

\subsection{Velocity anisotropy profile}
MAMPOSSt also allows for a wide variety of velocity anisotropy profiles for
each of the tracer components,
based on up to 3 parameters (inner anisotropy, outer anisotropy and
transition radius).
We consider the following models for the anisotropy profile, $\beta(r)$:
\begin{description}
  \itemsep 5pt

	\item [{\bf T}:] the \citet{Tiret+07} profile,
	\begin{equation}
	\beta_{\rm
          T}(r)=\beta_{0}+\left(\beta_{\infty}-\beta_{0}\right)\,\frac{r}{r+r_\beta}
        \ ,
        \label{betaTiret}
	\end{equation}
	\item [{\bf T}$_{\mathbf{0}}$:] the same as the Tiret profile, but with
          $\beta_0 = 0$.


	\item [{\bf gOM}:] the generalized Osipkov-Merritt model \citep{Osipkov79,Merritt85},
	\begin{equation}
	\beta_{\rm
          gOM}(r)=\beta_{0}+\left(\beta_{\infty}-\beta_{0}\right)\,\frac{r^2}{r^2+r_\beta^2}\ ,
        \label{betaOM}
	\end{equation}
\end{description}
where the usual Osipkov-Merritt anisotropy is recovered for $\beta_0=0$ and
$\beta_\infty=1$.
Note that the usual constant anisotropy
model can be retrieved as a singular case of the above T and gOM
models, when assuming $\beta_{0}=\beta_{\infty}$.

We generally assume that the anisotropy scale radius, $r_\beta$, matches the radius of
slope --2, $r_\nu$, of the tracer in consideration, which we call the
\emph{Tied Anisotropy Number Density} (TAND) assumption. This is indeed the case for
halos of dark matter particles \citep{Mamon+10}.

\subsection{Assumptions}
Our modeling assumes spherical symmetry for the visible and mass components,
neglecting any possible rotation or other non-radial streaming motions. With
our choice of regular clusters, we are in a better position to assume that
the galaxies are non-interacting tracers of the gravitational potential.

We fit our analytical
models to the total mass
profile, thus neglecting the contributions of the galaxy and gas components
to the cluster.
The three galaxy populations are considered as massless independent
tracers of the same potential while performing a joint likelihood analysis
with MAMPOSSt.  
This is the first time that such a joint analysis is performed considering
galaxy morphological types, and in particular the first study of the orbits of
S0 galaxies in clusters.

\subsection{Maximum physical radius in integrals}
We integrate the inversion of the JE (eq.~[\ref{sigmar}]) out to 120 Mpc and
the LOS integral of equation~(\ref{pofvz}) out to 40 Mpc. The former is integrated three times
further than the second, to ensure that the radial velocity dispersion is
obtained with sufficient accuracy for the LOS integral.

\subsection{Free parameters}

The following parameters can be set free in MAMPOSSt:
\begin{itemize}
\item logarithm of the mass normalization ($\log M_{200}$ or equivalently $\log r_{200}$);
\item logarithm of the mass scale radius (or equivalently of the
  concentration $c_{200} = r_{200}/r_\rho$);
\item inner slope of the mass density profile (from $-1.99$ to 0);
\item logarithms of the tracer scale radii (for each of the 3 galaxy types;
\item inner ($r=0$) and outer ($r\to\infty$) symmetrized velocity anisotropies
  (for each of the 3 galaxy types)
\begin{equation}
  \beta_{\rm sym} = 2\,{\left\langle v_r^2\right\rangle-\sigma_\theta^2 \over
    \left\langle v_r^2\right\rangle+\sigma_\theta^2} = {\beta \over
    1-\beta/2} \ ,
\label{betasym}
\end{equation}
which can be as low as $-2$ for circular orbits and as high as +2 for radial orbits, and
where $\beta_{\rm sym} \to \beta$ for $|\beta| \ll 1$ (we allow --1.8 to
1.8);
\item anisotropy transition radius $r_\beta$ (see eqs. [\ref{betaTiret}] and
  [\ref{betaOM}]) for each of the 3 galaxy types (unless we assume TAND).
\end{itemize}
This amounts to a maximum of 15 free parameters.
We also allow ourselves an extra mass component, treated as a massive tracer,
potentially adding 2 extra free parameters (but we then force an NFW cluster
mass model, thus subtracting the free inner slope, for a net single extra
parameter).

Given our lack of knowledge on the parameters,
we adopt flat priors for all parameters, except for the log scale radii of the E, S0,
and S tracer
density profiles, determined (externally) from the photometric data (Paper~I), for
which we adopt Gaussian priors (using our mean values of $\log r_\nu$ and
their uncertainties).

In our basic set of \numruns\ MAMPOSSt runs, we assumed that the cluster mass concentration
follows the ``$\Lambda$CDM'' relation found by \cite{Dutton&Maccio14} for
massive halos in dissipationless cosmological
$N$-body simulations in a Planck cosmology:
\begin{equation}
  \log c_{200} = 2.13 - 0.10\,\log
  \left({M_{200}\over \rm M_{\odot}}\right) \ .
  \label{cM}
\end{equation}
We assumed a Gaussian prior on this relation with 0.1 dex uncertainty.
We also performed extra MAMPOSSt runs with free cluster mass concentration, with different minimum and maximum allowed projected radii,
with different minimum number of galaxies per cluster in building the stacks, and
for the different stacks.

\subsection{Marginal distributions (MCMC)}

We determine the marginal distributions of the $k$ free parameters using the
Markov Chain Monte Carlo (MCMC) technique.
We use the public {\sc CosmoMC} code \citep{Lewis&Bridle02}
in the generic mode. {\sc CosmoMC} uses the Metropolis-Hastings algorithm to
probe the  distribution of posteriors (likelihoods times priors) in the
$k$-dimensional parameter space.
A chain of points in this space is initialized by selecting a point in $k$-dimensional parameter
space from a $k$-dimensional Gaussian centered on the $k$-dimensional
hyper-cube, with standard deviations
$\sigma_k = [{\rm Max}(\theta_k)-{\rm Min}(\theta_k)]/5$.
We then advance each chain, by moving from position
$\vec{\theta_{\rm old}}$ to $\vec{\theta_{\rm new}}$ using a
$k$-dimensional Gaussian \emph{proposal density} function,
with standard deviations equal $\sigma_k/2$, i.e. one-tenth of the allowed range of parameters.
Because this
proposal density function is symmetric between two consecutive steps,
the Metropolis-Hastings algorithm advances the position $\vec{\theta_{\rm
    old}}$ in $k$-dimensional
parameter space to 
with probability
\begin{equation}
  P = {\rm Min} \left[{p\left(\vec{\theta_{\rm new}}\right)\over
      p\left(\vec{\theta_{\rm old}}\right)},1 \right]
  \ ,
\end{equation}
where the $p$s are the posteriors.
In other words, the point is 
kept if the
posterior is greater than for the previous point. If the posterior of the
new point is lower, a random number is drawn and the new point is kept if the
ratio of posteriors (obviously between 0 and 1) is lower than the random
number. Otherwise, the new point is discarded and the \emph{walker} remains
on the previous point, whose weight is increased by unity.

We run 6 chains in parallel on an 8-core computer, 
each one run for $10\,000\,k$ steps in parallel. While the proposal density
function is initially circular in $k$ dimensions, {\sc CosmoMC} 
periodically updates it to the (elliptical-shaped) parameter covariances of
the previous elements of the chain.

We then discard the first $2000\,k$
steps, where the posterior distribution is dependent on the starting
points of the chains (the \emph{burn-in} phase).

We compute the radial profiles of mass density and of the velocity anisotropy of the
3 galaxy types from the marginal distributions of the free parameters (after
discarding the burn-in phase).

\section{Results}
\label{sec:results}

\begin{table*}[htb]
  \caption{MAMPOSSt runs for the {\tt sigv} stack, all with the
    concentration-mass relation of eq.~[\ref{cM}]}
  \begin{center}
\tabcolsep 4pt
\begin{tabular}{r@{\hspace{5mm}}lcccccccccccccr@{\hspace{2.5mm}}cc}
      \hline
      \hline
\multicolumn{1}{c}{model} & 
\multicolumn{2}{c}{mass model}
& \multicolumn{1}{c}{anis.} &
      \multicolumn{3}{c}{inner anisotropy}
      & & \multicolumn{3}{c}{outer anisotropy} & TAND
& $R^{-1}$ &  $-\ln {\cal L}_{\rm  MLE}$ &\multicolumn{1}{c}{\#}  & AIC & BIC       \\
\cline{2-3}
\cline{5-7}
\cline{9-11}
& cluster & BCG &
\multicolumn{1}{c}{model} & E & S0 & S & & E & S0 & S & & & & \multicolumn{1}{c}{free} \\
\multicolumn{1}{c}{(1)} & \multicolumn{1}{c}{(2)} & (3) & (4) & (5) & (6)
&\multicolumn{1}{c}{(7)} & & \multicolumn{1}{c}{(8)} & \multicolumn{1}{c}{(9)} & (10) & (11) & (12) & (13) & \multicolumn{1}{c}{(14)} &
      (15) & (16)  \\
\hline
{\bf \gNFWgTTAL} & gNFW & -- & T & F & F & F & & F & F & F & Y & 0.004 & {\bf 33526.28}& 12 & 67076.62 & 67154.32 \\ 
{\bf \BCGgT} & NFW & NFW & T & F & F & F & & F & F & F & N & \reditalics{0.065} & 33526.47 & 16 & 67085.05 & 67188.62 \\ 
{\bf \BCGgTTAL} & NFW & NFW & T & F & F & F & & F & F & F & Y & 0.012 & 33526.50 & 13 & 67079.08 & 67163.24 \\ 
{\bf \gNFWTLSgTTAL} & gNFW & -- & T & F & 0 & 0 & & F & F & F & Y & 0.005 & 33526.50 & 10 & \greenitalics{67073.05} & 67137.80 \\ 
{\bf \BCGggOMTAL} & NFW & NFW & gOM & F & F & F & & F & F & F & Y & 0.007 & 33526.55 & 13 & 67079.18 & 67163.34 \\ 
{\bf \gNFWggOMTAL} & gNFW & -- & gOM & F & F & F & & F & F & F & Y & 0.011 & 33526.79 & 12 & 67077.65 & 67155.34 \\ 
{\bf \gNFWgT} & gNFW & -- & T & F & F & F & & F & F & F & N & \reditalics{0.040} & 33526.91 & 15 & 67083.92 & 67181.01 \\ 
{\bf \BCGPSfourgTTAL} & NFW & PS4 & T & F & F & F & & F & F & F & Y & 0.005 & 33528.33 & 13 & 67082.74 & 67166.90 \\ 
{\bf \NFWgTTAL} & NFW & -- & T & F & F & F & & F & F & F & Y & 0.002 & \blueitalics{33528.36} & 11 & \blueitalics{67078.77} & 67150.00 \\ 
{\bf \NFWTLSgTTAL} & NFW & -- & T & F & 0 & 0 & & F & F & F & Y & 0.001 & 33528.41 & 9 & 67074.86 & 67133.14 \\ 
{\bf \gNFWTTAL} & gNFW & -- & T & 0 & 0 & 0 & & F & F & F & Y & 0.005 & 33528.41 & 9 & 67074.86 & 67133.14 \\ 
{\bf \NFWgT} & NFW & -- & T & F & F & F & & F & F & F & N & \reditalics{0.031} & 33528.50 & 14 & 67085.09 & 67175.72 \\ 
{\bf \NFWTTAL} & NFW & -- & T & 0 & 0 & 0 & & F & F & F & Y & 0.002 & 33528.54 & 8 & 67073.11 & 67124.92 \\ 
{\bf \BCGPSfourgT} & NFW & PS4 & T & F & F & F & & F & F & F & N & \reditalics{0.043} & 33528.55 & 16 & 67089.21 & 67192.77 \\ 
{\bf \NFWggOMTAL} & NFW & -- & gOM & F & F & F & & F & F & F & Y & 0.003 & 33528.92 & 11 & 67079.90 & 67151.12 \\ 
{\bf \NFWisoETTAL} & NFW & -- & T & 0 & 0 & 0 & & 0 & F & F & Y & 0.003 & 33529.20 & 7 & {\bf 67072.42}& 67117.76 \\ 
{\bf \EisoELTTAL} & E (free $n$) & -- & T & 0 & 0 & 0 && 0 & 0 & F & Y & 0.002& 33529.74 & 6 & 67073.50 & 67118.84 \\
{\bf \gNFWisoELT} & gNFW & -- & T & 0 & 0 & 0 & & 0 & 0 & F & N & 0.015 & 33529.90 & 8 & 67075.83 & 67127.64 \\ 
{\bf \gNFWisoELgTTAL} & gNFW & -- & T & 0 & 0 & F & & 0 & 0 & F & Y & 0.003 & 33530.03 & 8 & 67076.09 & 67127.90 \\ 
{\bf \NFWisoELT} & NFW & -- & T & 0 & 0 & 0 & & 0 & 0 & F & N & 0.007 & 33530.23 & 7 & 67074.48 & 67119.82 \\ 
{\bf \gNFWisoELTTAL} & gNFW & -- & T & 0 & 0 & 0 & & 0 & 0 & F & Y & 0.002 & 33530.27 & 7 & 67074.56 & \greenitalics{67119.90} \\ 
{\bf \NFWisoELgTTAL} & NFW & -- & T & 0 & 0 & F & & 0 & 0 & F & Y & 0.002 & 33530.35 & 7 & 67074.72 & 67120.06 \\
{\bf \EsixisoELTTAL} & E ($n$=6) & -- & T & 0 &
0& 0 && 0 & 0 & F & Y &0.002 & 33530.50 & 6 & 67073.02 & {\bf 67111.88} \\
{\bf \NFWisoELTTAL} & NFW & -- & T & 0 & 0 & 0 & & 0 & 0 & F & Y & 0.001 & 33530.68 & 6 & 67073.38 & \blueitalics{67112.24} \\ 
{\bf \cNFWisoELTTAL} & cNFW & -- & T & 0 & 0 & 0 & & 0 & 0 & F & Y & 0.003 & 33532.30 & 6 & 67076.62 & 67115.48 \\ 
{\bf \HisoELTTAL} & H & -- & T & 0 & 0 & 0 & & 0 & 0 & F & Y & 0.001 & 33534.44 & 6 & 67080.90 & 67119.76 \\ 
{\bf \NFWisoELgOMTAL} & NFW & -- & gOM & 0 & 0 & 0 & & 0 & 0 & F & Y & 0.002 & 33537.09 & 6 & 67086.20 & 67125.06 \\ 
{\bf \NFWiso} & NFW & -- & iso & 0 & 0 & 0 & & 0 & 0 & 0 & --- & 0.002 & 33538.52 & 5 & 67087.05 & 67119.44 \\ 
{\bf \NFWisoESTTAL} & NFW & -- & T & 0 & 0 & 0 & & 0 & F & 0 & Y & 0.001 & 33539.46 & 6 & 67090.94 & 67129.80 \\ 
{\bf \NFWisoLSTTAL} & NFW & -- & T & 0 & 0 & 0 & & F & 0 & 0 & Y & 0.002 & 33539.56 & 6 & 67091.14 & 67130.00 \\ 
\hline
\end{tabular}
  \end{center}
  \label{tab:runs}

  \noindent  Notes: The runs are listed in decreasing maximum likelihood
  (really posterior), i.e. increasing $-\ln {\cal L}_{\rm MLE}$. The columns are as follows.
  (1): run number;
  (2): cluster mass model (i.e. `E' for Einasto, `H' for Hernquist);
  (3): brightest cluster galaxy mass model (`PS4' for $n=4$ Prugniel \& Simien
  (1997), which is a good approximation to the deprojection of a S\'ersic (1968) model);
  (4): velocity anisotropy model (`iso' for isotropic,
  `T' for generalized Tiret et al. and `gOM'
  for generalized Osipkov-Merritt);
  (5-7): inner velocity anisotropy for E, S0 and S galaxies (`F' = free, 0 =
  isotropic);
  (8-10): outer velocity anisotropy ($\beta$) for E, S0 and S galaxies (`F' = free, 0 =
  isotropic);
  (11): velocity anisotropy radius tied to tracer scale radius? 
  (12): MCMC convergence criterion ($R^{-1}<0.02$ is considered as
  properly converged, worse convergence runs are shown in \reditalics{red italics});
  (13): minus log (maximum likelihood estimate, which really is a maximum posterior);
  (14): number of free parameters;
  (15): corrected Akaike Information Criterion;
  (16): Bayes Information Criterion.
  The best values for $-\ln {\cal L}_{\rm MLE}$, AIC and BIC are highlighted in {\bf bold}, while
  \blueitalics{blue italics} and \greenitalics{green italics} respectively
  highlight the best NFW and gNFW models that do not have
  an extra BCG component.
\end{table*} 
\nocite{Prugniel&Simien97}
\nocite{Sersic68}

\subsection{Model comparison}
\label{sec:modelcomp}

\subsubsection{Preamble}
Most studies employ a single set of priors and show their results. Some will
consider a few extra choices for their priors. Here, we have choices to make
on the inner slope of the cluster mass profile, an additional mass profile
for a possible BCG, fixed or free inner and anisotropy profiles for all three
components.
This led us to consider a large number of sets of priors.

Table~\ref{tab:runs} displays the
\numruns\
MAMPOSSt runs on our WINGS
clusters, stacked according to their velocity dispersions ({\tt sigv}), and sorted in
decreasing \emph{maximum likelihood estimate} (MLE), i.e. increasing  $-\ln
{\cal L}_{\rm MLE}$ (column 3).
Our values of ${\cal L}_{\rm MLE}$ are really posteriors, but are close to
likelihoods since all of our priors are flat (within a wide range), except for Gaussian
priors on the tracer and cluster mass log scale radii (roughly 0.1 dex -- see
Table~\ref{tab:numbers} --- and
exactly 0.1 dex, respectively).

We mainly considered mass priors using the NFW or gNFW models. But, we later
added 2 priors with the Einasto mass model, with either free index or fixed
to $n$\,=\,6, as found for $\Lambda$CDM haloes \citep{Navarro+04}.
We did not run more Einasto models, given
that the $n$=6 Einasto is very similar to the NFW model, while  free
index Einasto models  resemble the gNFW models of given inner slope. Indeed, we
found that the mass profile of the $n$=6 Einasto fits the NFW one to 8.5 per
cent relative precision in the range 0.135 to 13.5$\,r_{-2}$ (0.03 to
3$\,r_{200}$), equally spaced in $\log\,r$, for
typical cluster concentrations (the index $n=4.4$ provides the best fit --
4.8 per cent relative precision -- to
the NFW mass profile in this radial range). Similarly, 
the mass profiles of Einasto models with free indices (up to $n=25$) fit those of
given gNFW models to better than 8.5 per cent rms relative precision in the
same range of radii (6.1 per cent for $\gamma\geq-1.9$).

\subsubsection{Bayesian evidence methods}
Using different priors can lead to different results, so one has to be careful
in analyzing MAMPOSSt results.
The runs leading to the highest likelihood ${\cal L}_{\rm MLE}$ naturally
tend to have the largest number of free parameters (Table~\ref{tab:runs}).
But one can ask whether the
addition of extra free parameters improves the likelihood significantly, or
whether one is over-fitting the data instead.
For this, we use both the
\emph{Akaike Information Criterion} (AIC, \citealp{Akaike73}):
\begin{equation}
  {\rm AIC} = -2 \ln {\cal L}_{\rm MLE} + 2\,N_{\rm pars} 
  \label{AIC}
\end{equation}
and the
\emph{Bayes Information Criterion} (BIC, \citealp{Schwarz78}):
\begin{equation}
  {\rm BIC} = -2 \ln {\cal L}_{\rm MLE} + \ln N_{\rm data}\, N_{\rm pars} \ .
\label{BIC}
\end{equation}
Given our data sample with $N_{\rm data} = 4682$ (for the {\tt sigv} stack),
each extra parameter must 
decrease $-\ln {\cal L}_{\rm MLE}$ by 4.23 to lead to a better (lower) BIC
value, while with AIC it must decrease by only 1.

According to \cite{Kass&Rafferty95}, when a model has BIC lower than another
model by 6 (10, 2) units, one can conclude 
that there is strong (very strong, positive) evidence in favor of the former one.
Since BIC penalizes  extra parameters much more than AIC,
BIC seems preferable to AIC
when our model is built with a small number of parameters.
Equations~(\ref{AIC}) and (\ref{BIC}) lead to
$\Delta {\rm AIC} = \Delta {\rm BIC} - (\ln N_{\rm data} - 2)\,\Delta N_{\rm
  pars}$. Therefore, the
condition for strong AIC evidence for the simpler model
compared to a another one with  $\Delta N_{\rm pars}$ extra free parameters, given our
data sample sizes 
(Table~\ref{tab:numbers}), is
$\Delta{\rm AIC} > 6 - 6.45\,\Delta N_{\rm pars}$, i.e. $\Delta\rm AIC>-0.45$
compared to a more complex model with a single extra parameter.
Conversely, a more complex model is strongly favored if
$\Delta\rm BIC < -6$, leading to
$\Delta\rm AIC < -6 -6.45\,\Delta N_{\rm pars}$, i.e. $\Delta\rm AIC <
-12.45$ compared to a simpler model with one less free parameter.

\subsubsection{Preferred models}
The models with the highest likelihoods from the MCMC analysis
(lowest $-{\cal L}_{\rm MLE}$) are complex models with more free parameters.
Among our \numruns\ models, the models ranked first and fourth in likelihood
(models~\gNFWgTTAL\ and \gNFWTLSgTTAL)
all have a cluster mass profile that is steeper (i.e. gNFW with $\gamma <
-1$) than NFW at inner radii.
Note that the free Einasto model with free velocity anisotropy with TAND
(not shown in Table~\ref{tab:runs}) has an even (slightly) lower $-\ln {\cal
  L}_{\rm MLE}$ than model~\gNFWgTTAL.
In comparison, our models \BCGgT\ and \BCGgTTAL,
which have
an additional mass
component for a central BCG, also fit well as they rank second and third.

The 7-parameter model~\NFWisoETTAL\ (NFW
cluster with isotropic ellipticals and T$_0$ anisotropy with TAND for S0s
and spirals) has the lowest AIC.
The best BIC
is reached for the 6-parameter model~\EsixisoELTTAL, with $n$=6 Einasto mass density,
isotropic orbits for ellipticals and lenticulars, and T$_0$ anisotropy
with TAND for spirals.\footnote{We found an even better BICs with the
  5-parameter priors of NFW or $n$=6 Einasto mass and  velocity
  anisotropy model $\beta(r) = (1/2)\,r/(r+r_\beta)$ proposed by
  \cite{Mamon&Lokas05b}, which is a special case of the T model, with
  $\beta\to 1/2$ at large radii. But this
  model has no physical basis, for example the anisotropy at $r_{200}$ 
increases with halo mass \citep{Ascasibar&Gottlober08,Lemze+12}.}
  This model is also the 2th best
model for AIC.

\subsubsection{Rejection of mass models}
There is strong BIC evidence that
Einasto
model~\EsixisoELTTAL\
is preferable to all other mass models,
except the equivalent NFW model \NFWisoELTTAL, model~\cNFWisoELTTAL\ (cored NFW mass),
and (marginally)  model~\NFWisoETTAL\ 
(identical to model~\NFWisoELTTAL, but with free outer
anisotropy for S0 galaxies).

In particular, there is strong evidence ($\Delta \rm BIC > 7$) favoring
the $n$=6 Einasto or NFW 
cluster mass density profiles of models~\EsixisoELTTAL\ and
\NFWisoELTTAL\ compared to the
\cite{Hernquist90} mass profile (model~\HisoELTTAL, with identical velocity
anisotropy).
This conclusion is unchanged if we relax the concentration-mass relation of
equation~(\ref{cM}).

There is also strong BIC evidence against replacing the $n$=6 Einasto and NFW
cluster mass models 
with a free Einasto index model or a gNFW  (free inner slope) model, as both
models (\EisoELTTAL\ and \gNFWisoELTTAL) have higher BICs by over 7 for the
same velocity anisotropy priors. 
The evidence against a gNFW cluster mass model is even stronger ($\Delta\rm
BIC > 10$, i.e. very strong) in
moderately complex anisotropy models \gNFWTTAL\ vs. \NFWTTAL, but the
evidence against gNFW is only weak for more complex anisotropy models,
e.g. \gNFWgTTAL\ vs. \NFWgTTAL\ and \gNFWTLSgTTAL\ vs. \NFWTLSgTTAL.
And while the best AIC is reached for NFW cluster model~\NFWisoETTAL,
the 2nd best model for AIC
is gNFW
model~\gNFWTLSgTTAL,
which has $\Delta\rm AIC$ nearly --2 relative to the analogous NFW
model~\NFWTLSgTTAL.
Hence, the case against gNFW is less clear with AIC than with BIC.

Similarly, there is very strong BIC evidence against the need for a specific BCG component in
comparison to a single NFW cluster mass model, with $\Delta \rm BIC=13$
between models~\BCGgTTAL\ and \NFWgTTAL\ as well as between
models~\BCGgT\ and \NFWgT.
The posteriors and BIC values are even worse (higher) when assuming that
the BCG mass follows the stellar mass, with an
$n$=4 
\cite{Prugniel&Simien97} model (an excellent approximation to the
deprojection of the
\citealp{deVaucouleurs48} surface density model):
as seen by comparing models~\BCGPSfourgTTAL\ and \BCGgTTAL.
There is also strong evidence in favor of the gNFW cluster mass model
compared to the NFW cluster plus NFW BCG (which has an extra free parameter),
with $\Delta\rm BIC\simeq 9$ 
between models~\BCGgTTAL\ and \gNFWgTTAL\ and $\Delta\rm BIC > 7$ between
models~\BCGgT\ and \gNFWgT.
These conclusions are unaltered when using AIC in place of BIC.

\subsubsection{Rejection of velocity anisotropy models}
Moreover, according to its returned BIC values, MAMPOSSt shows no need for
complex velocity anisotropy models.
Indeed, there is strong BIC evidence that model~\NFWisoELTTAL\
with NFW mass profile and T anisotropic outer orbits for the spirals (and isotropic
orbits for E and S0 galaxies) is preferable to
a) isotropic
orbits for the spirals as well as the E and S0 galaxies (model~\NFWiso),
b) anisotropic outer orbits only for
ellipticals (model~\NFWisoLSTTAL)
or lenticulars (model~\NFWisoESTTAL),
c) anisotropic outer orbits for all three morphological types (model~\NFWTTAL),
d) anisotropic inner orbits only for spirals
(model~\NFWisoELgTTAL),
e) the generalized gOM anisotropy model with inner
isotropic orbits for the spirals (model~\NFWisoELgOMTAL).
There is also strong evidence against the
need for allowing a free anisotropy radius for the spirals (model~\NFWisoELT)
instead of TAND (and very strong evidence against free anisotropy radius
comparing the more complex anisotropy
models~\NFWgTTAL\ and \NFWgT, both with NFW cluster mass).

Similarly, compared to the lowest BIC model with gNFW mass profile
(model~\gNFWisoELTTAL, with isotropic orbits for ellipticals and S0s and
T$_0$ anisotropy for spirals),
there is strong evidence against all variations on the velocity
anisotropy, i.e.
a) allowing for outer anisotropy for E and S0 galaxies (in addition
to spirals, model~\gNFWTTAL),
b) allowing for fully anisotropic models for all morphological types
(model~\gNFWgTTAL),
and
c) freeing the transition radius of the spiral anisotropy profile
(model~\gNFWisoELT).
For more complex anisotropy models, there is no preference for the T
model compared to gOM (e.g. comparing models~\gNFWgTTAL\ and \gNFWggOMTAL\ for
gNFW,
or models~\BCGgTTAL\ and \BCGggOMTAL\ for an extra NFW BCG).
Finally, there is strong evidence that the anisotropy radius is close to the
TAND value (models~\gNFWgTTAL\ vs. \gNFWgT\ for gNFW mass, 
\BCGgTTAL\ vs. \BCGgT\ for an extra NFW BCG, and
\BCGPSfourgTTAL\ vs. \BCGPSfourgT\ for an $n=4$ S\'ersic BCG). 

AIC is more forgiving than BIC for extra parameters.
Its preferred anisotropy is with the relatively simple model~\NFWisoETTAL, with isotropic
elliptical orbits and T$_0$ anisotropy for S0s and spirals (both with
TAND).
However, the 2nd best
non-Einasto model
for AIC is model~\gNFWTLSgTTAL, where only
the inner orbits of S0s and spirals are fixed to isotropic. The 3rd best
model for AIC is model~\NFWTTAL, which is intermediate in its complexity,
with inner isotropy and free outer anisotropy for all 3 morphological types.
It also fails to distinguish between T and gOM anisotropy
(both of which involve the same number of free parameters) and also prefers
TAND.

\subsubsection{Summary of model comparison}
In summary, while AIC slightly prefers the NFW mass model over gNFW and the $n$=6
Einasto mass model over the free one, BIC strongly
rejects the more complex gNFW and free Einasto 
mass models (except for complex velocity anisotropy models).
Both AIC and BIC point to isotropic elliptical orbits and radial outer spiral
orbits, but AIC prefers radial orbits for the lenticulars, while BIC finds moderate
evidence against anisotropic outer velocities for S0s. 
Both AIC and BIC prefer the anisotropy radii to be set by TAND, and fail to
distinguish between T and gOM anisotropy.

\subsection{Best-fitting parameters}
\label{sec:bestfit}

We now focus on just a few of the MAMPOSSt runs, by considering the highest
likelihood model (model~\gNFWgTTAL), the strongest AIC evidence
(model~\NFWisoETTAL), 
the
NFW model with strongest  BIC evidence
(model~\NFWisoELTTAL), 
the
model with gNFW mass with strongest BIC evidence
(model~\gNFWisoELTTAL), and the 2-component mass model with the strongest AIC
and BIC evidences (model~\BCGgTTAL).

\begin{figure}[ht]
  \centering
    \includegraphics[width=1.01\hsize,viewport=14 0 560 570]{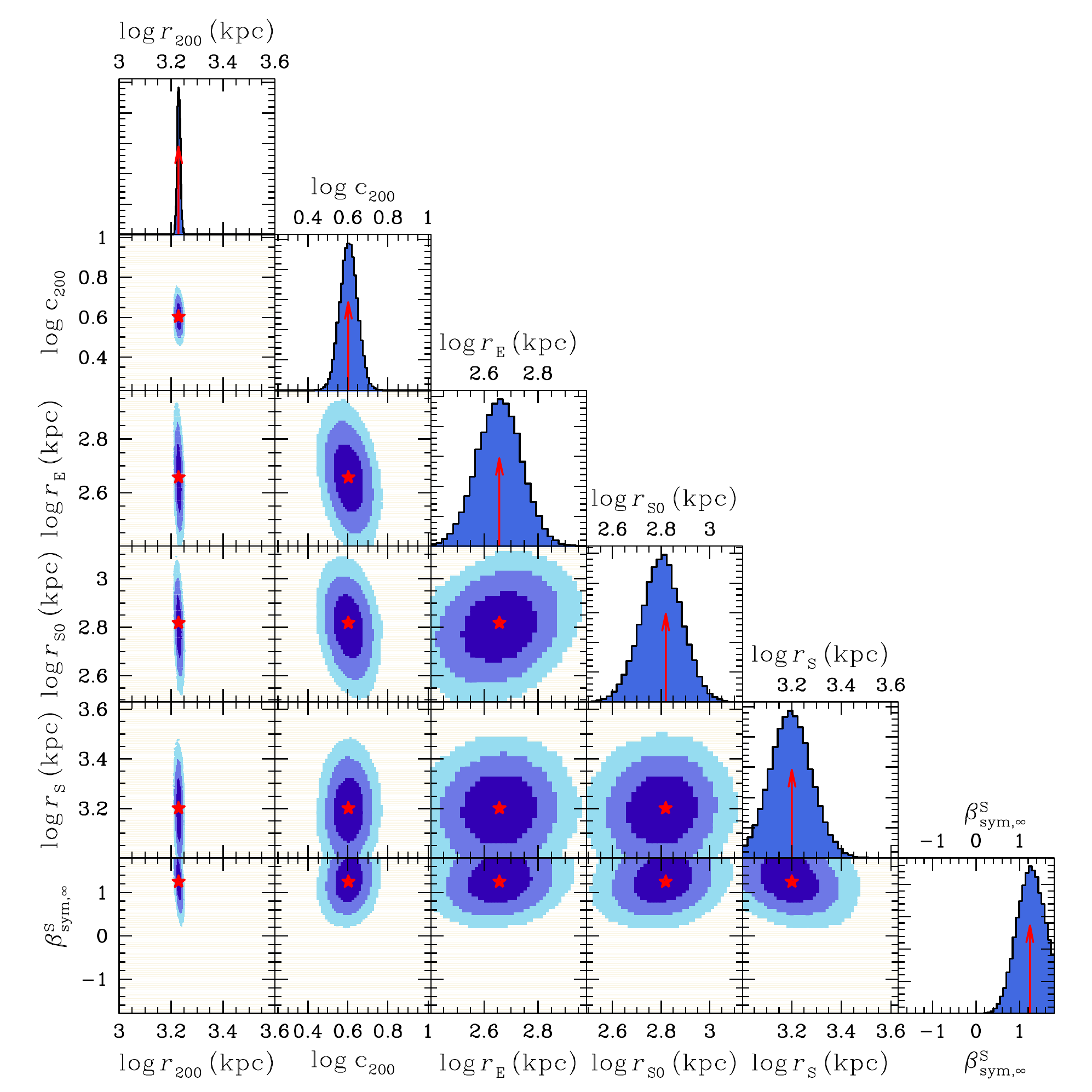}
  \caption{MAMPOSSt marginal distributions (diagonal panels) and covariances
    for model~\NFWisoELTTAL\
    (the
    non-Einasto
    model with the strongest BIC evidence) in the
    {\tt sigv} stack.
    The \emph{red stars} and 
    \emph{arrows} show the parameters with the highest likelihoods. The
    priors are flat within the panels and zero outsize, except for the tracer
    NFW scale radii $r_{\rm E}$, $r_{\rm S0}$, and $r_{\rm S}$, for which
    they are Gaussians with means in the middle and extending to
    $\pm3\,\sigma$ on the panel edges.  }
  \label{fig:mosaicBIC}
\end{figure}
Figure~\ref{fig:mosaicBIC} shows the MAMPOSSt MCMC posterior marginal
distributions and covariances for model~\NFWisoELTTAL.
All parameters are well fit within their allowed range (except that the outer
anisotropy of spirals can reach the physical limit of pure radial
orbits).

The lower left panel of Fig.~\ref{fig:mosaicBIC} shows that the `virial' radius
is anti-correlated with the outer anisotropy of the spiral
population (recall that E and S0 galaxies are assumed here isotropic).
The mass-anisotropy degeneracy is more acute on the anisotropy 
(0.5 on $\beta_{\rm sym}$ amounting to 0.11 dex on $\left\langle v_r^2\right\rangle^{1/2}/\sigma_\theta$)
than on the `virial' mass ($3\times
0.006 = 0.017$ dex). This precise measurement of the cluster mass
with little influence from the anisotropy (of the spirals) is
a consequence of the mass inversion of NFW models being most insensitive to
the anisotropy near the `virial' radius (\citealp{Mamon&Boue10}, their fig.~3) where
cluster mass is measured.\footnote{This ``sweet spot'' at the virial radius
  where mass is most independent of velocity anisotropy is  a blessing from
  nature to use the cluster mass function derived from spectroscopic surveys
  as a cosmological tool.}

Figure~\ref{fig:mosaicgNFW} in Appendix~\ref{sec:figures}
is similar to Figure~\ref{fig:mosaicBIC}, but for
model~\gNFWisoELgTTAL,
which
is the gNFW model with the 3rd lowest BIC
model (same as model~\NFWisoELTTAL,
except that it
allows for
free inner slope and
free inner anisotropy for the spirals).
This figure indicates that the inner density slope is weakly correlated with the
mass normalization (left panel of 3rd row) and concentration of the mass
profile (2nd panel of 3rd row), as well as with the
outer spiral anisotropy (3rd panel of bottom row), and also correlated with the outer
spiral anisotropy (3rd panel of bottom row).

Figure~\ref{fig:mosaicgNFW2} in Appendix~\ref{sec:figures} shows the marginal
distributions and covariances for model~\gNFWTTAL, with gNFW mass
and T$_0$ anisotropy for all morphological types. 
One notices that the
outer anisotropies of the 3 types are correlated.

\begin{table}[ht]
  \caption{Best-fitting parameters and uncertainties from MAMPOSSt for models~\gNFWgTTAL\ (most likely),
\gNFWisoELTTAL\
(strongest BIC evidence among
models with gNFW mass),
and
\NFWisoELTTAL\
(strongest BIC evidence
non-Einasto
model).}
  \begin{center}
    \tabcolsep 1.5pt
    \begin{tabular}{lccrrr}
      \hline
      \hline
      Parameter & unit & range & \multicolumn{1}{c}{model~\NFWisoELTTAL}
      & \multicolumn{1}{c}{model~\gNFWisoELTTAL} & \multicolumn{1}{c}{model~\gNFWgTTAL} \\
      \hline
      mass & & & \multicolumn{1}{c}{NFW} & \multicolumn{1}{c}{gNFW} & \multicolumn{1}{c}{gNFW} \\
      \multicolumn{3}{l}{velocity anisotropy} & \multicolumn{1}{c}{T$_0$ (S)}
      & \multicolumn{1}{c}{T$_0$ (S)} & \multicolumn{1}{c}{T} \\
      \hline
$\log r_{200}$ & kpc & \ \ 3.0 -- 3.6 & 3.23$\pm$0.01 & 3.23$\pm$0.01 & 3.19$\pm$0.02 \\
$\gamma$ & \multicolumn{1}{c}{--} & --2 -- 0 & \multicolumn{1}{c}{--} &
      $-1.39_{-0.33}^{+0.50}$ & $-1.64_{-0.27}^{+0.74}$ \\
$\log r_{\rm E}$ & kpc & 2.69$\pm$0.10 & 2.66$\pm$0.08 & 2.67$\pm$0.08& 2.66$\pm$0.08 \\
$\log r_{\rm S0}$ & kpc & 2.81$\pm$0.11 & 2.82$\pm$0.08 & 2.82$\pm$0.08& 2.79$\pm$0.08 \\
$\log r_{\rm S}$ & kpc & 3.31$\pm$0.10 & 3.20$\pm$0.08 & 3.19$\pm$0.08& 3.27$\pm$0.08 \\
$\beta_{\rm sym,0}^{\rm E}$ & \multicolumn{1}{c}{--} & --1.8 -- 1.8 & \multicolumn{1}{c}{--} & \multicolumn{1}{c}{--}& --0.66$\pm$0.44 \\
$\beta_{\rm sym,0}^{\rm S0}$ & \multicolumn{1}{c}{--} & --1.8 -- 1.8 & \multicolumn{1}{c}{--} & \multicolumn{1}{c}{--}& --0.18$\pm$0.39 \\
$\beta_{\rm sym,0}^{\rm S}$ & \multicolumn{1}{c}{--} & --1.8 -- 1.8 & \multicolumn{1}{c}{--} & \multicolumn{1}{c}{--}& --0.08$\pm$0.17  \\
$\beta_{\rm sym,\infty}^{\rm E}$ & \multicolumn{1}{c}{--} & --1.8 -- 1.8 & \multicolumn{1}{c}{--} & \multicolumn{1}{c}{--}& 1.19$\pm$0.55 \\
$\beta_{\rm sym,\infty}^{\rm S0}$ & \multicolumn{1}{c}{--} & --1.8 -- 1.8 & \multicolumn{1}{c}{--} & \multicolumn{1}{c}{--} & 0.95$\pm$0.41 \\
$\beta_{\rm sym,\infty}^{\rm S}$ & \multicolumn{1}{c}{--} & --1.8 -- 1.8 & 1.24$\pm$0.31 & 1.11$\pm$0.32 & 1.61$\pm$0.33 \\
      \hline
      \end{tabular}
\end{center}

  \noindent Notes: the parameters are uniformly distributed in the given
ranges, except for the scale radii of the E, S0 and S galaxies, for which the
mean and uncertainty are given and MAMPOSSt assumes Gaussian priors with
dispersion $\sigma$ equal to the uncertainty and cut at $\pm3\,\sigma$.
The quoted values for the 3 models are the MLE estimates and
$(p_{84}-p_{16})/2$ estimates from the MCMC chains, where $p_i$ are the
$i$th percentiles.
Models~\gNFWisoELTTAL\ and \NFWisoELTTAL\ are isotropic for E and S0 galaxies.
  \label{tab:fits}
\end{table}

Table~\ref{tab:fits} shows, for models~\gNFWgTTAL,
\gNFWisoELTTAL, and \NFWisoELTTAL,
the MLE values and the
uncertainties from the 
marginal distributions derived from the MAMPOSSt MCMC.
For model~\NFWisoELTTAL\ with an NFW mass model, the `virial' radius is very well measured
leading to a MLE value of $r_{200} = 1690\pm20\,\rm kpc$,
i.e. with an uncertainty of only 0.005 dex. This
value of $r_{200}$ is consistent with the value 
$1749\pm64\,\rm kpc$ given in Paper~I using the less accurate (see
\citealp{Old+15}) Clean method (which assumes the NFW mass model), as it
should be. The gNFW model~\gNFWisoELTTAL\ leads to $r_{200} = 1675\pm23\,\rm
kpc$, still consistent with the Clean value, while the gNFW
model~\gNFWgTTAL\ leads to $r_{200} = 1507\pm59 \,\rm kpc$, significantly
smaller than the Clean value.

For models~\gNFWisoELTTAL\ and \gNFWgTTAL, the inner slope is consistent with the --1 value
for NFW.

When it is a free parameter, the mass
concentration of model~\NFWisoELTTAL\
is $c_{200} =3.8\pm0.4$
Given the mass at $r_{200}=1698\pm24\,\rm kpc$ of $10^{14.8} \,\rm M_\odot$,  
our concentrations are consistent with the values found
for relaxed $\Lambda$CDM halos ($c_{200} = 4.4$ according to
\citealp{Dutton&Maccio14} -- see eq.~[\ref{cM}] --  or 4.0 according to \citealp{Child+18})
and for relaxed clusters of galaxies using weak lensing
($c_{200} \simeq 4.2$ \citealp{Okabe&Smith16}, which is the median of the 13
measurements within a mass range of 0.3 dex, see fig.~12 of
\citeauthor{Child+18}).
We will further discuss the concentration-mass relation in Sect.~\ref{sec:rhocomplitt}.

Moreover our concentration (set free) for model~\NFWisoELTTAL\
leads to
$\log(r_\rho/{\rm kpc}) = 2.65\pm0.05$
in comparison with
$\log (r_{\rm E}/{\rm kpc}) = 2.67\pm0.08$,
$\log (r_{\rm S0}/{\rm kpc}) = 2.82\pm0.08$, and
$\log (r_{\rm S}/{\rm kpc}) = 3.19\pm0.08$.
  Thus, the elliptical galaxies appear to follow the mass, while the distribution of S0s
  is very slightly (one-third) but quite significantly more extended. In contrast,
  the distribution of 
  spirals is nearly 4 times more extended than that of the mass or of the
  ellipticals. We will return to this issue in Sect.~\ref{sec:profiles}.

\subsection{Goodness of fit}

\begin{figure}[ht]
    \includegraphics[width=\hsize]{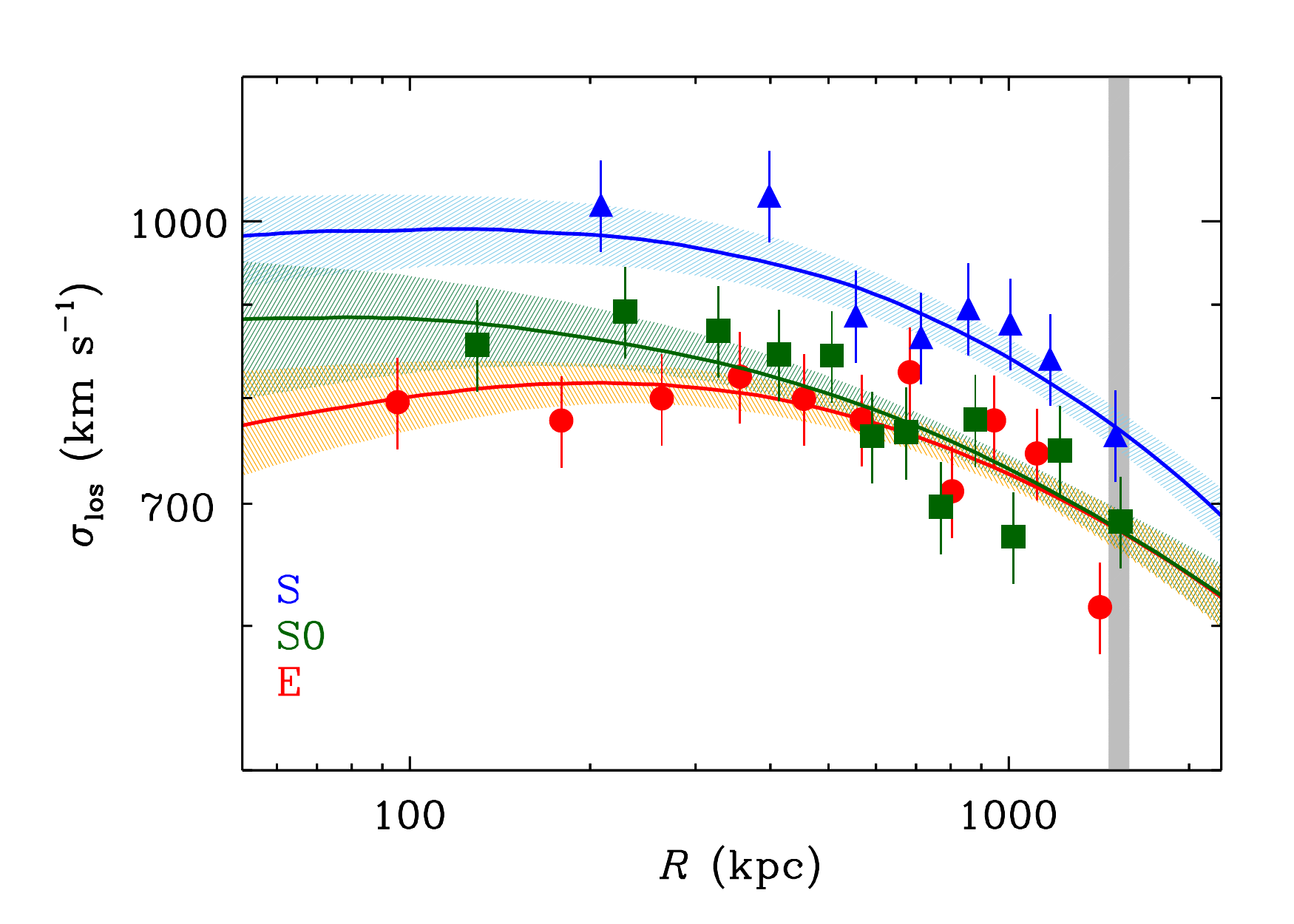}
  \caption{Best-fit line-of-sight velocity dispersion profiles for model~\gNFWgTTAL\
    (gNFW with free T outer anisotropy with TAND) in the {\tt
      sigv} stack.
    The symbols are the data (150 galaxies per radial bin) as in
    Fig.~\ref{fig:sigmalos}, while the \emph{curves} and \emph{shaded regions} respectively
    show the median predictions of model~\gNFWgTTAL\ and their MCMC uncertainties. 
    The \emph{vertical grey shaded region}
    represents $r_{200}$ and its MCMC uncertainty.}
  \label{fig:siglosmodel6}
\end{figure}

Figure~\ref{fig:siglosmodel6} shows the LOS velocity dispersion profiles for
the elliptical, lenticular and spiral galaxies predicted from
model~\gNFWgTTAL\ (gNFW mass with T anisotropy for the 3 morphological
types, all with TAND). The
MAMPOSSt model predictions reproduce very well the data.

\subsection{Radial profiles}
\label{sec:profiles}

\begin{figure}[ht]
  \centering
  \includegraphics[width=0.95\hsize,viewport=5 16 555 735]{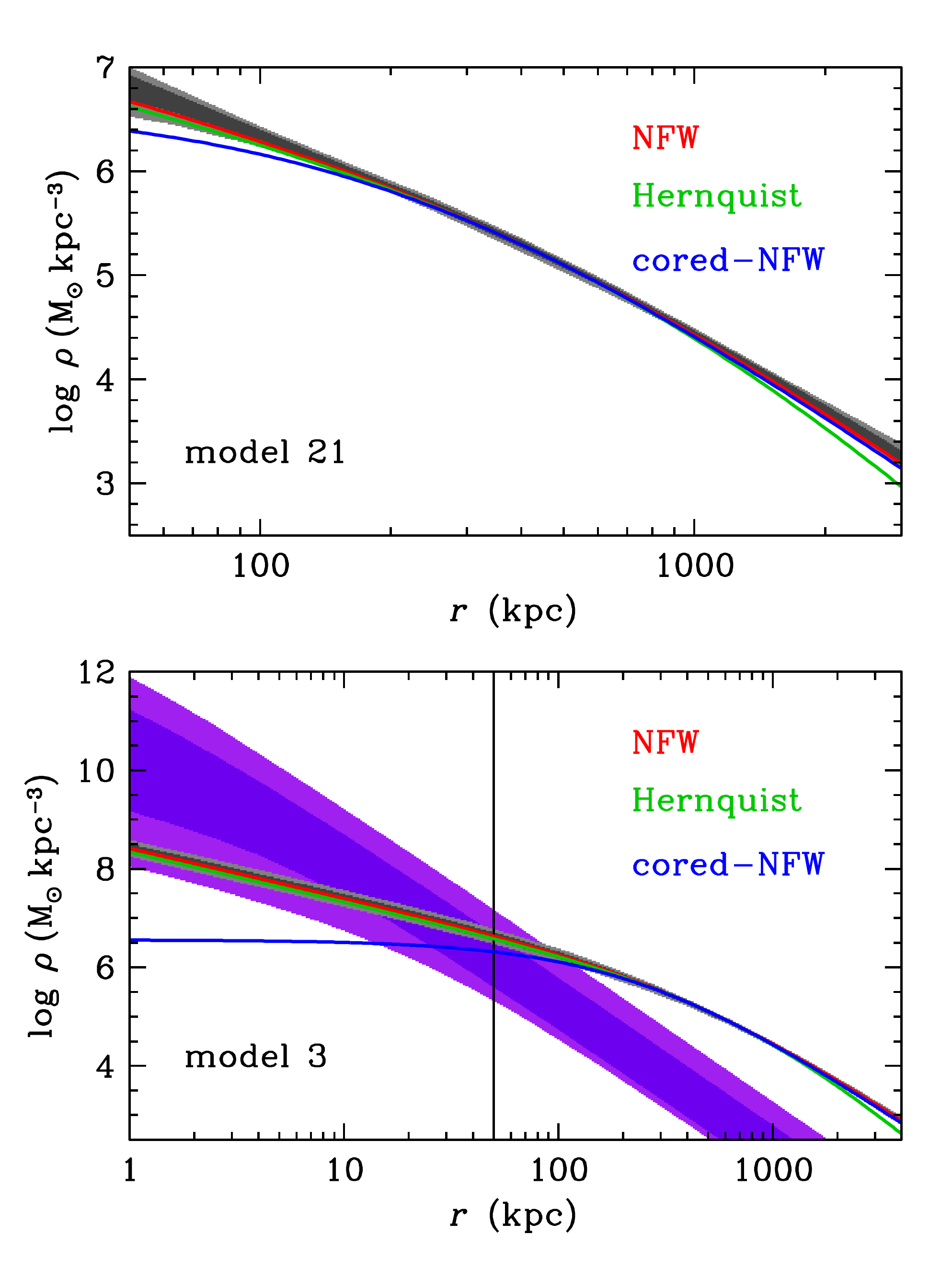} 
  \caption{Radial mass density profiles for
    models~\gNFWisoELTTAL\ (gNFW with isotropic orbits for the ellipticals and S0s,
    and T$_0$ velocity anisotropy for the spirals, \emph{top}) and
    \BCGgTTAL\ (NFW cluster + NFW BCG, with T velocity
    anisotropy, \emph{bottom}) in the {\tt sigv} stack. In both models, the anisotropy
    radius is tied to the scale radius of the galaxy distribution.
    The \emph{shaded regions} show the MAMPOSSt constraints for the cluster (\emph{light}
    and \emph{dark grey}) and the BCG (\emph{light} and \emph{dark purple}),
    where the \emph{light} and \emph{dark zones} respectively delimit 90\% and 68\% confidence
    intervals. The \emph{curves} are the predictions from various analytical
    models, normalized to have the mass scale radii and the same density at the scale radius,
    simply to guide the eye.
    The scale of the bottom panel is different, and the curves to the left of
    the \emph{vertical line}, denoting the minimum considered projected
    radius, are extrapolations.
  }
\label{fig:rhoofr}
\end{figure}

We now show the radial profiles of mass density, mass over number density,
and velocity anisotropy.
These profiles were computed in radial bins of width 0.2 dex. Extracting the free parameters from
1001 random chain elements (after the burn-in phase) among the $6$
(chains) $\times$ (10\,000--2000) $\times$ (\# free parameters),
i.e. typically half a million or more chain elements, we computed the set of
three radial
profiles at each radial bin.

Figure~\ref{fig:rhoofr} displays the mass density profiles for
models~\gNFWisoELTTAL\ (gNFW)
and \BCGgTTAL\ (NFW + NFW for BCG).
In the
top panel, a gNFW model was assumed by MAMPOSSt, and the density
profile prefers to be steeper than NFW,
but not significantly  ($\gamma=-1.51\pm0.42$ according to
Table~\ref{tab:fits}).
Only 85\% of
all chain elements past burn-in produce $\gamma < -1$.
Figure~\ref{fig:rhoofr} and the constraints on the inner slope from
Table~\ref{tab:fits} for model~\gNFWisoELTTAL\
both suggest that the cNFW model (blue) is ruled out.
However, as seen in Table~\ref{tab:runs},
model~\cNFWisoELTTAL, which is the same as model~\NFWisoELTTAL,
replacing NFW by cNFW, leads to 
${\rm Min}(-\ln {\cal L})$ only 1.6 higher than for model~\NFWisoELTTAL.
Considering the cNFW model to be a physical one, it has the same
number of parameters as the NFW model and its BIC is only 3.2
higher than that of model~\NFWisoELTTAL. So, one cannot reject the cored NFW model for
clusters with our WINGS data.

The bottom panel of Fig.~\ref{fig:rhoofr} shows the mass density
profile for model~\BCGgTTAL\ (NFW cluster + NFW BCG).
MAMPOSSt was not able to constrain well the BCG mass
profile given the minimum allowed projected radius of 50 kpc. MAMPOSSt
prefers a BCG with a tiny scale radius, which is not physical.
  
\begin{figure}[ht]
  \centering
  \includegraphics[width=0.95\hsize,viewport=0 30 550 736]{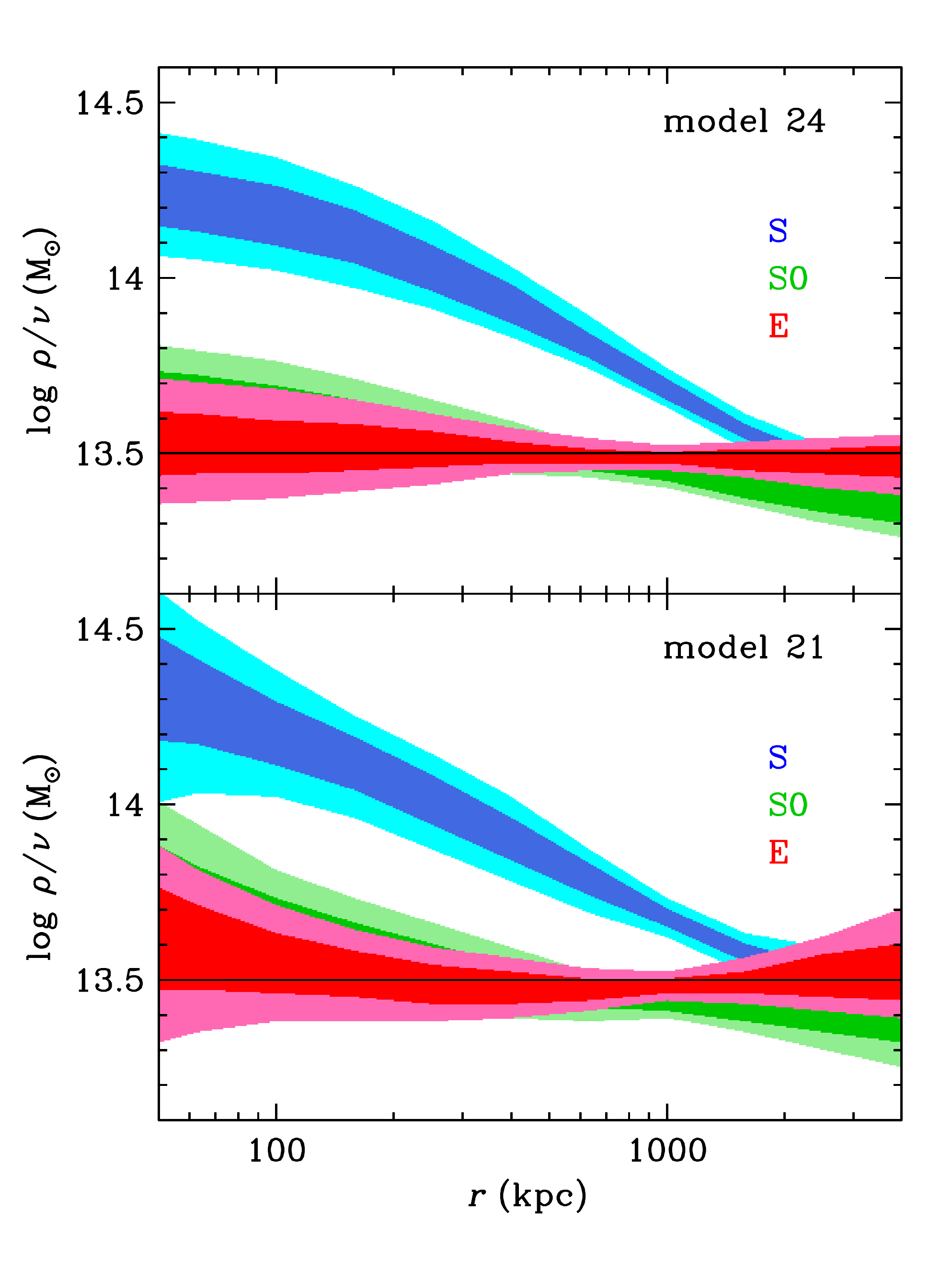} 
  \caption{Radial profiles of mass over number density ratios from MAMPOSSt for models~\NFWisoELTTAL\
    (NFW mass profile, \emph{top}),
and
\gNFWisoELTTAL\ (gNFW mass profile, \emph{bottom}), for E (\emph{red}), S0 (\emph{green}), and S
(\emph{blue}) galaxies,  for the {\tt sigv} stack.
The normalization is explained in equations~(\ref{Ntot})--(\ref{NprojtildeNFW}). 
    The \emph{horizontal lines} are shown to highlight how well mass follows number.
  \label{fig:movern}}
\end{figure}
One may also wonder which morphological type has a number density profile closest to
the mass density profile.
Figure~\ref{fig:movern} displays the ratios of mass density over number
density for the 3 morphological types for models~\NFWisoELTTAL\ and
\gNFWisoELTTAL.
We normalize the number density profiles by eliminating $N(r_\nu)$ between equation~(\ref{nuofr})
and the average number of galaxies of given morphological type per cluster,
$N_{\rm tot}$,
between the minimum and maximum allowed projected
radii, $R_{\rm min}$ and $R_{\rm max}$, which we model as
\begin{eqnarray} 
  N_{\rm tot} &=& N_{\rm p}(R_{\rm max}) - N_{\rm p}(R_{\rm min})
  \label{Ntot}\\
  &=& N(r_\nu)\,\left[\widetilde N_{\rm p}(R_{\rm max}) -
    \widetilde N_{\rm p}(R_{\rm
      min})\right] \ ,
  \label{Ntot2}
\end{eqnarray} 
where
\begin{equation}
  \widetilde N_{\rm p}(X) = {1\over \ln2-1/2}\,{C^{-1}(1/X) \over
    \sqrt{|X^2-1|}}
  +\ln \left  ({X\over 2}\right)
\label{NprojtildeNFW}
\end{equation}
for the NFW model, with $N_{\rm p}(1) = (1-\ln 2)/(\ln2-1/2)$ and where
$C^{-1}$ is given in equation~(\ref{Cinv}). 

Recall that the NFW model was assumed for the number density profiles of the 3
morphological types and that these number density profiles were obtained from
fits to the photometric data, and thus do not suffer from any spectroscopic incompleteness.
In both panels of Fig.~\ref{fig:movern}, the
elliptical galaxies trace almost perfectly the mass, S0 galaxies are nearly as good 
mass tracers as ellipticals (but somewhat more extended), while spirals are much
more extended. In runs with free concentration, the ellipticals trace even
better the mass.
Since model~\gNFWisoELTTAL\ (bottom panel)
is based on a gNFW mass profile, there is
less agreement between the NFW number density profile of the ellipticals and
the gNFW mass profile, but ellipticals remain the
best tracers of mass
among the 3 morphological types. Indeed, the mass over elliptical number
density ratio is  nearly consistent with being constant
(horizontal line), although there may be a need for extra mass in the
BCG.

\begin{figure*}[ht]
  \centering
  \includegraphics[width=0.45\hsize,viewport=0 20 570 560]{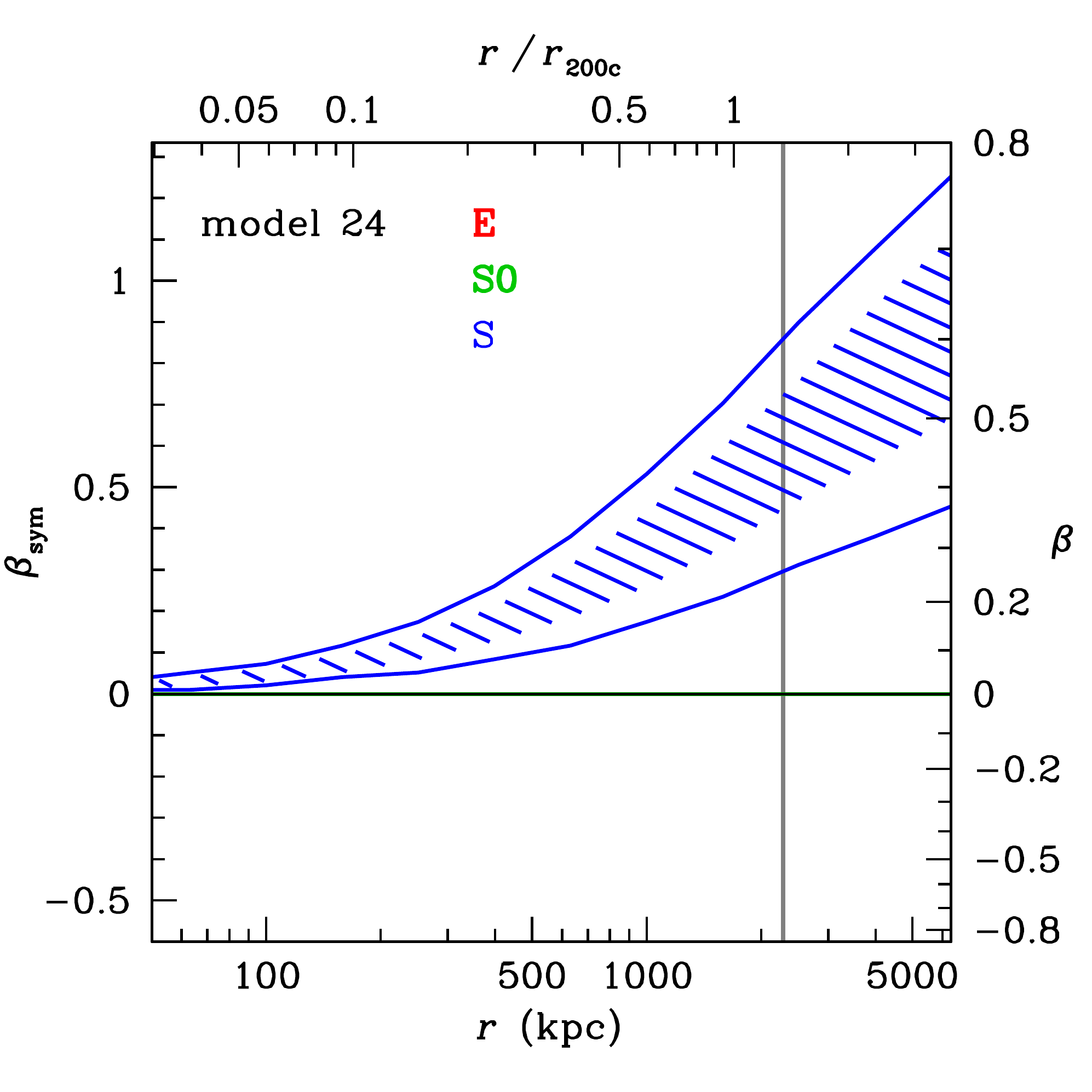}
  \includegraphics[width=0.45\hsize,viewport=0 20 570 560]{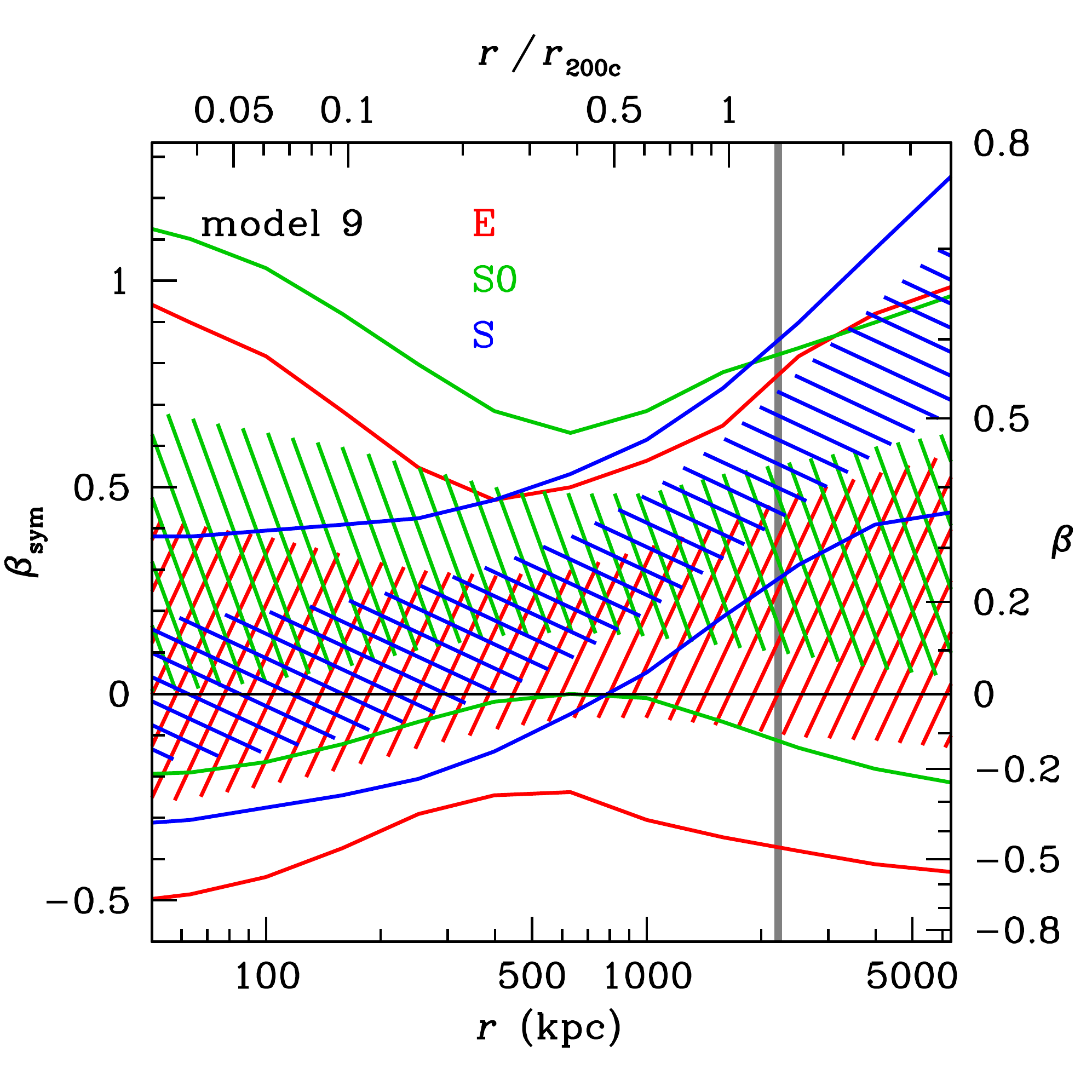}
  \includegraphics[width=0.45\hsize,viewport=0 20 570 560]{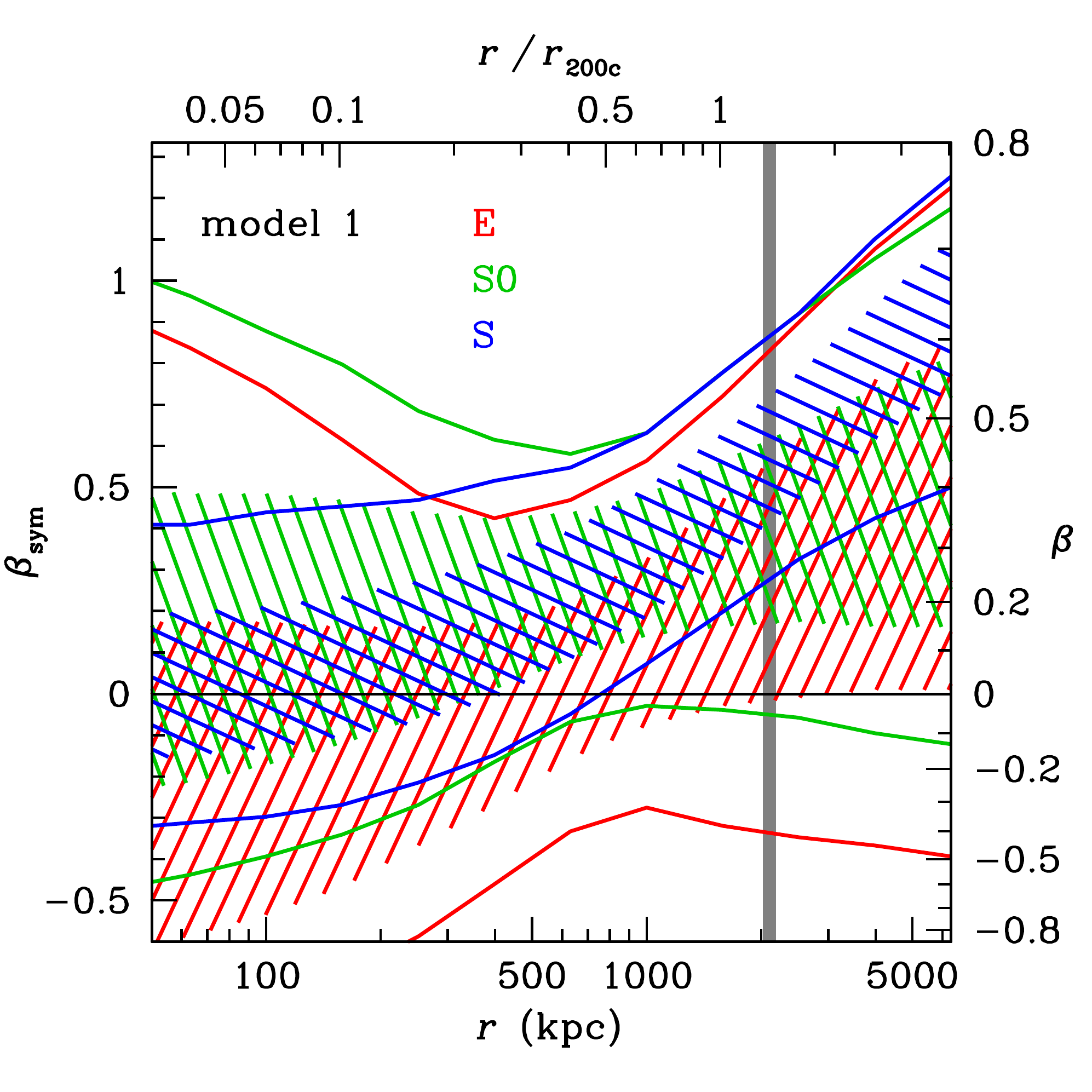}
  \includegraphics[width=0.45\hsize,viewport=0 20 570 560]{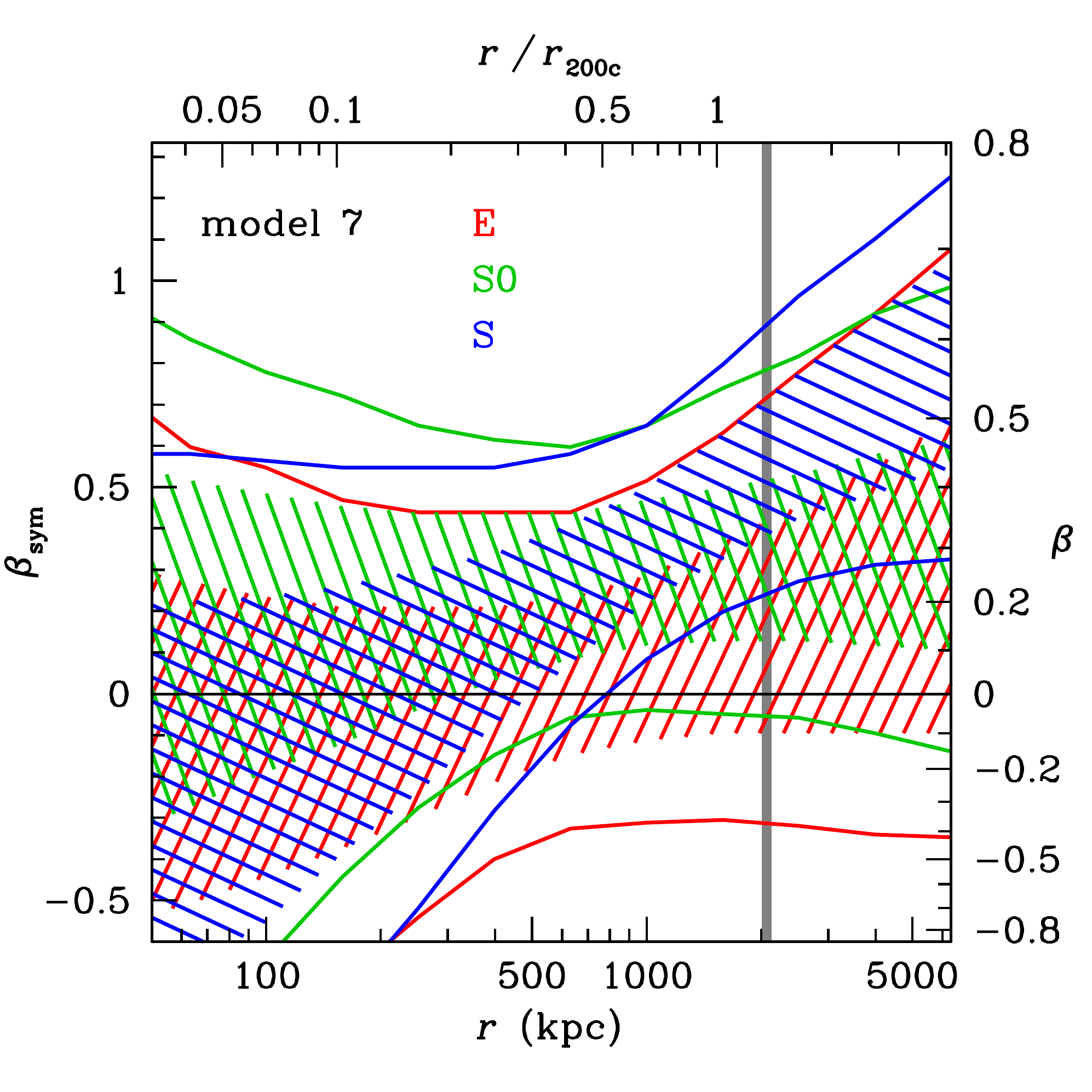} 
  \caption{Velocity anisotropy (eq.~[\ref{betasym}])
    profiles of the E, S0 and S galaxies from MAMPOSSt for
    models~\NFWisoELTTAL\
    (NFW with T-TAND anisotropy for spirals only, \emph{upper left}),
    \NFWgTTAL\ (NFW
    with T-TAND anisotropy, \emph{upper right}),
    \gNFWgTTAL\ (gNFW
    with T-TAND anisotropy, \emph{lower left}),
    and
    \gNFWgT\ (gNFW
    with T anisotropy and free anisotropy radius, \emph{lower right}),
 for the {\tt sigv} stack.
    The \emph{hashed} regions indicate  the 68\% confidence zone, while
    the \emph{curves  } display the 5th and 95th percentiles.
    The \emph{thick vertical grey line} indicates the position of
    $r_{100} = 1.35\,r_{200}$, which is close to the theoretical virial
    radius, where its width shows the uncertainty on $\log r_{200}$. 
  }
  \label{fig:betaofr}
\end{figure*}

Figure~\ref{fig:betaofr} displays the anisotropy profiles for
models~\NFWisoELTTAL, \NFWgTTAL, \gNFWgTTAL, and \gNFWgT.
Model~\NFWisoELTTAL\ (our best
non-Einasto
model in terms of BIC), which assumes
isotropic orbits for the E and S0 galaxies and inner 
isotropy also for the spirals, shows that the spiral galaxies clearly have radial
orbits in the outer regions of clusters.
The other 3 models, with fully free priors on inner and outer velocity
anisotropy, confirm that spiral galaxies have increasingly radial orbits at
large distances. Early-type galaxies show moderately radial outer orbits for
these 3 models, but all consistent with isotropy.
The similarity in the anisotropy profiles between models~\gNFWgTTAL, and
\gNFWgT, which only differ in that the latter has a free anisotropy radius, 
confirms that this radius is close to the scale radius of the
tracer density as in the TAND approximation.
The inner and outer anisotropies are displayed in Table~\ref{tab:anisotropy}.

\subsection{Outer versus inner velocity anisotropies}

\begin{figure*}[ht]
  \centering
  \includegraphics[width=0.45\hsize,viewport=0 30 570 556]{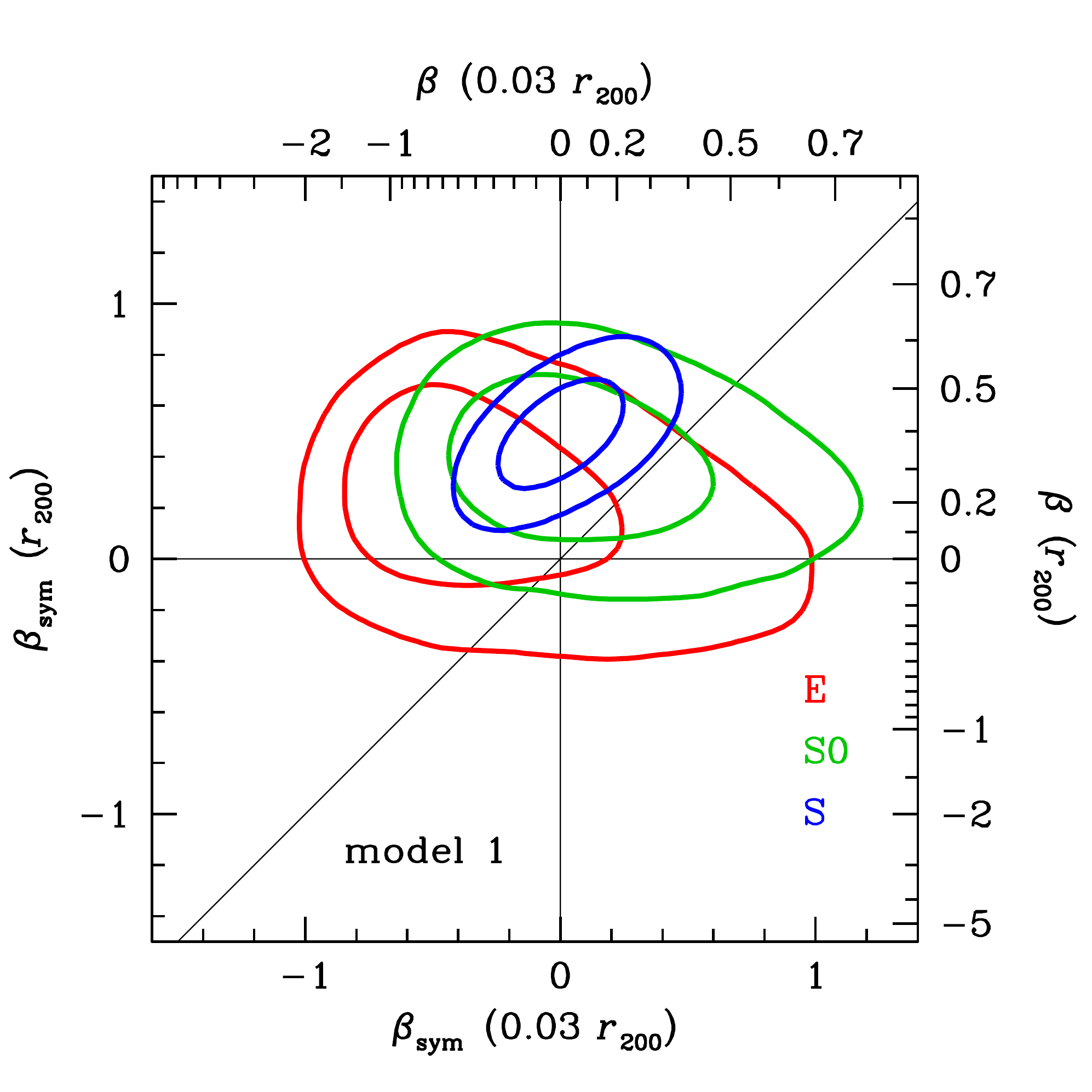}
\quad
  \includegraphics[width=0.45\hsize,viewport=0 30 570 556]{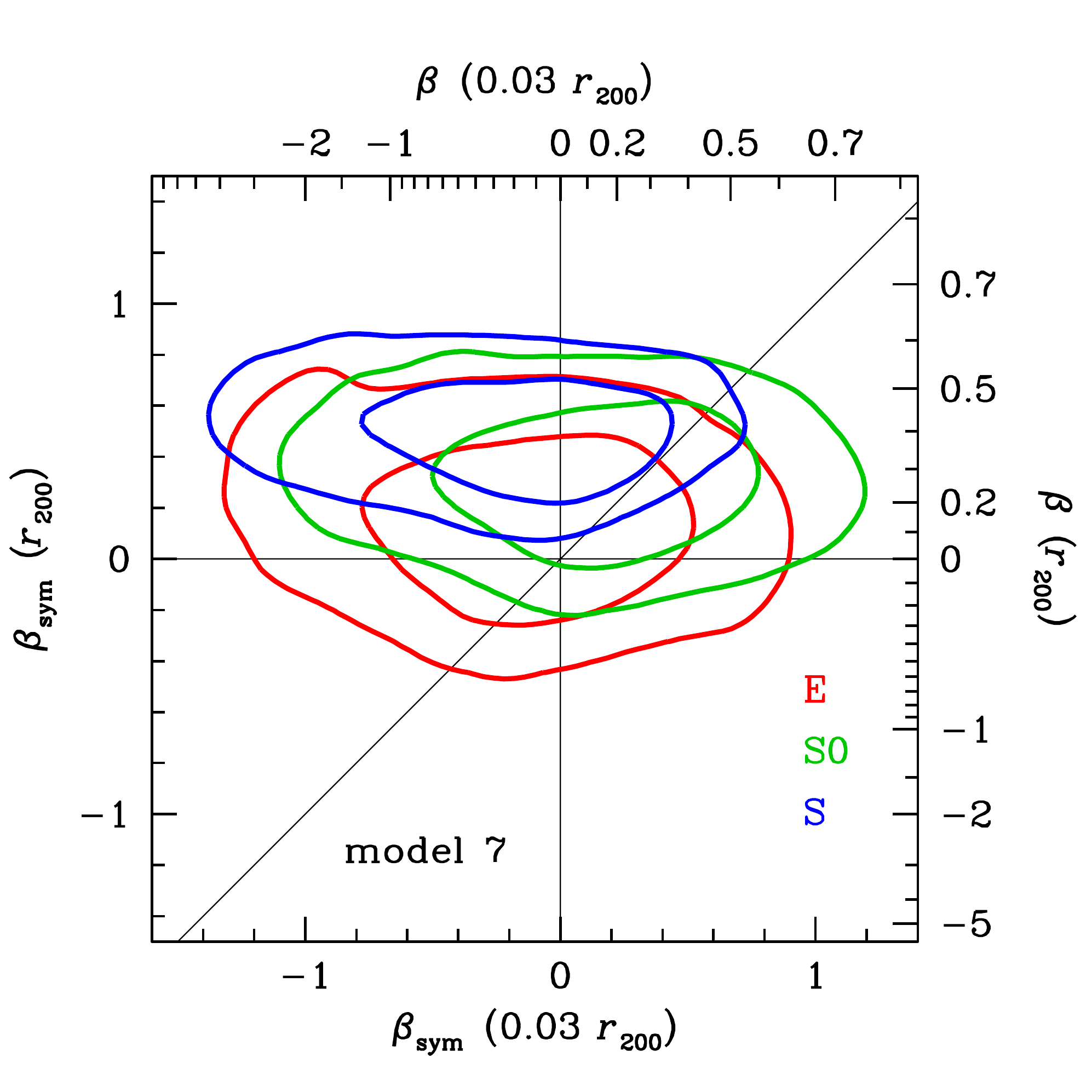}
  \includegraphics[width=0.45\hsize]{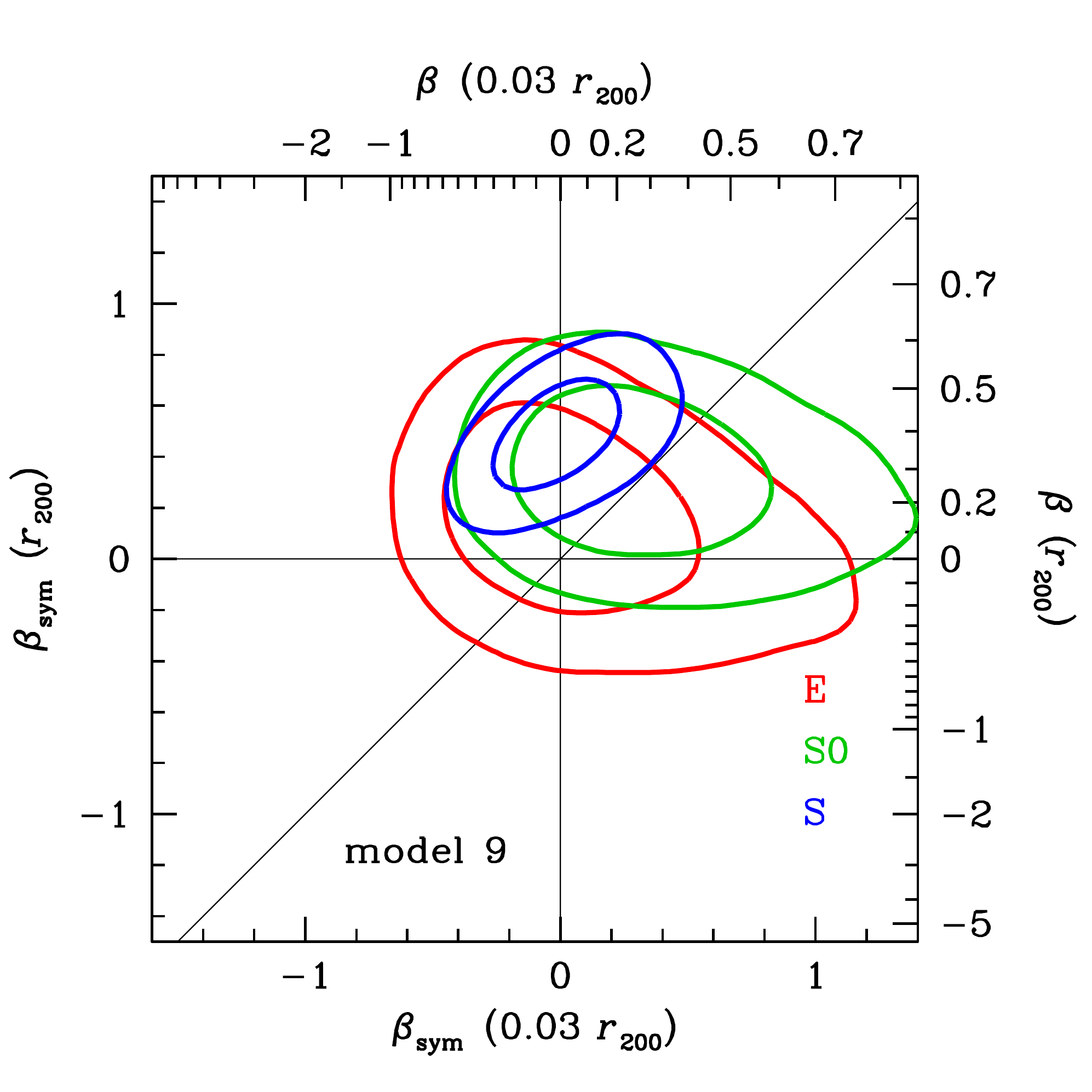}
  \includegraphics[width=0.45\hsize]{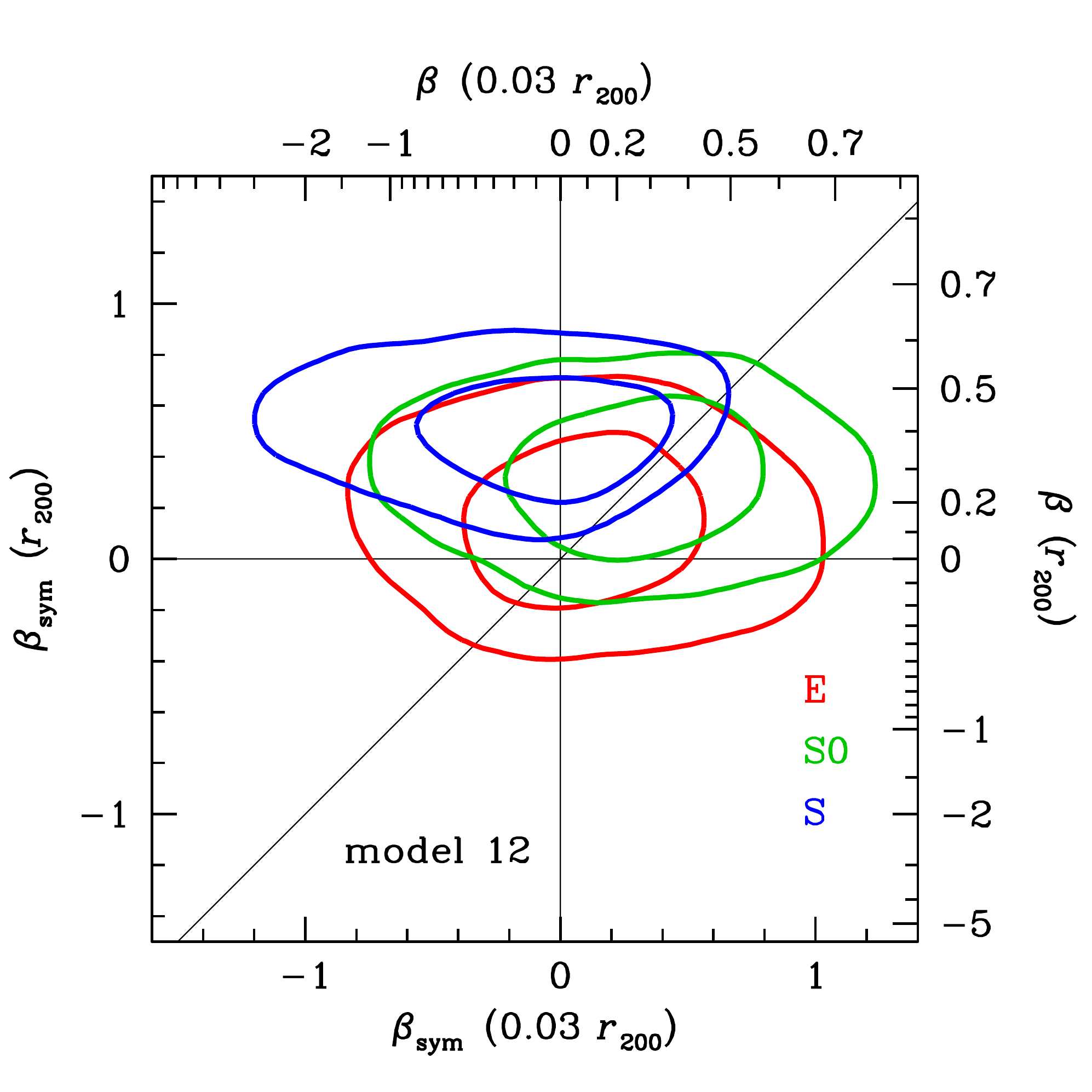}
  \caption{Outer ($r = r_{200}$) vs. inner ($r=0.03\,r_{200}$) velocity
    anisotropy 
    (eq.~[\ref{betasym}])
    from MAMPOSSt
    for the 3
    morphological types, for models~\gNFWgTTAL\ (gNFW-T anisotropy with TAND),
    \gNFWgT\ (gNFW-T),
    \NFWgTTAL\ (NFW-T-TAND), and
    \NFWgT\ (NFW-T),
    for the {\tt sigv} stack.
    The contours are 68 and 95\% confidence (with pixel
    resolution $\Delta \beta_{\rm sym}=0.03$ and smoothed with a Gaussian
    of $\sigma=2$ pixels). } 
  \label{fig:betabeta}
\end{figure*}

\begin{table}[ht]
  \caption{Velocity anisotropies from MAMPOSSt at $0.03\,r_{200}$ and at $r_{200}$}
  \begin{center}
    \tabcolsep 2pt
    \begin{tabular}{r@{\ \ \ }llccrcc}
      \hline
      \hline
      \multicolumn{1}{l}{model} & mass & $\beta(r)$ & TAND & type & 
      \multicolumn{2}{c}{
        $\beta_{\rm sym}$
        }
      & $\nearrow$ \\
\cline{6-7}
        & & & & &       \multicolumn{1}{c}{$(0.03\,r_{200})$} &
\multicolumn{1}{c}{$(r_{200})$} & \\
(1) & \multicolumn{1}{c}{(2)} & (3) & (4) & (5) & \multicolumn{1}{c}{(6)} & (7) & (8) \\
        \hline
      \gNFWgTTAL & gNFW & T & Y & E\ \, & --0.33$_{-0.29}^{+0.49}$ & 0.22$\pm$0.24 & 0.81 \\
       &  &  &  & S0 & 0.06$_{-0.30}^{+0.41}$ & 0.36$\pm$0.20 & 0.73 \\
      &  &  &  & S\ \, & 0.01$\pm$0.17 & 0.48$\pm$0.14 & 1.00 \\
      \\
      \BCGgTTAL & 2NFW & T & Y & E\ \, & --0.12$_{-0.31}^{+0.39}$ &
      0.14$\pm$0.26 & 0.70 \\
      & & & & S0 & 0.21$_{-0.31}^{+0.40}$ & 0.30$\pm$0.20 & 0.58 \\
      & & & & S\ \, & --0.02$\pm$0.17 & 0.47$\pm$0.14 & 0.99 \\
      \\
      \gNFWgT & gNFW & T & N & E\ \, & --0.11$_{-0.46}^{+0.39}$ & 0.14$\pm$0.21 & 0.71
      \\
      &  &  &  & S0 & 0.16$_{-0.49}^{+0.37}$ & 0.31$\pm$0.19 & 0.63 \\
      &  &  &  & S\ \, & --0.15$_{-0.53}^{+0.36}$ & 0.49$\pm$0.15 & 0.97 \\
      \\
      \NFWisoELTTAL & NFW & T$_0$ & Y & S\ \, & 0.02$\pm$0.01 & 0.50$\pm$0.12 &
      1.00 \\
      \hline
    \end{tabular}
  \end{center}

  Notes: The columns are:
  (1): model number (see Table~\ref{tab:runs});
  (2): mass model (2NFW stands for NFW + NFW (BCG));
  (3): anisotropy model;
  (4): is the anisotropy radius set equal to the tracer scale radius (TAND)?
  (5): morphological type;
  (6): $\beta_{\rm sym}$ (eq.~[\ref{betasym}]) at $r = 0.03\,r_{200}$;
  (7): $\beta_{\rm sym}$ at $r = r_{200}$;
  (8): fraction of anisotropy profiles that increase with radius.
  The uncertainties are $(p_{84}-p_{16})/2$, where $p_i$ are the $i$th percentiles.
  \label{tab:anisotropy}
\end{table}

Table~\ref{tab:anisotropy} illustrates the details of the anisotropy for 4 models.
If we free the inner
and outer anisotropies (but tie the anisotropy radii to the scale radii, model~\gNFWgTTAL),
we find that the anisotropy at  $r_{200}$ of
the E and S0 galaxies are also typically somewhat radial
(see also Table~\ref{tab:fits} and
Figs.~\ref{fig:betaofr} and
\ref{fig:betabeta}). 
However, 
the uncertainties on outer anisotropy
are much larger (almost double in $\beta_{\rm sym}$ for E vs. S)
for these early-type galaxies compared to
spirals (Figs.~\ref{fig:betaofr} and \ref{fig:betabeta}), which explains why
BIC evidence prefers having radial outer orbits 
for the spiral population only.
Hence, there is only marginal evidence that the S0 population has 
radial outer orbits, while the moderately radial orbits of ellipticals is
not statistically significant (Table~\ref{tab:anisotropy}).
The outer anisotropies of the elliptical and S0 galaxies are less radial
when the anisotropy radii are set free (model~\gNFWgT) compared to the
analogous TAND model~\gNFWgTTAL.

The inner anisotropies of the 3 morphological types are always consistent
with isotropy (Table~\ref{tab:anisotropy}), where
the uncertainties for
the spiral population are much smaller for the TAND assumption.
But a close inspection of Table~\ref{tab:anisotropy} indicates that
elliptical galaxies have slightly tangential inner values of anisotropy, as
expected from
our quick look at the LOS velocity
dispersion profile (Fig.~\ref{fig:sigmalos}). However, this tangential
anisotropy is not statistically significant.

Figure~\ref{fig:betabeta} provides a clearer way to view the anisotropy
profiles, by plotting the value at $r_{200}$ as a function of the value at
$0.03\,r_{200}$. We restrict these plots to models with free inner and outer
anisotropies for all morphological types.
When the anisotropy radius is forced to the scale radius, as favored by
the Bayesian (both AIC and BIC) evidence (top panel of Fig.~\ref{fig:betabeta}),
the 95 percent confidence contours for $\beta(r_{200})$ for spirals are
above zero for all values of the inner anisotropy (at $0.03 r_{200}$), which
is almost the case for S0 galaxies, but not the case for ellipticals.
Moreover, the 95th percent confidence level is always in the direction of
increasingly radial  anisotropy for the spirals, which is not the case for
the S0s and ellipticals.
Also, the inner anisotropy of the ellipticals is somewhat tangential (though
not significantly),
while those of the S0s and spirals appear to be even more isotropic.

On the other hand, by freeing the anisotropy radii
(bottom panel of
Fig.~\ref{fig:betabeta}), the outer anisotropies become independent of the
inner values, for all 3 morphological types.
The free vs. fixed anisotropy radii have a stronger effect on the contours of
outer vs. inner anisotropy than does the mass model.
Nevertheless, 
only spiral galaxies show clearly radial anisotropy at $r_{200}$
(Table~\ref{tab:anisotropy}).
The lack of
correlation between inner and outer anisotropies is probably due to the 
wide range of anisotropy radii allowed by the data.
Indeed, while the log anisotropy radii (in units of kpc) are allowed to span
between 1 and 4, the uncertainty on the best-fit anisotropy radii for the non-TAND
runs are typically as high as 1 dex for all 3 morphological types.
Nevertheless, as for the TAND case, spirals are the sole morphological type for which the orbits
systematically become more radial from the inner regions to near the virial
radius (last column of Table~\ref{tab:anisotropy}, and as seen by the 
contours for spirals lying above the oblique line in Fig.~\ref{fig:betabeta}).

\section{Discussion}
\label{sec:discuss}
This work represents the largest analysis of velocity anisotropy in
cluster galaxies and the first to distinguish the orbits of ellipticals,
spirals and lenticulars using a Bayesian model to predict the distribution of
these 3 morphological types in PPS.
We have constructed a stacked cluster, which helps us avoid departures from
spherical symmetry, although it introduces artificial phase mixing.

Our conclusions depend on our choice of priors. We have presented
\numruns\
choices
of priors (and tried many more). We can restrict our
conclusions to the simpler set of priors that lead to the highest Bayesian
evidence measures (within $\Delta$BIC = 6 of the lowest BIC), or we can
analyze the detailed radial profiles expected from the 
models that reach the highest likelihoods (really posteriors), although their BIC Bayesian
evidence is so high that they can be strongly rejected relative to the lowest
BIC model.

\subsection{Mass density profiles of WINGS clusters}

\subsubsection{General trends}
Our highest BIC Bayesian evidence is reached for models~\EsixisoELTTAL\ and
\NFWisoELTTAL, where the mass profile is $n$=6 Einasto or NFW with isotropic
velocities for ellipticals and S0s, while for the spirals
they are isotropic at the center and fairly radial in the outer regions of
clusters (Table~\ref{tab:runs}).
There is strong Bayesian evidence against the Hernquist
model~(\HisoELTTAL), whose
outer slope is steeper (--4) then that of NFW (--3).
There is only positive (but not strong) evidence against a
cored NFW model (relative to model~\NFWisoELTTAL).
The case against the gNFW and free index Einasto models is less clear with AIC evidence: the lowest
AIC value is reached for an NFW model, but the 2nd lowest
among non-Einasto models 
is for a gNFW model.

On the other hand, our best fitting models
prefer a mass profile with a free inner slope of order
$-1.6\pm0.4$, which is marginally consistent with the --1 slope of the NFW
model (Table~\ref{tab:fits} and Figs.~\ref{fig:mosaicgNFW} and
\ref{fig:mosaicgNFW2}), and rejected by BIC evidence.
High likelihoods are also attained by summing an NFW model for the cluster
with another smaller NFW model for a central BCG (Table~\ref{tab:runs}).
But surprisingly, the BCG would require a very high concentration NFW model
that is essentially a --3 power law in the innermost regions of the 
cluster that we analyze (Fig.~\ref{fig:rhoofr}). The BCG dominates the
cluster within the inner 20 kpc, i.e well inside the minimum projected radius
for which we are confident of
our cluster centers (assumed to be at the BCG location) before we stack them.
We thus simply do not have enough tracers to constrain the mass
density profile within this radius.

\subsubsection{Robustness}
\label{sec:robustrho}

We ran several models for the two other cluster stacks ({\tt Num} and {\tt tempX}).
We found that for model~\gNFWisoELTTAL\ (which uses gNFW for the cluster mass), we find that
the constraints on the inner slope range from
$\gamma=-1.4_{-0.3}^{+0.5}$ for {\tt sigv}
to $-1.7_{-0.2}^{+0.3}$ for {\tt tempX}
and $-1.8_{-0.1}^{+0.4}$ for {\tt Num}.
The steeper inner mass density slopes for the {\tt tempX} and {\tt Num}
stacks lead to different Bayesian evidence for or against gNFW:
Indeed, comparing model~\NFWisoELTTAL\ (NFW) to \gNFWisoELTTAL, we find 
that
AIC prefers gNFW for {\tt Num} ($\Delta\rm AIC = AIC(gNFW) - AIC(NFW) = -5.9$)
and {\tt tempX}  ($\Delta\rm AIC = -2.5$),
whereas it slightly favors NFW for {\tt sigv} ($\Delta\rm AIC = 1.2$).
However, BIC does not favor gNFW despite the much steeper
gNFW inner slopes:
whereas there is  strong BIC evidence against gNFW with {\tt sigv} 
($\Delta\rm BIC> 7$), there is still positive evidence against gNFW
with {\tt tempX} ($\Delta\rm BIC = 3.7$)  and {\tt Num}
($\Delta\rm BIC=0.5$).

\begin{figure}[ht]
  \centering
  \includegraphics[width=0.85\hsize,viewport=0 30 550 730]{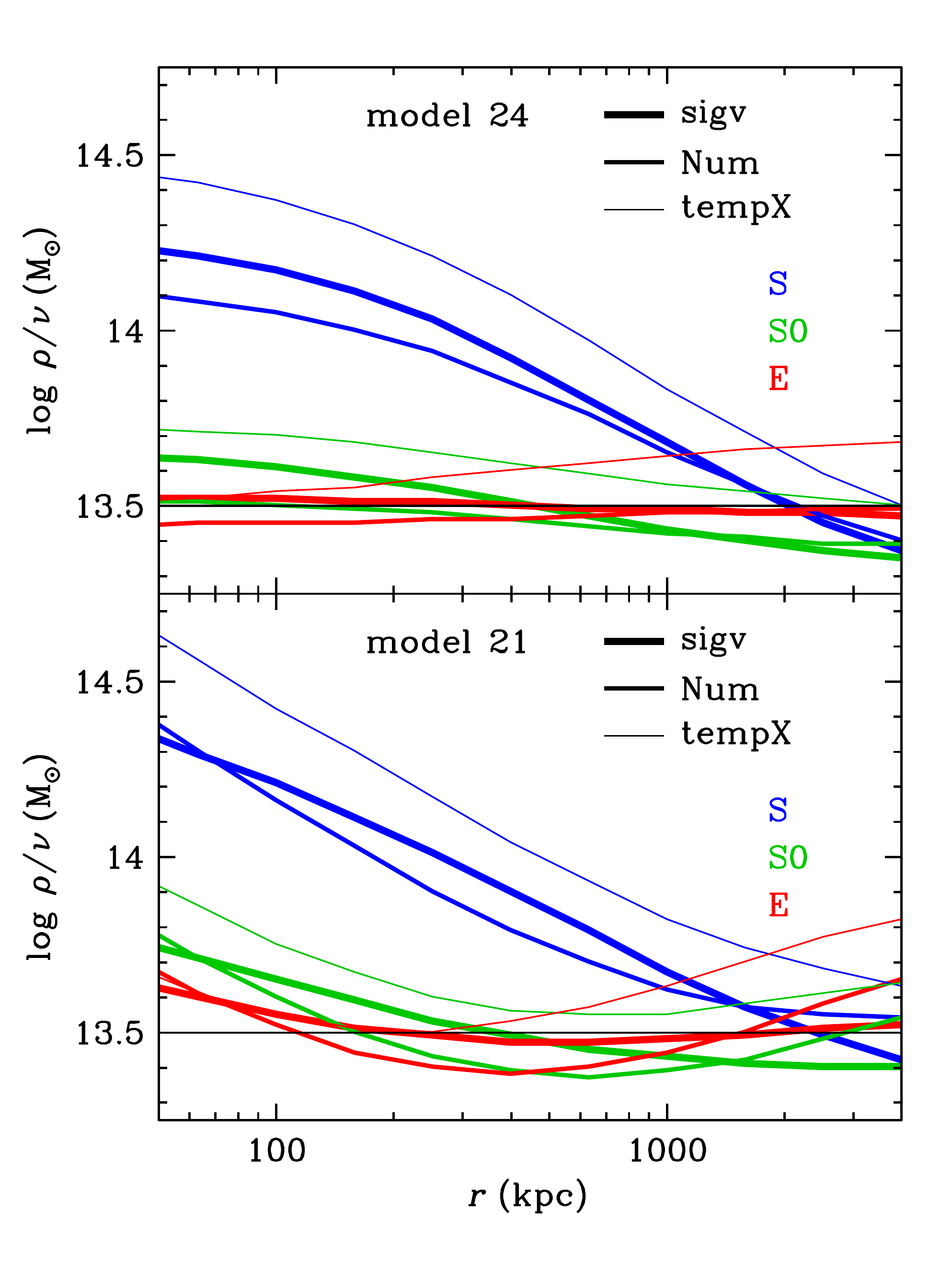} 
  \caption{Same as Fig.~\ref{fig:movern} (but only median trends are shown,
    for clarity),
    for different cluster stacking
    methods, again for models~\NFWisoELTTAL\ and \gNFWisoELTTAL.}
  \label{fig:rhoovernu_stacks}
\end{figure}

We also test the robustness of the radial 
profiles of mass over number density for the 3 morphological types to the
choice of stack.
Figure~\ref{fig:rhoovernu_stacks} shows that for model~\NFWisoELTTAL,
the (NFW) mass density profile is almost exactly proportional to the elliptical number
density profile for all 3 stacks, and close to proportional to the S0 number
density profile, while the spirals trace poorly the mass profile,
as they are more extended, hence $\rho/\nu$ is more concentrated.

For model~\gNFWisoELTTAL\ (which is the same as model~\NFWisoELTTAL,
but with gNFW mass instead of NFW), ellipticals and S0s show a U-shaped
mass-over-number profile. A close look indicates that
the elliptical number density profile
traces slightly better the  mass density profile (within $r_{200}$)
then do the S0 galaxies, except for
the {\tt tempX} stack where the two types trace the mass with similar
accuracy. Again, the spiral galaxies trace poorly the mass
profile.

\subsubsection{Comparison with other work}
\label{sec:rhocomplitt}
In comparison, combining weak lensing at large radii, strong lensing at
intermediate radii and stellar kinematics at low radii to study 7 regular
clusters, \cite{Newman+13a} 
deduced that the total mass density profile is close to a gNFW with inner
slope $-1.2\pm0.1$, while the dark matter follows a gNFW with a shallow
slope of $-0.5\pm0.2$ \citep{Newman+13b}.
Their total mass profile is consistent with ours (NFW for lowest BIC model~\NFWisoELTTAL\
as well as $\gamma=-1.4_{-0.3}^{+0.5}$ for model~\gNFWisoELTTAL).

\begin{figure}[ht]
  \includegraphics[width=\hsize]{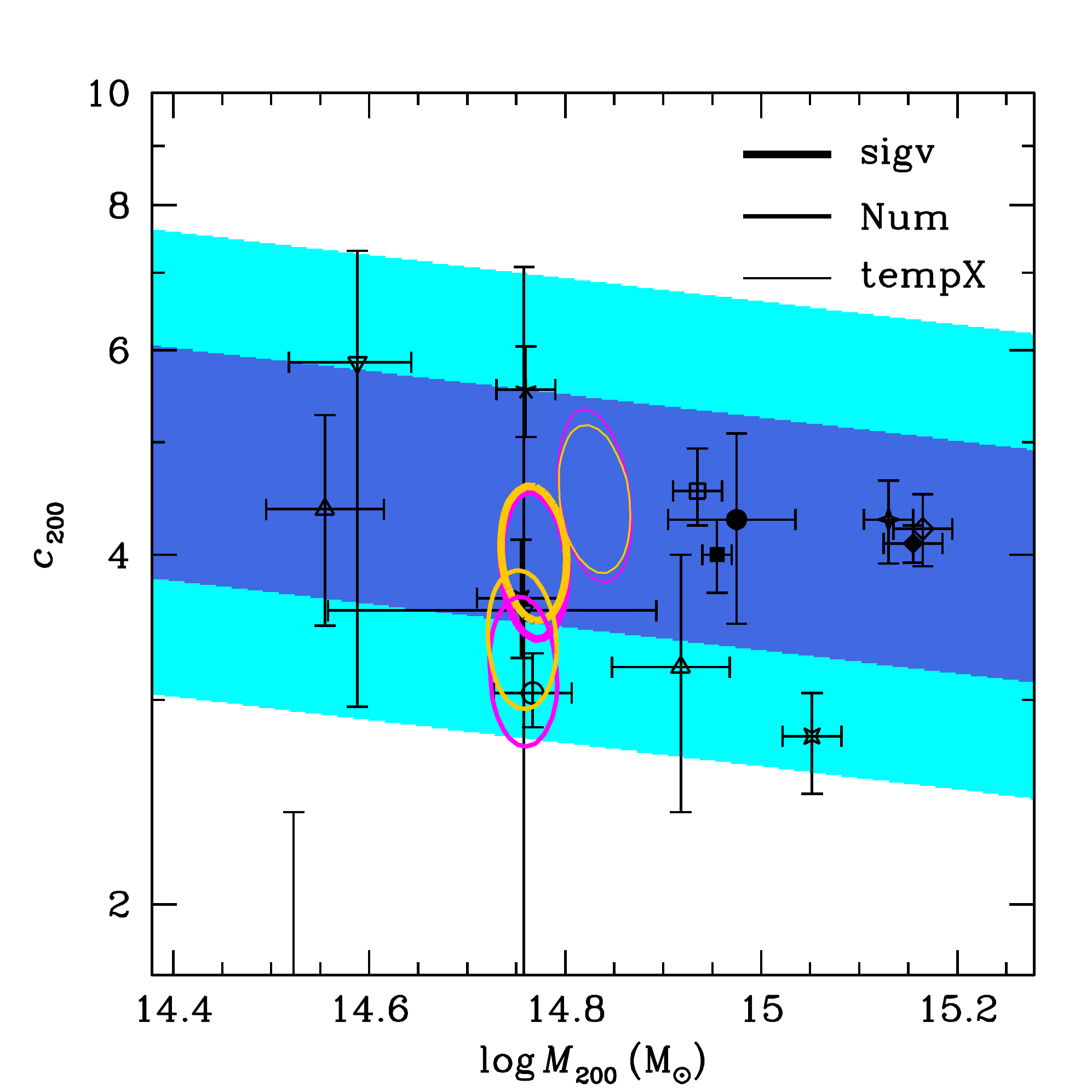}
  \caption{Cluster mass-concentration versus mass for the 3 stacks
    with model~\NFWisoELTTAL. The \emph{magenta contours} indicate the run
    with free concentration (flat prior $0 < \log c_{200} < 1$), while the
    \emph{yellow contours} display 
  the run with the $\Lambda$CDM relation of equation~(\ref{cM}), highlighted
  in the \emph{shaded regions} for 1 and $2\,\sigma$ constraints.
  The contours are 68\% constraints.
  The points are the weak-lensing analyses by
  Johnston et al. (2007) (\emph{triangles}),
  Mandelbaum et al. (2008) (\emph{downwards triangles}),
  Okabe et al. (2010) (\emph{curly squares}),
  Oguri et al. (2012) (\emph{cross}),
  Okabe et al. (2013) (\emph{open square}),
  Sereno \& Covone (2013) (\emph{open circle}),
  Umetsu et al. (2014) (\emph{curly diamond}),
  Umetsu et al. (2016) (\emph{diamond}),
  Okabe \& Smith (2016) (\emph{filled square}),
  Umetsu \& Diemer (2017) (\emph{filled diamond}),
  and Cibirka et al. (2017) (\emph{filled circle}), all corrected to be
  $(1+z)^{0.38} c_{200}$ following Child et al. (2018).
  The error bar at the bottom is from Mandelbaum et al. (2008).
  }
  \label{fig:cM}
\end{figure}
\nocite{Johnston+07}
\nocite{Mandelbaum+08}
\nocite{Okabe+10}
\nocite{Okabe+13}
\nocite{Okabe&Smith16}
\nocite{Oguri+12}
\nocite{Sereno&Covone13}
\nocite{Umetsu+14}
\nocite{Umetsu+16}
\nocite{Umetsu&Diemer17}
\nocite{Cibirka+17}
\nocite{Child+18}

Figure~\ref{fig:cM} shows the constraints on the concentration-mass relation
obtained with NFW model~\NFWisoELTTAL. Interestingly, with free concentration
(flat prior $0 < \log c_{200} < 1$), the contours match well the $\Lambda$CDM
relation of \cite{Dutton&Maccio14} given in equation~(\ref{cM}), especially
for stacks {\tt tempX} and {\tt sigv}. It is not
surprising that folding in this relation as a prior, we recover similar
contours, simply closer to the relation itself.
The MAMPOSSt analysis of the cluster kinematics matches well the concentrations obtained by weak lensing,
except for one (fairly old) weak lensing study that strongly underestimates
the concentration.
\cite{Biviano+17} performed MAMPOSSt analysis of 49 WINGS clusters,
individually, and found a cluster-to-cluster scatter in concentration that
was greater than the uncertainties returned from MAMPOSSt ($\simeq$0.3 dex)
and from the effect of range of cluster masses ($\simeq$0.3 dex dispersion) combined with
the $-0.1$ slope of the concentration-mass relation (leading to a dispersion
of 0.03 dex).  Their best fit at the median
cluster mass yields $c_{200} = 3.34$, close to the value of our {\tt Num} stack
(Fig.~\ref{fig:cM}). Our other two stacks lie within the confidence band of
the concentration-mass relation of \cite{Biviano+17}.

With our assumptions of NFW mass and number density profiles, we find that
our lowest NFW BIC
model~\NFWisoELTTAL\ indicates that the elliptical population follows the mass, while the
lenticulars are slightly
less concentrated and the spirals are much less concentrated (top panel of
Fig.~\ref{fig:movern}).
These results can be compared to the mass traced by the  red
(\citealp{vanderMarel+00} for CNOC clusters), early spectral type
(\citealp{Biviano&Girardi03} for 
a stack of 43 2dFGRS clusters), and non-BCG E and S0 (\citealp{Katgert+04}
for ENACS clusters) galaxies.
The much weaker  concentration of the spiral population agrees with the much
weaker concentration of the blue galaxies relative to the red ones observed
by \cite{Collister&Lahav05}.

Allowing for a gNFW mass model, the scale-radius (radius of slope --2), hence
concentration of the mass remains
consistent with the corresponding values of the elliptical population, but
not of the lenticulars or spirals
(bottom
panel of Fig.~\ref{fig:movern}). There is a discrepancy between mass and
elliptical number at
very small radii (bottom panel of Fig.~\ref{fig:movern}) because of the
steeper rise with decreasing radius of the gNFW mass profile compared to the
NFW number profile of the ellipticals.
Since the BCG contributes a large stellar mass at the
cluster center, we expect that the E stellar mass density profile
follows the total mass density profile even better than does the E
number density profile.

Finally,
it is surprising that the
distribution of ellipticals, which may be assimilated to the dwarf
spheroidals orbiting the Milky Way, follows the dark matter in contrast with
the subhalos in the Aquarius dark matter-only simulations
\citep{Springel+08}.
This discrepancy might be attributed to the missing
dissipative gas in Aquarius, and the uncertain link between the subhalo and
satllite radial distributions given the uncertain radially-dependent link between the minimum subhalo  and
galaxy stellar masses. One could also blame the NFW assumption for the E, S0 and S
radial dirtibutions, but these are consistent with the data \citep{Cava+17}.

\subsection{Velocity anisotropy profiles of WINGS clusters}

\subsubsection{General trends}

For the
single component
mass models, all anisotropy models that differ from that
of T$_0$ 
anisotropy for the spirals and isotropy for the E and S0 galaxies are
strongly rejected by BIC Bayesian evidence, with the sole exception of the case
(model~\NFWisoETTAL, the best one using AIC evidence) 
where S0 galaxies have outer anisotropy as do the spirals
(Table~\ref{tab:runs}).
But the Bayesian evidence of model~\NFWisoELTTAL\ (with outer anisotropy only for the
spirals) against model~\NFWisoETTAL\ is $\Delta \rm BIC=5.5$, i.e. ``positive'' and
almost ``strong'' ($\Delta\rm BIC>6$).
Hence, we have good confidence that the spiral
population has radial orbits at $r_{200}$, which are significantly
more radial than at $0.03\,r_{200}$.
(e.g. Table~\ref{tab:anisotropy}). On the other hand, while early-type
galaxies appear to prefer radial orbits at the `virial' radius, the trend is
not statistically significant:
The BIC Bayesian evidence suggests that there is no need for radial outer
anisotropy of the E, and marginally so for S0 populations, while AIC evidence prefers to also
have radial outer orbits for the S0s.

\subsubsection{Robustness}
\label{sec:robanis}
\begin{figure}[ht]
  \includegraphics[width=\hsize,viewport=0 30 570 550]{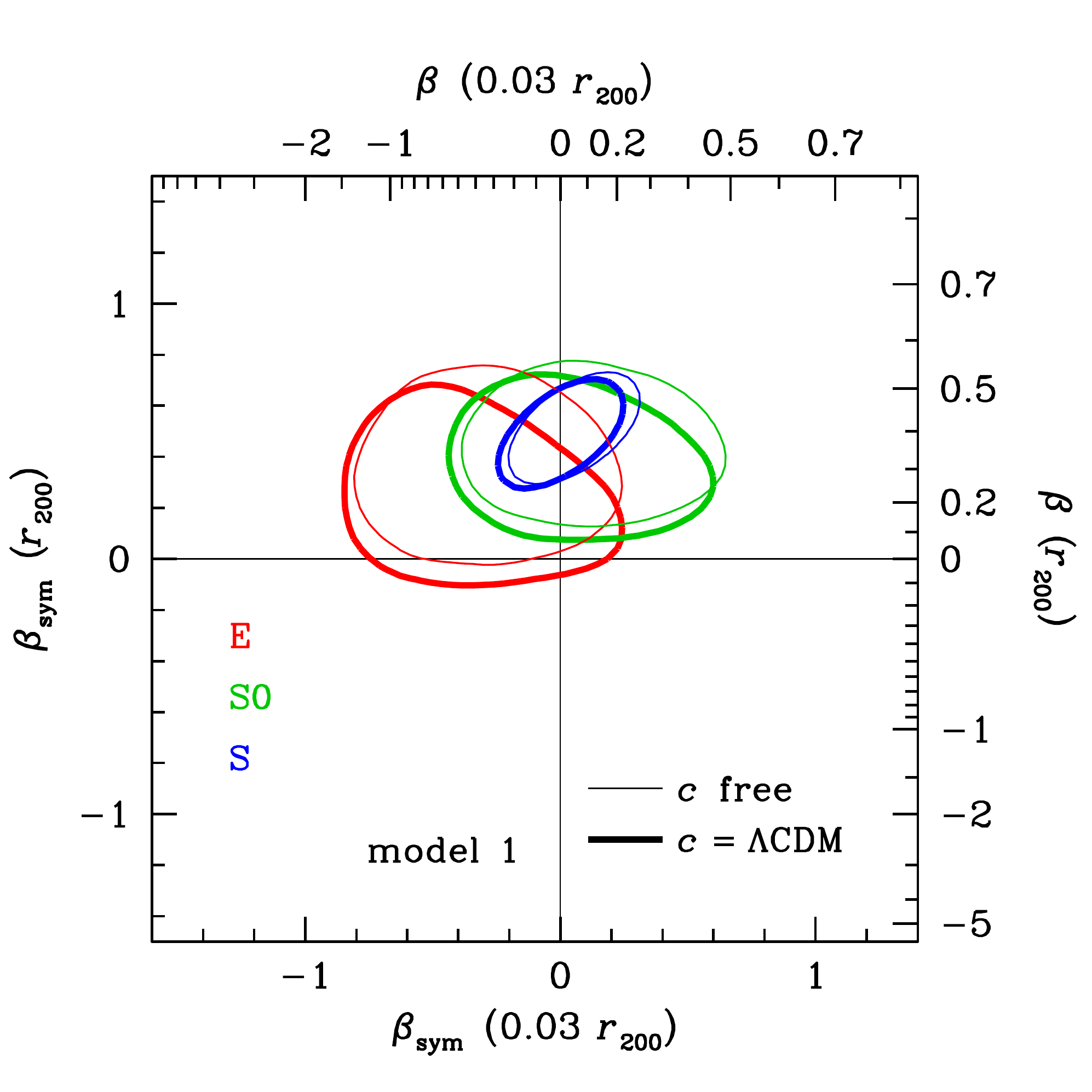}
  \includegraphics[width=\hsize,viewport=0 18 570 576]{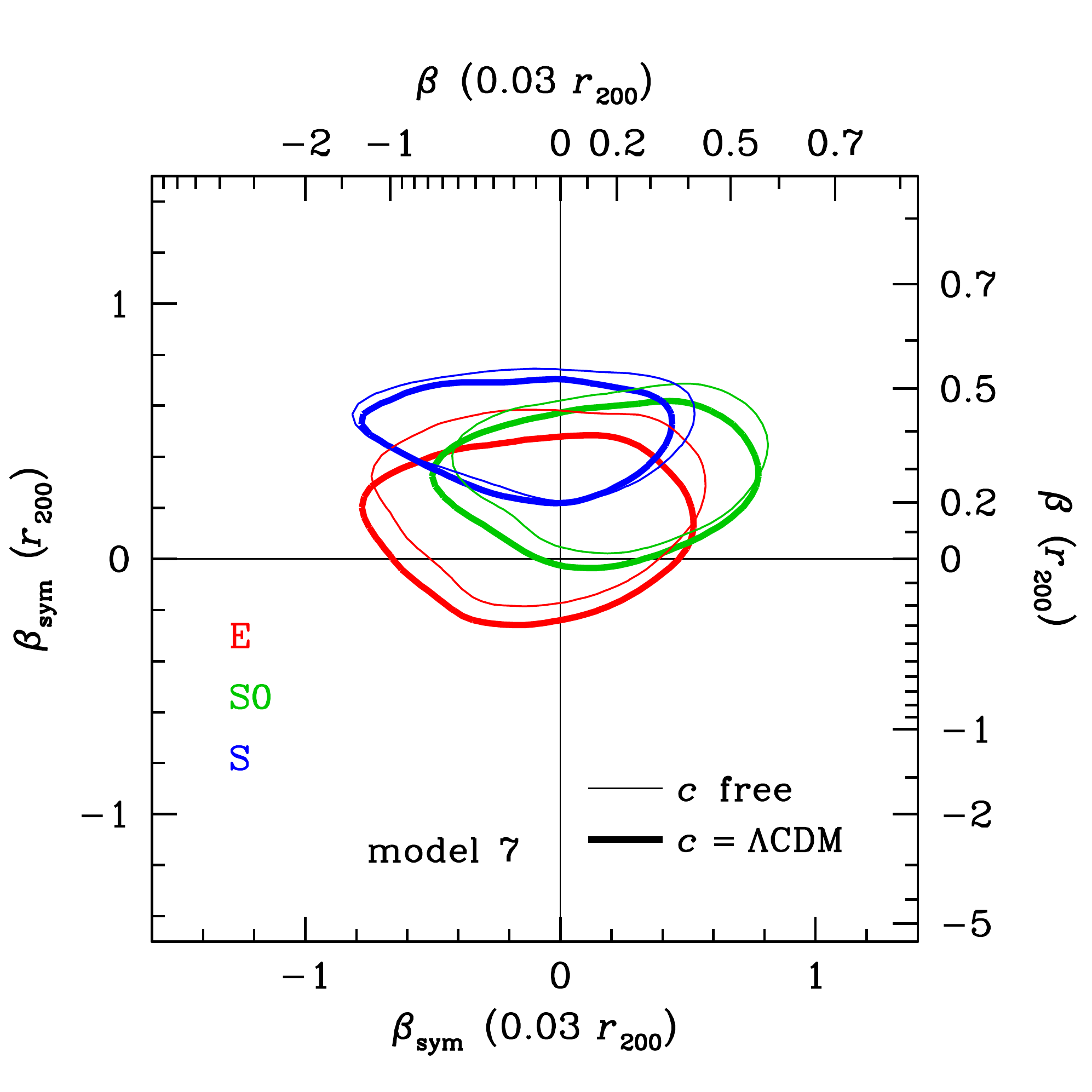}
  \caption{
    Same as Figure~\ref{fig:betabeta} comparing the priors on the cluster
    concentration for models~\gNFWgTTAL\ (TAND, \emph{top}) and
    \gNFWgT\ (free anisotropy radius, \emph{bottom}).
    Only 68\% contours are shown.
  }
  \label{fig:betabeta_cofM}
  \end{figure} 
We now test the robustness of our results on anisotropy.
Figure~\ref{fig:betabeta_cofM} shows the effect on the inner and outer
velocity anisotropies of moving from free mass concentration (thin
contours) to the $\Lambda$CDM concentration (eq.~[\ref{cM}], thick contours).
Since the concentrations obtained when they have wide (free) priors end up in
the realm of the $\Lambda$CDM concentration-mass relation (Fig.~\ref{fig:cM}),
there is virtually no difference between the anisotropies obtained with free
or $\Lambda$CDM concentrations. 

\begin{figure}[ht]
  \includegraphics[width=\hsize,viewport=0 30 570 555]{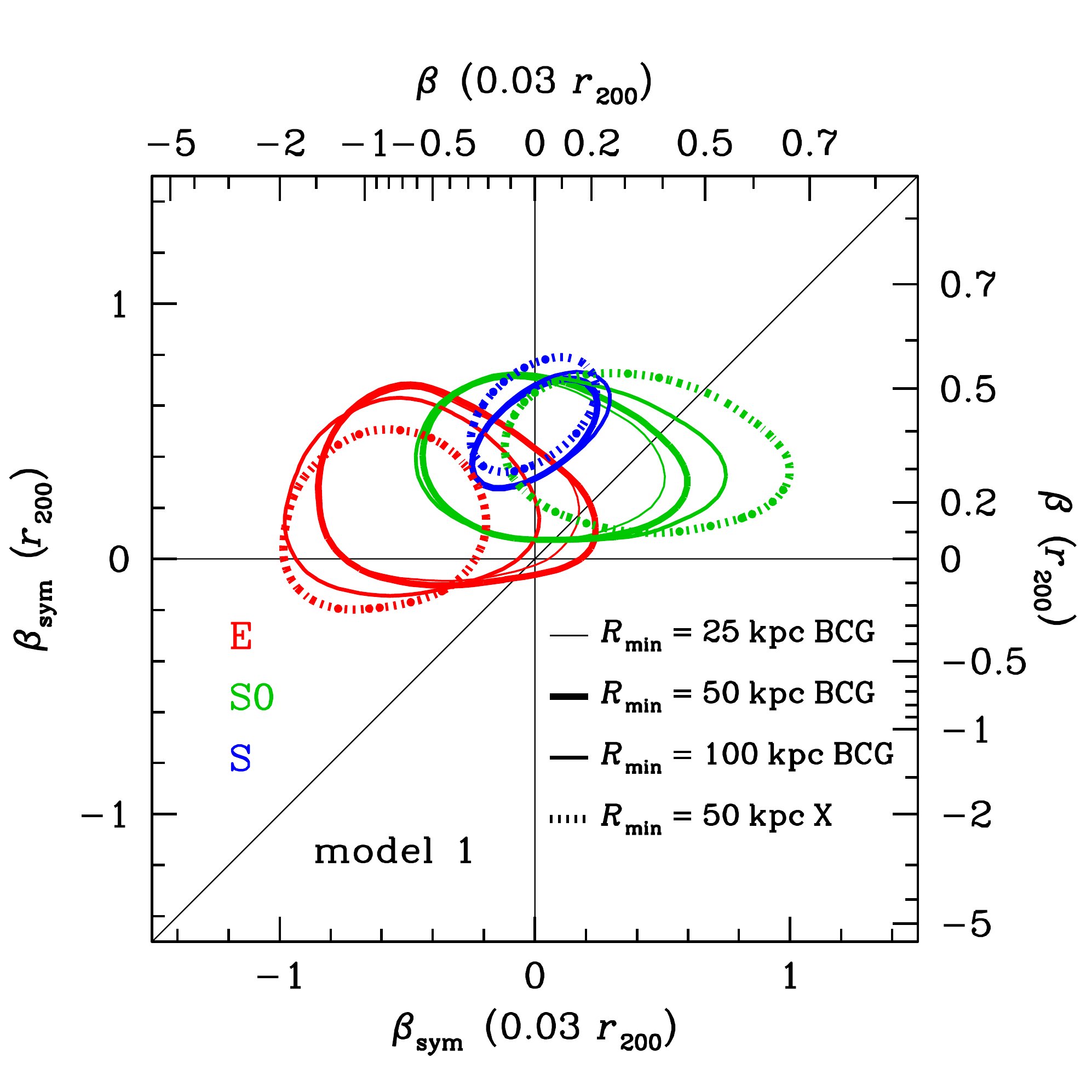}
  \includegraphics[width=\hsize,viewport=0 18 570 576]{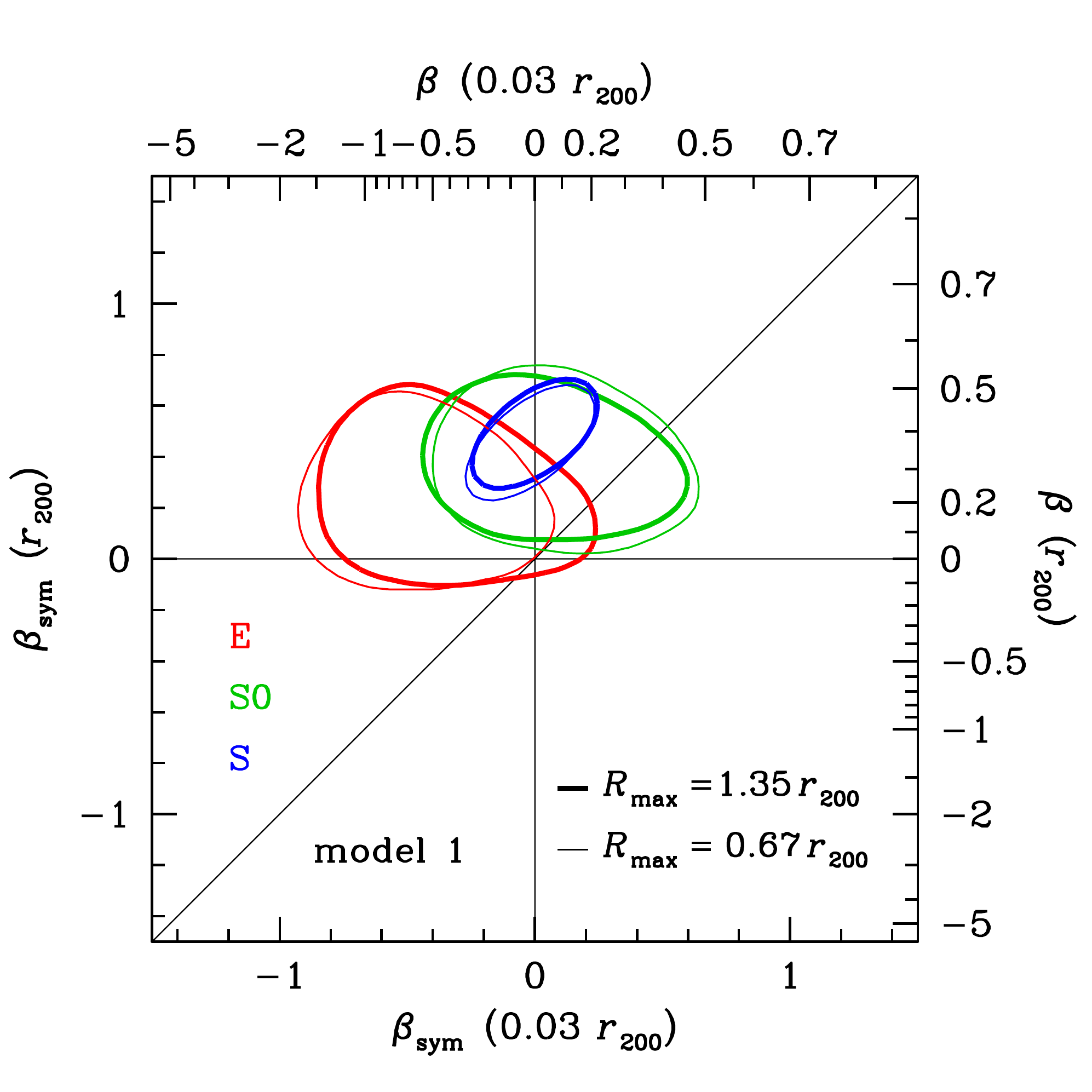}
  \caption{{\bf Top:} Same as Figure~\ref{fig:betabeta_cofM} (for
    model~\gNFWgTTAL) varying the minimum projected 
    radius for galaxy selection: 25 (\emph{thin}), 50 (\emph{medium}, our
    standard case), and 100 kpc  (\emph{thick}). The \emph{thick dotted
      contour}
    is for the case where clusters were stacked after centering on their X-ray positions instead of their BCGs, with $R_{\rm min} = 50\,\rm kpc$. 
  {\bf Bottom:} Same as top panel, for the maximum projected radius for galaxy selection: $0.67
  r_{200} \simeq r_{100}/2$ (\emph{thin}) and $1.35 r_{200} \simeq r_{100}$
  (\emph{thick}, our standard case).}
  \label{fig:Rminmax}
\end{figure}
%
Figure~\ref{fig:Rminmax} indicate that our results are fully robust to
our choice of minimum and maximum projected radii.
Figure~\ref{fig:Rminmax} also highlights the effect of changing the
definition of the individual cluster centers before the stacking.
Indeed, the BCGs can be displaced from the cluster center
(e.g. \citealp{Skibba+11}), although centering on the BCG
 or the X-rays leads to cuspier cluster profiles than using
the barycenter \citep{Beers&Tonry86}.
The figure shows that centering clusters on X-rays instead of BCGs hardly affects the velocity
anisotropy of the spirals, but leads to more tangential (radial) inner anisotropy for
the ellipticals (S0s). 

\begin{figure}[ht]
  \includegraphics[width=\hsize,viewport=0 18 570 555]{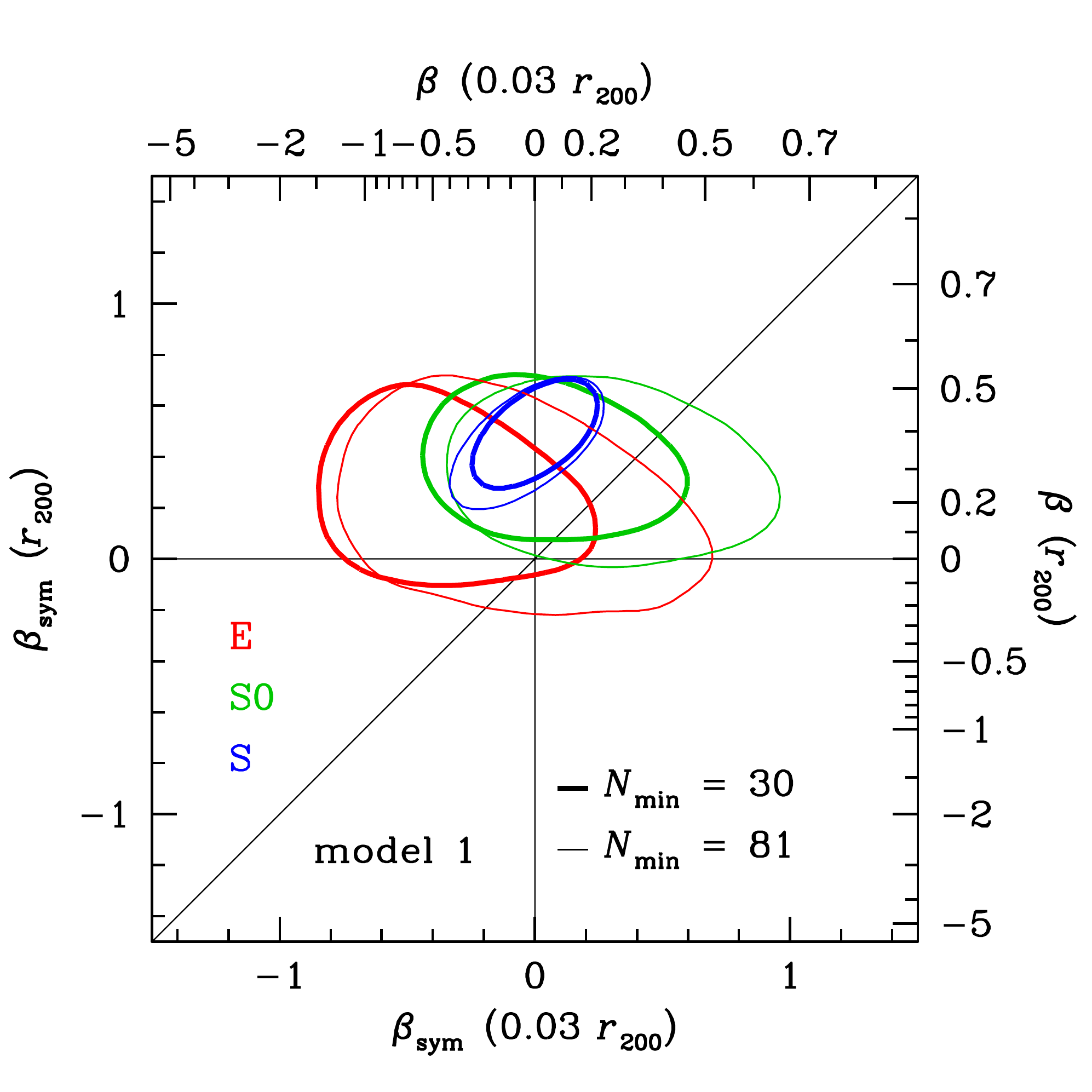}
  \caption{Same as Figure~\ref{fig:Rminmax} comparing the minimum number of
    members in individual clusters used for the stacks: 30 (\emph{thick}) and
    81  (\emph{thin}).}
  \label{fig:Nmin}
\end{figure}
Figure~\ref{fig:Nmin} compares the outer vs. inner
velocity
anisotropies when we
change the minimum number of member galaxies in clusters that we stack.
The orbits of spirals are virtually unaffected by the minimum number of
cluster members, whereas the ellipticals and S0s both allow somewhat more
radial inner orbits with 81 minimum members per cluster.

\begin{figure}[ht]
  \includegraphics[width=\hsize,viewport=0 18 570 555]{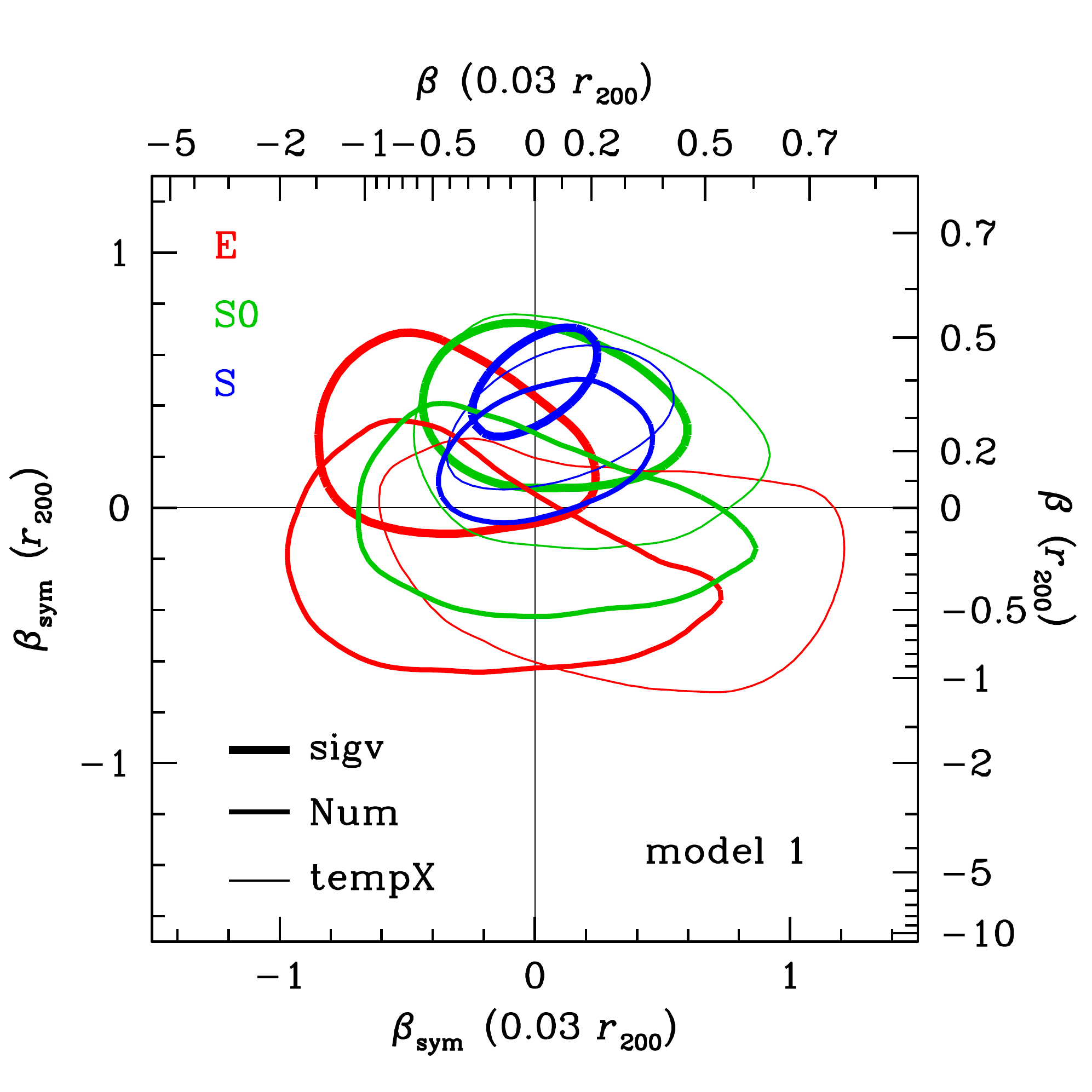}
  \caption{Same as Figure~\ref{fig:Rminmax} for 3 different stacks of
    clusters according to different estimates of the virial radii of
    individual clusters:
    our standard Clean method based on the distribution  of
    galaxies in PPS ({\tt sigv}, \emph{thick}),
    and a richness in PPS method ({\tt Num}, \emph{medium}),
and the mass - temperature relation from X-ray measurements ({\tt tempX},
\emph{thin}).
}
  \label{fig:stacks}
\end{figure}
Figure~\ref{fig:stacks} compares the outer vs. inner
anisotropies from
stacks computed using three different methods to estimate the $r_{200}$ radii
of the individual clusters (see Paper~I for details).
The outer orbits of S0s are radial for {\tt sigv}, quasi-radial for {\tt
  tempX}, and isotropic for {\tt Num}.
For ellipticals, the outer orbits are only slightly radial for {\tt sigv},
but isotropic for the other two stacks.

In the {\tt sigv} and {\tt Num} stacks, the ellipticals show signs of tangential
inner anisotropy, while they do not in the {\tt tempX} stack, which is
consistent with isotropic velocities for the ellipticals at all radii.
The {\tt Num} stack shows isotropic outer velocities for the S0s, while the
{\tt tempX} and especially our standard {\tt sigv} stacks indicate radial outer
orbits.
On the other hand, the radial outer orbits of the spirals are robust to the
stacking method (but the strongest radial anisotropy is seen in the {\tt
  sigv} stack).

\subsubsection{Comparison with previous studies}

\begin{figure*}[ht]
  \centering
  \includegraphics[width=0.48\hsize,viewport=0 18 570 556]{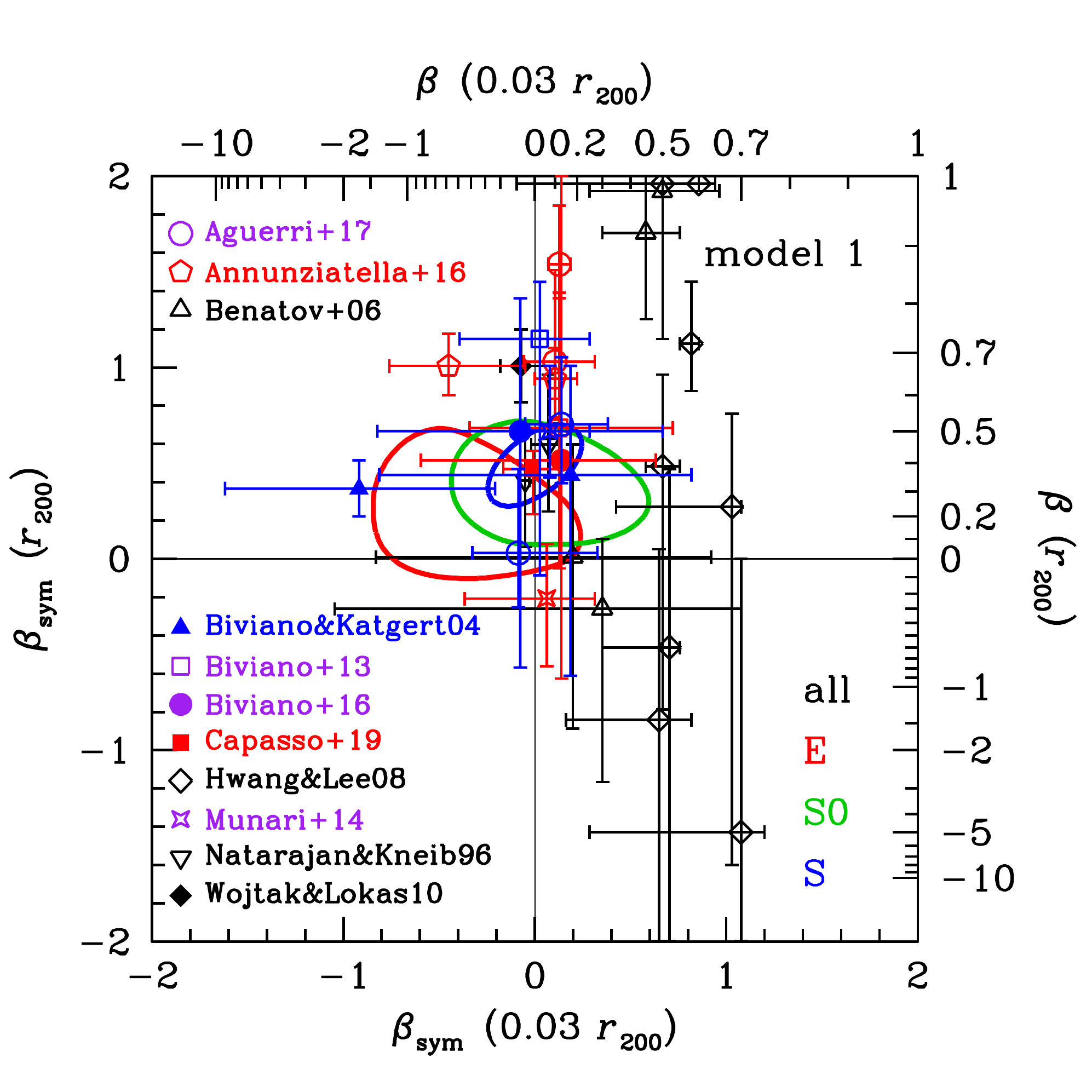}
  \quad
  \includegraphics[width=0.48\hsize,viewport=0 18 570 556]{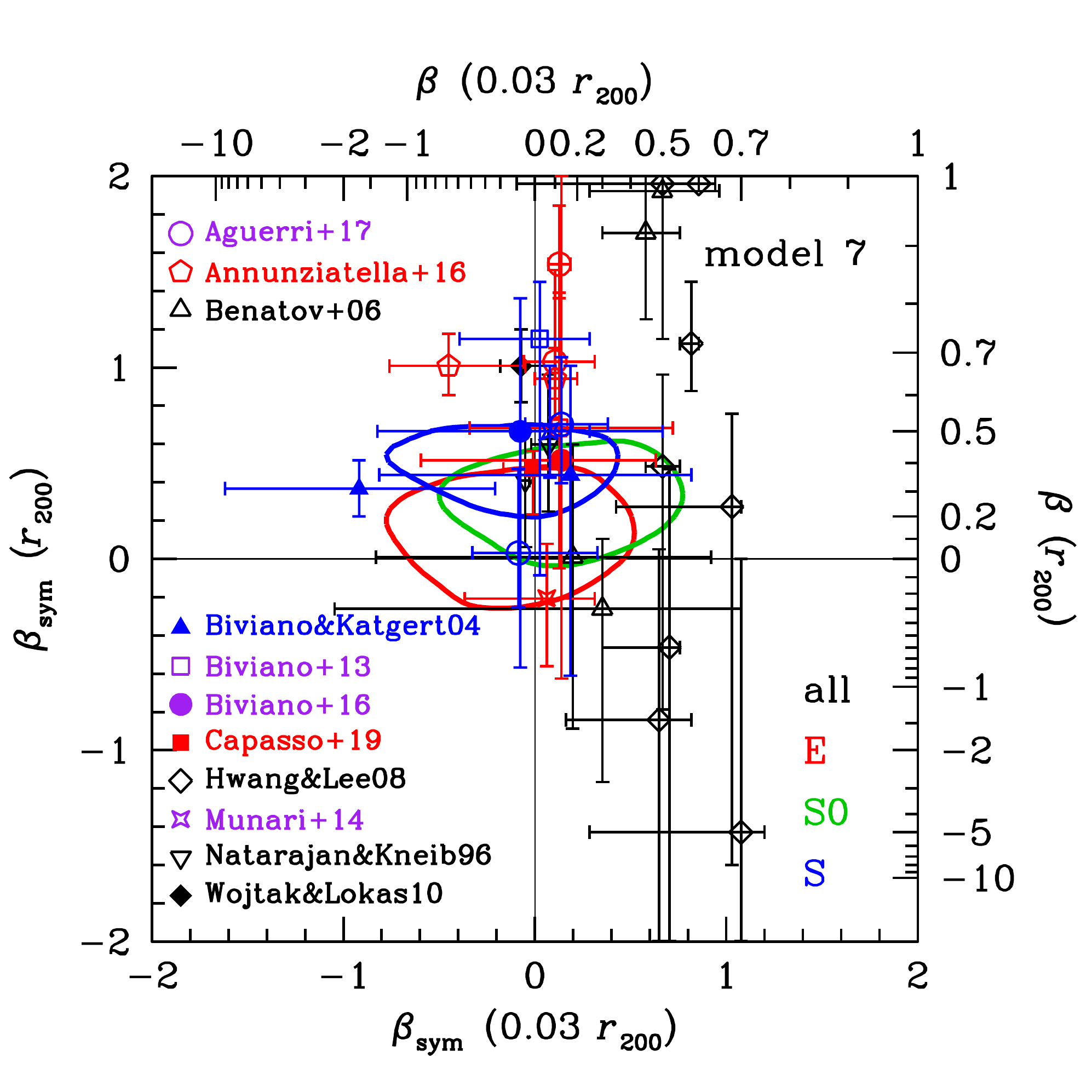}
  \includegraphics[width=0.48\hsize,viewport=0 18 570 556]{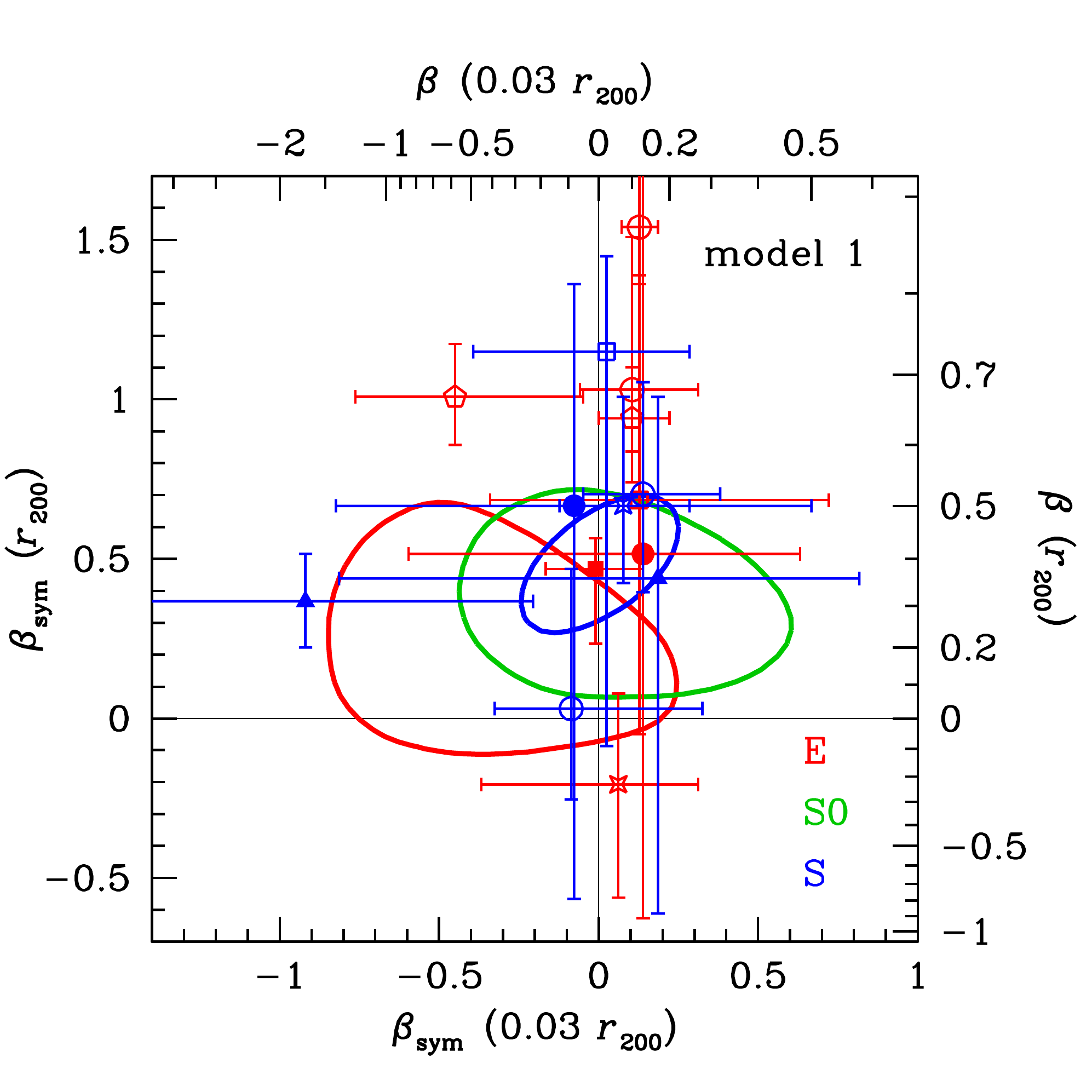}
  \quad
  \includegraphics[width=0.48\hsize,viewport=0 18 570 556]{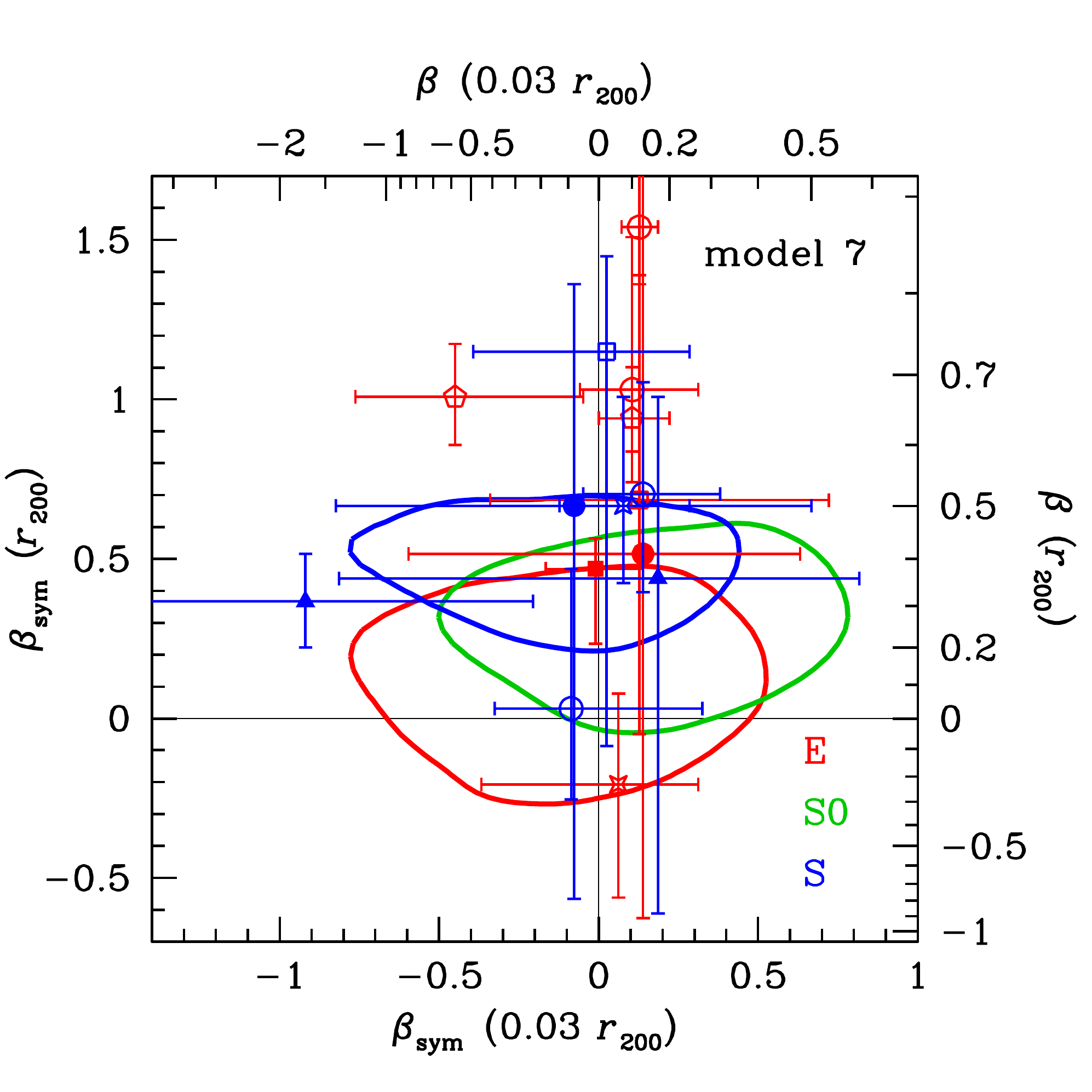}
  \caption{Comparison with previous measurements (\emph{symbols}) of the
    velocity anisotropy (eq.~[\ref{betasym}]) from 
    MAMPOSSt (68\% confidence \emph{contours}) 
    at $0.03\, r_{200}$ and 
    $r_{200}$ for models~\gNFWgTTAL\ (gNFW, T, TAND, \emph{left}) and
    \gNFWgT\ (same as \gNFWgTTAL\
    but with free anisotropy radii,
    \emph{right}), all for the {\tt sigv} stack.
    The \emph{top} panels show the comparison with the full literature over the
    entire possible range of anisotropies, while the \emph{bottom} panels
    show zooms, restricted to previous works that differentiated between galaxy
    types (with the same symbols meanings as in the top panel).
    The \emph{open} and \emph{filled symbols} respectively correspond to
    single clusters and stacks of clusters. The symbols are color-coded by
    the galaxy population: \emph{red} for early-type or passive, \emph{blue} for late-type
    or star forming, and \emph{black} for all galaxies.
    In the legend (top panels), the \emph{purple symbols} denote studies that separately
    analyzed both passive (or red) and 
    star forming (or blue) galaxies.
    The \emph{vertical} and \emph{horizontal lines} respectively indicate isotropic inner
    and outer velocities.
    The two symbols for Biviano \& Katgert are for the Sa and Sb
    spirals ($\beta(0.03\,r_{200}) = 0.2$) and for the later type spirals
    ($\beta(0.03 r_{200}) = -1.7$). The two for Annunziatella et al. refer to
    low ($\beta(0.03\,r_{200}) \approx -0.6$) and
    high ($\beta(0.03 r_{200}) \approx 0.1$) mass galaxies.
    Finally, the two blue symbols for Aguerri et al. are for all
    ($\beta(r_{200}) \approx 0.5$) and dwarf ($\beta(r_{200}) \approx 0$) galaxies.
  }
  \label{fig:complitt}
\end{figure*}

Figure~\ref{fig:complitt} compares our constraints on inner and outer
anisotropy with those from the literature.
We first compare to
previous studies that did not separate
galaxies into different classes.
The two anisotropy measures of Abell~2218 by \cite{Natarajan&Kneib96} are
consistent with our anisotropies.
However, the inner anisotropies of the 6 clusters
measured by \cite{Hwang&Lee08} are much more radial than we (or others)
found.
\cite{Aguerri+17} and \cite{Benatov+06} 
found much lower inner radial anisotropy for Abell~85 and Abell~2199,
respectively, than \citeauthor{Hwang&Lee08}.
Still, \citeauthor{Benatov+06} and \cite{Wojtak&Lokas10} both find a very wide
range of anisotropies at 
$r_{200}$, sometimes perfectly radial.

\cite{Biviano+13} found very radial anisotropy at
$r_{200}$ for the star forming galaxies in MACS~J1206, which is
consistent
with our analysis given the important uncertainty of their
anisotropy.
The
inner and outer anisotropies of the passive galaxies agree with those of our
early-type galaxies.
The results of \cite{Munari+14} on red and blue galaxies in Abell~2142
are marginally consistent with
our respective results on E+S0 and S galaxies (their anisotropy of the red
galaxies at $r_{200}$
appears too tangential).
There is tension between the very radial ($\beta > 0.55$) anisotropy of the
passive galaxies at $r_{200}$ found by \cite{Annunziatella+16} for Abell~209
(for both low and high stellar mass).
On the other hand, the inner and outer isotropy of passive
galaxies found by \cite{Capasso+19} is consistent with the orbits we find for
E and S0 galaxies.
The anisotropies found in Abell~85
by  \cite{Aguerri+17} for blue
galaxies are
consistent with ours, except that the outer isotropy that they found for their blue dwarf
galaxies is in some tension with our results for spirals. On the other hand,
they found significantly more radial orbits for red galaxies in Abell~85 compared to
us for early types galaxies.

The discrepancies in orbital anisotropies between studies of
individual clusters and our stacked analysis may be caused by a possible
diversity of clusters, either natural or modulation with mass or redshift
(the alternative is that the methods and/or priors were
different). It is therefore interesting to compare the anisotropies from the
analyses of stacked or clusters or from joint analyses of clusters.

\citeauthor{Wojtak&Lokas10}
performed a joint analysis of their 31 relaxed clusters that shows a small
range of radial 
anisotropies. They found significantly
more radial orbits at the virial radius (for which they used the rather large
value of 7 times the mass
scale radius) than our anisotropies at $r_{200}$ averaged over the 3
morphological types (Table~\ref{tab:anisotropy} and Fig.~\ref{fig:complitt}).

Our results
for spirals agree with the analysis of ENACS clusters by
\cite{Biviano&Katgert04} for early-type spirals (Sa and Sb), while they are 
marginally inconsistent with their results for late-type spirals, for
which they
found
tangential inner anisotropies.
Note that for these Sa and Sb galaxies,
\citeauthor{Biviano&Katgert04} found $\beta(r)$ to increase and
then decrease.
Our results are consistent with those for
$z$$\sim$1 GCLASS clusters of \cite{Biviano+16},  who found that
both passive (i.e. E and S0) and star forming (S) galaxies show isotropic
orbits inside and radial orbits outside.

The uncertainties on anisotropy are always greater at inner radii than at the
`virial' radius
(e.g. Table~\ref{tab:anisotropy}), which suggests that there may be a greater
range of anisotropies deep inside clusters rather than near their virial
radius. This appears to contradict the wide range of outer anisotropies found
by \cite{Benatov+06}, \cite{Wojtak&Lokas10}, as well as from
\cite{Wojtak&Mamon13} for the satellites of galaxies that may or may not be
brightest group galaxies. On the other hand, it is consistent with the work of
\cite{Annunziatella+16} who find that the inner anisotropy of passive
galaxies (i.e. ellipticals and S0s) depends on their stellar mass, while their
outer anisotropy does not.

\subsubsection{Infall}

These constraints on inner and outer anisotropy help understand the
mechanisms and timescales for the transformation of morphological types
for the quenching of star formation (comparing the orbital
anisotropy of star forming vs. passive populations). 

The simplest view is that spiral galaxies fall onto clusters on nearly radial
orbits, and are fairly rapidly transformed into
S0 and E galaxies as they orbit through the cluster. Such
morphological transformation may occur
through processes of galaxy merging (in the cluster envelope,
\citealp{Mamon92}, or in infalling groups), galaxy harassment from numerous
minor flybys \citep{Moore+96}
or starving the galaxy of its supply of infalling gas either by tidal
stripping \citep*{Larson+80} or by ram pressure stripping \citep{Gunn&Gott72}.
In this picture of fairly rapid morphological transformation,
assuming a monolithic evolution of clusters (as in the onion-ring model of
\citealp{Gott75}), galaxies that enter the cluster later (as the spirals)
will lie at larger radii. This picture is confirmed in cosmological
N-body simulations (e.g. fig.~11 of \citealp{Haines+15}).
It is also confirmed by the much higher scale radius (lower
number concentration) observed for spirals
relative to the S0s and ellipticals (Paper~I and Table~\ref{tab:fits}).
The rapid transformation of spirals into early-type morphologies might imply
that infalling spirals may not have time to exchange energy and acquire
angular momentum from the other cluster galaxies, hence they should not
isotropize. Indeed, we find that the outer anisotropy of spirals  is greater
than that of ellipticals
(Fig.~\ref{fig:stacks}).

The signs of some radial outer anisotropy for the lenticular and possibly even the
elliptical galaxies may indicate that, at the virial radius, the early-type
galaxies are a
mixture of an isotropized virialized population with 
 other early-type galaxies that are infalling for the first time (mostly
 the central galaxies and quenched satellites in galaxy groups).
 This is consistent with the narrower range of
 outer anisotropies of spirals relative to that of ellipticals and S0s
 (Figs.~\ref{fig:betabeta} and \ref{fig:stacks}). As one moves to
smaller physical radii, galaxies  first entered the
cluster at earlier times, and thus has had more time to isotropize (see
Sect.~\ref{sec:toiso} below), which
would explain the positive gradients in $\beta(r)$. However, the kinematical
evidence for this natural scenario is thin: only spiral galaxies show
statistical evidence of increasingly radial anisotropy
profiles as one moves from $0.03\,r_{200}$ to $r_{200}$
(Table~\ref{tab:anisotropy}).

\subsubsection{Isotropization}
\label{sec:toiso}

At small radii, the great majority of early-type galaxies is expected to have
entered the cluster sufficiently long ago to have been morphologically
transformed from their spiral progenitors (again, as in the onion model of cluster
growth of \citealp{Gott75}).
Should early-type galaxies isotropize or retain the radial orbits of their
spiral progenitors?

The natural way for them to isotropize is by two-body
relaxation with other galaxies. The typical timescale for two-body relaxation
roughly scales as $N/(8 \ln N)$ times the
orbital time (eq. [4.9] of \citealp{Binney&Tremaine87}), which for NFW models
is never less than e (i.e. 2.718) times the crossing 
time (at any radius).
However, this formula assumes that the system is self gravitating, whereas
galaxies in clusters account for a small portion of the cluster mass. In
other words, the dominant dark matter in cluster leads to much greater
galaxy velocities than expected from their number and mean mass.

In appendix~\ref{sec:relax}, we perform a more precise quasi-analytical measurement of the
two-body relaxation time of galaxies infalling into clusters. For the
relatively low median
galaxy mass of our sample ($10^{10.0}\,\rm M_\odot$, Sect.~\ref{sec:sample}),
mean number of galaxies 
per cluster in our sample (87), and for reasonable choices of the pericenter
radius and the ratio of apocenter to pericenter, we find
(Fig.~\ref{fig:relax}) that at the very least 30 orbits are required to
isotropize.
According to
fig.~B1 of \citeauthor{Tollet+17}, who considered a growing NFW cluster,
it takes 3 Gyr for a galaxy to move from
pericenter 
to its second apocenter (assuming hereafter that $r_{\rm apo}/r_{\rm
  vir}(t_{\rm apo}) = 3.5$), hence over 4 Gyr, between the last two pericentric
passages. These orbital times were shorter at early times, but the number of
orbits is expected to be less than a dozen.
This suggests that
two-body relaxation suffered by a single galaxy is insufficient to
explain the isotropy of the early-type galaxies in the inner regions (or
further out in some of the stacks).

Galaxies also lose energy by their encounters with other galaxies and
especially with dark matter particles (since dark matter dominates the
cluster mass distribution). The dynamical friction (DF) time is of order $[M(r)/m]/\ln(1+M(r)/m)$
times the orbital time \citep{Mamon95_Chalonge,Jiang+08}, and the ratios of
cluster mass $M(r)$ to galaxy subhalo mass $m$ are too high for
DF to be effective (especially since tidal stripping of
subhalos by the cluster gravitational field leads to much lower subhalo masses). However, DF affects the
groups that fall into clusters, hence the galaxies within these groups will
lose their radial velocities by DF on their host groups.
Moreover, the
infalling groups will be distorted by the cluster tidal field, and this tidal
heating will lead to tidal braking, i.e. the transfer of orbital energy into
internal energy. However, simulations indicate that galaxies bounce out of
clusters to 1 to 2.5 virial radii \citep{Mamon+04, Gill+05}, suggesting that
not all galaxies lose their orbital energy by the DF and tidal
braking of their host groups.

Since clusters do not evolve monolithically, but
grow by
mergers, galaxies may see their orbits perturbed by the rapidly varying
gravitational potential. This violent relaxation \citep{LyndenBell67} 
occurring during major
cluster mergers should transfer angular momentum from the other cluster into
the galaxies, leading to
isotropization. According to cosmological $N$-body simulations (figure 3
of \citealp*{Fakhouri+10}), 
cluster-mass halos typically undergo 0.8 major mergers since $z = 1$ (7 Gyr
for the cosmology of the simulation studied). At $z=1$, the Hubble constant
is 1.75 times greater, hence the orbital time is 1.75 times shorter, i.e. a
little over 2 Gyr. Thus, since $z=1$, roughly one-third of clusters undergo a
major merger, hence one-third of galaxies in our stacked cluster would have
gone through rapid isotropization. However, this fraction is an overestimate
in our case, because we selected our cluster
sample to be composed of regular clusters, thus avoiding clusters that have
gone through recent major mergers --- although they may have suffered major mergers
in the fairly recent past. We further discuss irregular clusters in
Sect.~\ref{sec:perspectives}. 

Finally, the inner isotropy of galaxy orbits may be the consequence of the
artificial phase mixing that is inherent in our stacked cluster, although
many studies of individual clusters also find isotropic inner orbits
(Fig.~\ref{fig:complitt}).

\subsubsection{The inner isotropy of spiral galaxies: a selection effect due
  to single orbits?}
\label{sec:betainS}

It is more difficult to explain why the inner orbits of spirals are
isotropic.
The time for spirals to  morphologically
transform into lenticulars or ellipticals should be at least as large as the
quenching time for star formation, which is expected to be slow for massive
spirals within clusters.
This was first shown by \cite*{Mahajan+11}, who compared
the distributions of galaxies in PPS with predictions from
cosmological simulations, and concluded that star
formation in infalling galaxies is only quenched around the time when these
galaxies cross the virial radius of the cluster on their first passage out
of the cluster.
According to fig.~B1 of \cite{Tollet+17}, this occurs $\sim$\,3\,Gyr after
pericenter or $\sim$\,4\,Gyr after cluster entry, while \citealp{Wetzel+13}
find (their fig.~8)
4.5 Gyr since cluster entry for our median stellar and halo masses).
Therefore, 
the mean radial velocities of the spiral population should
be near zero, and the MAMPOSSt analysis is based on a valid Jeans equation.
Also, the inner anisotropy of the spiral population should be roughly as
radial as the outer anisotropy, in contrast to the isotropic inner velocities
that we find for the spirals, for all 3 stacks.

However, many of our galaxies are not so massive and may be quenched at pericenter.
The simplest explanation for the isotropic inner orbits of spirals would then be
that spiral morphologies are destroyed at or before their first pericentric
passage in the cluster. But there may not be sufficient time for the cluster
to alter the morphologies of infalling spirals.
Indeed, if ram pressure stripping is at the origin of morphological
transformation of spirals into S0s (with depleted disks), the timescale for
such morphological 
transformation should be at least the gas consumption time, which is
typically 2 Gyr \citep{Bigiel+11}. If, instead intermediate mergers are the
cause of transforming spirals into S0s (by bloated bulges), 
the timescale for violent relaxation during
the merger 
should be of order of a few
internal galaxy crossing times, roughly 1~Gyr.
This is comparable to the time of $\sim$\,$1.3\,\rm Gyr$ from entry
through the cluster virial radius to the first pericenter
(see fig.~B1 of \citealp{Tollet+17}).
It is difficult to imagine that spirals begin their morphological
transformation as soon as their first entry into the cluster virial sphere.

On the other hand, spiral morphologies may be transformed before their return
into the cluster on their 2nd passage.
Given the typical orbital times of 4 Gyr today (see Sect.~\ref{sec:toiso}),
and that orbital times scale
as the age of the Universe, galaxies that reach their 2nd pericenter today
have had an orbit lasting $\sim$3 Gyr. 
Therefore, present-day spirals should have time to complete their morphological
transformation in a single orbit.

\begin{figure}[ht]
  \centering
  \includegraphics[width=0.65\hsize,viewport=42 182 597 736]{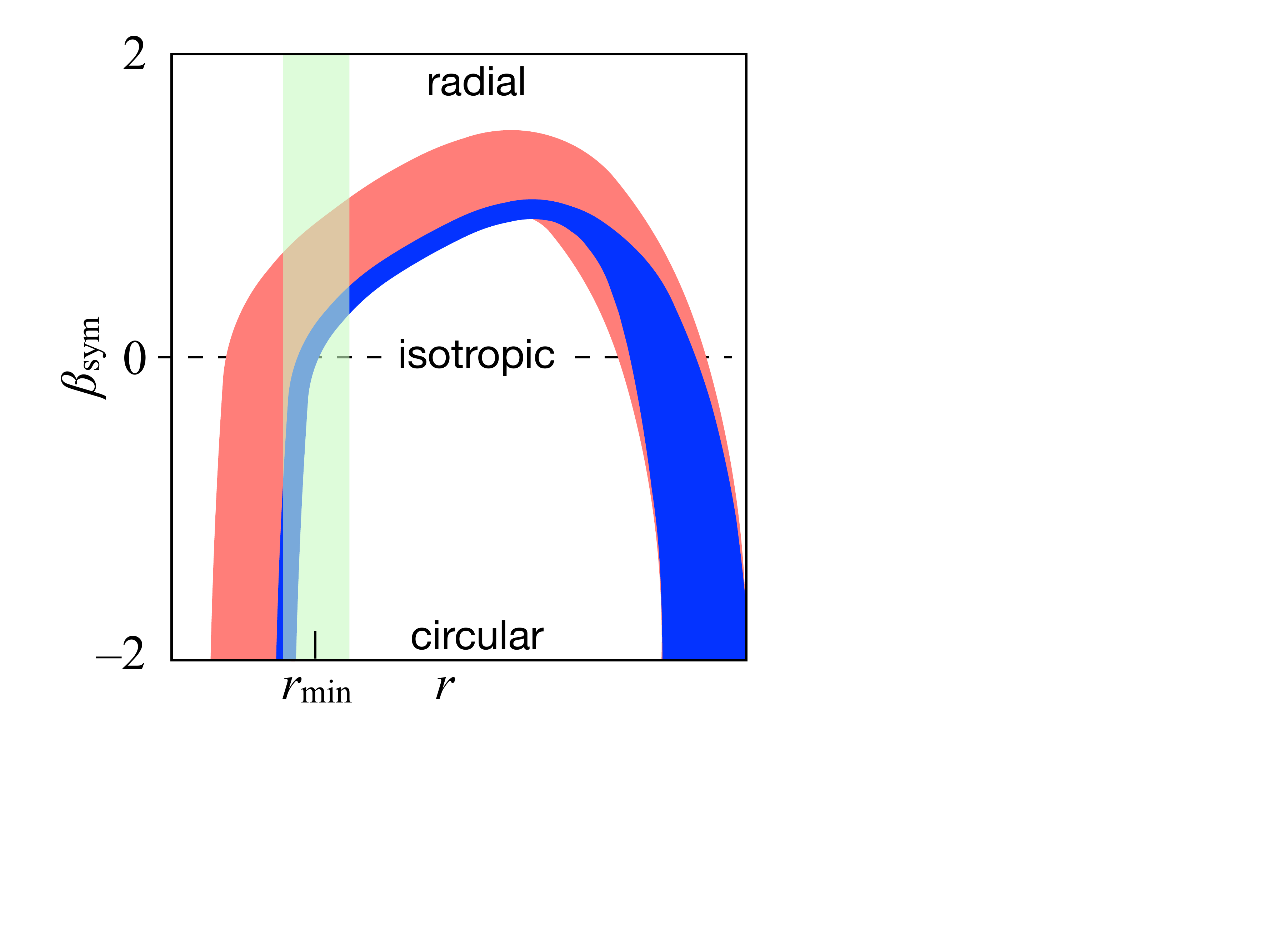} 
\caption{Illustration of a wide (\emph{salmon}) and narrow (\emph{blue}) range
  of pericenters, leading to different mixes of velocity anisotropies   at $r
  = r_{\rm min}$ (\emph{light green}): 
in the case of  a narrow range of pericenters just below $r_{\rm min}$, the
orbits at $r_{\rm min}$ cannot be radial, leading
to less radial velocity anisotropy at $r_{\rm min}$.
At pericenter and apocenter, orbits are necessarily
  circular, hence $\beta_{\rm sym}=-2$. }
\label{fig:orbits}
\end{figure}

If spirals orbit only once around the clusters with
their original morphology, their range of log pericenters will be much narrower
than if they orbit many times. Those that fell in the cluster long
ago at very small pericenters (given their small apocenters, as was the
virial radius of the cluster's most massive progenitor) should have different
morphologies now.\footnote{If spirals lose their morphology in a single
  orbit, their number density profile is no longer an NFW model as assumed,
  but a similar model truncated at small radii. This does not affect our
  analysis since the integrations in
  MAMPOSSt are outwards.}
When the range of (log) pericenters is wide, the velocity anisotropy at a
given radius $r$ is dominated by the orbits
with pericenters much
smaller than $r$, which are near radial at $r$.
But in the limit of a unique pericenter, the velocity anisotropy of
spirals would be full tangential (circular) at $r=r_{\rm peri}$, rapidly increasing with
radius to the radial values caused by infall (if the apocenters were all
equal, one would return to circular at $r=r_{\rm apo}$).
This rapid transition in the velocity anisotropy profile is seen in recent hydrodynamical
simulations \citep{Lotz+18}.
But, as illustrated in Fig.~\ref{fig:orbits},
if the spirals only orbit once through the cluster, the radial orbits at $r$
contribute less in comparison to the quasi-circular obits for
$r \ga r_{\rm  peri}$, leading to a more isotropic velocity distribution at
$r$.

Therefore,
while early-type galaxies may owe their inner isotropic orbits to
isotropization from violent relaxation, as well as DF and
tidal braking (all over a few orbits),
the quasi-isotropic orbits in late-type galaxies may simply be a
selection effect leading to a narrow range for their pericenters, thus
missing the contribution of radial orbits at the radius of study.

At high redshift, the orbital times are
shorter, while the morphological transformation times should be roughly the
same. Hence high redshift spirals may survive several orbits in their
clusters, and we would then predict that the inner velocity anisotropy of
spirals in high-redshift clusters will be somewhat radial. This is indeed
found in the mass-orbit analysis of \cite{Biviano&Poggianti09}. However, one
could argue that at high redshift, regular quasi-spherical clusters do not yet
exist and galaxy motions are set by the more filamentary geometry of proto-clusters. 

\subsubsection{The somewhat tangential inner orbits of ellipticals: tidal selection
  effects from BCGs?}
\label{sec:betainE}
There are signs of preferentially tangential inner orbits for the elliptical
population (Figs.~\ref{fig:betaofr}, \ref{fig:betabeta}, and
\ref{fig:complitt}).
Admittedly, the evidence is weak (Table~\ref{tab:anisotropy}) and is only
seen for the {\tt sigv} stack.
This is in agreement with the evidence from a much smaller
galaxy system, the Milky Way. Indeed, using proper motions from the 2nd
release of the Gaia astrometric mission, the 3D orbits of the dwarf
spheroidals  of the Milky Way have been reconstructed. The left panel of
figure D.3 of \cite{GaiaCollaborationHelmi+18} indicates that the apocenters
of all the 9 dwarfs, except 
Leo II, are less than 3 times their pericenters. This implies
somewhat tangential
velocities, as demonstrated in Appendix~\ref{sec:apoperi}, where we found
$r_{\rm apo}/r_{\rm peri} = 3.79\pm0.01$ for an isotropic Hernquist model (the
differences with the NFW model are negligible within the radial extent of the dwarf
spheroidals).

The somewhat tangential orbits of ellipticals may be caused by those
with small pericenters being tidally stripped by the BCG to the point that their masses
fall below the limit of our sample (see \citealp{Annunziatella+16}),
which we call \emph{tidal selection}.
For example, returning to the Milky Way, globular clusters, which are much
more compact than dwarf spheroidals, have much more elongated 3D orbits
\citep{GaiaCollaborationHelmi+18}, which presumably is also caused by tidal selection. 

\subsubsection{The orbits of S0 galaxies: a clue to their formation?}
Finally, one may wonder whether the velocity anisotropy profiles of S0
galaxies provide clues to their formation.
Our analysis points to the S0 population having intermediate outer orbits relative
to the ellipticals and spirals for the {\tt sigv} stack.
This is not only seen in our complex models (Figs.~\ref{fig:betaofr},
\ref{fig:betabeta}, and \ref{fig:complitt} and Table~\ref{tab:anisotropy}),
but also from the simpler models, where Bayesian evidence favors radial outer
orbits for the spirals, but marginally disfavors S0 outer
anisotropy, while it strongly disfavors outer anisotropy of the ellipticals.
A closer look at 
Figures~\ref{fig:betabeta} and
\ref{fig:complitt} reveals that, for the {\tt sigv} stack,
the positions of S0s in outer vs. inner
velocity anisotropy space lie closer to those of the spirals, in particular
because of signs of tangential inner anisotropy of the ellipticals.

But the position of S0s relative to the E and S galaxies depends on the stack.
As seen in Figure~\ref{fig:stacks} displaying model~\gNFWgTTAL,
in the {\tt Num} stack S0s display isotropic outer orbits, the ellipticals
prefer slightly tangential outer orbits (but consistent with isotropy), while
the spirals show mildly radial outer orbits, and the S0s appear to lie
somewhat closer 
to the ellipticals than to the spirals. On the other hand, in the {\tt tempX}
stack, the S0s appear to lie closer to the spirals, in particular for their
outer orbits.

In fact, the 68\% contours of the S0s fully encompass those of the spirals for the {\tt sigv}
and {\tt tempX} stacks, but this is not seen in the {\tt Num} stack, nor
between the other types in any stack.
We quantified the correspondence of the velocity anisotropies between
morphological types
of a given stack by computing the Pearson correlation coefficient $C_{1,2}$ between
two
types as
\begin{equation}
  C_{1,2} = {\left\langle (f_1 - \left\langle f_1\right\rangle)\,(f_2 -
    \left\langle f_2\right\rangle) \right\rangle
    \over \sigma(f_1) \,\sigma(f_2) }
  \label{corr}
\end{equation}
where $f_i$ is the fraction of MCMC points (after burn-in) for type $i$ that lie in a
given cell of $\beta_{\rm sym}(r_{200})$ vs. $\beta_{\rm sym}(0.3\,r_{200})$
and equation~(\ref{corr}) is estimated over the entire set of cells.
\begin{table}[ht]
  \caption{Pearson correlation coefficients of inner-outer anisotropies
    between morphological types (for model~\gNFWgTTAL)}
  \begin{center}
    \begin{tabular}{lccc}
      \hline
      \hline
      stack & E vs. S0 & E vs. S & S0 vs. S \\
      \hline
          {\tt sigv} & 0.51 & 0.23 & 0.69 \\
          {\tt Num} & 0.72 & 0.17 & 0.50 \\
          {\tt tempX} & 0.44 & 0.18 & 0.82 \\
          \hline
    \end{tabular}
  \end{center}
\label{tab:correlation}
\end{table}
The Pearson coefficients between pairs of morphological types, displayed in
Table~\ref{tab:correlation}, for model~\gNFWgTTAL, imply that S0 galaxies have
orbits closer to the spirals for the {\tt sigv} and {\tt tempX} stacks, while
their orbits are closer to ellipticals for the {\tt Num} stack. The orbits
of spirals and ellipticals are the furthest apart for all 3 stacks.

The difference in inner anisotropy between S0s and ellipticals in the
{\tt sigv} stack may be
the consequence of the possible shorter timescale for  transformation from S0 to E  than for
tidal stripping, in which case
the BCGs would have less time to tidally strip the S0s than the ellipticals, and
therefore ellipticals on radial orbits are more subject to the selection
effect against radial inner orbits (Sect.~\ref{sec:betainE}) than are S0s.

S0s appear to form relatively late in clusters, the fraction of S0s
is 0.1 at $z$\,$\sim$\,1
\citep{Smith+05,Postman+05}, i.e. 8 Gyr ago, while our sample has $\sim 40\%$
of S0s (see table~\ref{tab:numbers}).
Roughly speaking, one then expects that the great majority of the S0s were
formed over 2 Gyr ago, which is the rough time scale for S$\to$S0 evolution.
One could 
then argue that orbital isotropization should take longer than the S$\to$S0
evolution
otherwise, S0 orbits should have mostly
isotropized and resemble those of ellipticals more than those of spirals.

Finally, our results on S0s should be taken with caution, because the S0
morphological class is notoriously difficult to cleanly classify for ranges
of inclinations and apparent magnitudes
(e.g. \citealp{vandenBergh09}) and may thus be contaminated by both spirals
and ellipticals.

\subsection{Perspectives}
\label{sec:perspectives}
A larger dataset
is needed to obtain better constraints on the
orbital anisotropy of galaxies of different morphological types.
This dataset should be more complete in projected radius and in stellar mass.
This ought to lead to smaller differences in the orbital anisotropies
inferred using different stacking methods.

It would also be interesting to compare our results for regular clusters with
those on irregular clusters. Indeed, since irregular clusters are merging
galaxy systems,
the orbits of galaxies are affected by the changes in the gravitational
potential. This should
cause violent relaxation,
leading to
orbit isotropization, which should occur on the timescale of the
cluster-cluster merger, which is of the order of the orbital timescale for
galaxies that have recently entered their cluster. According to Paper~I, the
scale radii of the ellipticals and S0s in irregular clusters are double those
in (comparable mass) regular clusters, while the  scale radii of the spiral
population is
somewhat smaller than in regular clusters. Thus, while in regular clusters
spirals had scale radii 4 (3) times greater than those of ellipticals (S0s),
the scale radii of the 3 morphological types are much more similar in
irregular clusters. Perhaps the violent relaxation in irregular clusters
perturbs the orbits more
efficiently than the different histories of the 3 morphological types
differentiate the orbits. 
  
\section{Conclusions}
\label{sec:concl}

We ran the MAMPOSSt mass/orbit modeling algorithm
on 3 stacks of
54 regular, nearby ($z$\,$\approx$\,0.05)
clusters from the WINGS sample,
composed of up to 4682 galaxies located between $0.03\,r_{200}$ and $r_{200}$,
split between ellipticals, lenticulars, and spirals (including irregulars).
MAMPOSSt is a Bayesian method that jointly fits the distribution of galaxies
of these 3 morphological types in projected phase space, fitting for the
shape of the total mass profile on one hand and of the 3 velocity anisotropy
profiles on the other.
We ran MAMPOSSt with \numruns\ different  sets of priors. Our results for the {\tt
  sigv} stack are as follows:

\begin{itemize}
  \item There is no compelling evidence for a mass density profile steeper
    than NFW or $n$=6 Einasto at $0.03\,r_{200}$
    (in fact a cored-NFW profile is only weakly rejected), even though our
    highest likelihoods are reached with total density profiles that are
    steeper than NFW (inner slope of roughly $-1.5\pm0.5$) or with an NFW profile for
    the BCG in addition to an NFW profile for the remaining part of the
    cluster.
  \item An outer slope as steep as --4 (Hernquist model) is ruled out.
  \item The concentration of the mass distribution, when set free,
    is
    consistent with those in massive halos within dissipationless
    cosmological $N$-body simulations as well as with those measured in
    similar-mass clusters using weak gravitational lensing.
  \item The number density profile of elliptical galaxies traces very well the total mass
      density profile, while that of S0s only marginally does so and
      the spiral one clearly does not.
  \item The velocity anisotropy of spirals rises from near isotropic in the
    inner regions to mildly radial ($\beta\simeq 0.45\pm0.08$) at $r_{200}$.
  \item The velocity anisotropy of the lenticulars also rises from near
    isotropic in the inner regions to somewhat less radial ($\beta =
    0.31\pm0.17$) at $r_{200}$ than for the spirals.
  \item The velocity anisotropy of the ellipticals is consistent with
    isotropic anywhere, even though the highest likelihoods are reached for
    slightly tangential inner orbits and mildly radial anisotropy
    ($\beta=0.19\pm0.25$) at $r_{200}$.
    \item BIC Bayesian evidence (which prefers
    simpler models), favors isotropy everywhere for the  ellipticals and S0s,
    but does not strongly reject having outer radial anisotropy for the S0s,
    which is actually the preferred model using AIC evidence.
  \item Bayesian evidence (both BIC and AIC) suggests that the anisotropy radius (transitioning
    from the lowest to highest values) is not different from the scale radius
    of the considered morphological type.
  \item For simple priors, Bayesian evidence favors mild increases to the velocity anisotropy
    (T model) compared to the sharp increase of the generalized
    Osipkov-Merritt model. For complex priors, the two models lead to similar
    likelihoods.
\end{itemize}

Some of these conclusions are marginally different for the other two stacks:
\vspace{-0.5\baselineskip}
\begin{itemize}
\item There is marginal evidence for a steeper inner mass density than NFW
  with the {\tt Num} and {\tt
  tempX} stacks.
\item The outer anisotropy of spirals is less pronounced.
  \item The outer orbits of E and S0 galaxies are consistent with being
    isotropic for {\tt Num}, and ellipticals also show isotropic outer orbits
    for {\tt tempX}, while they are are moderately radial with {\tt sigv}
    (although not favored by Bayesian evidence). 
    \item There is no weak evidence of tangential inner orbits for
      ellipticals.
      \item S0 orbits resemble more those of spirals for the {\tt sigv} and
        {\tt tempX} stacks and more those of ellipticals for the {\tt Num} stack.
\end{itemize}
\vspace{-0.5\baselineskip}
The velocity anisotropies of the 3 morphological types provide important
clues to their transformations as they orbit clusters.

The very large radial extent of spiral galaxies suggests that they are infalling. 
Such infall should lead to fairly radial outer orbits for spirals
(as seen in
the {\tt sigv} stack, but less so in the other two).
Near $r_{200}$, E and S0 galaxies should be a mix of the virialized
(isotropized) population and the infalling members, hence with less radial
orbits than the spirals.
The inner isotropy of the early-type galaxies cannot be produced by two-body
relaxation, which is too slow. One possibility is that inner isotropy
of the E and S0 galaxies is the consequence of violent
relaxation occurring during major mergers of clusters, which appear to occur
at a sufficient rate. Alternatively, galaxies may lose their orbital energy
by a combination of dynamical friction and tidal braking suffered by the host
groups that they may live in.  

The inner isotropy of  spirals  cannot be explained in this manner,
because spirals should be transformed into S0s over an orbital time (as
confirmed by their much wider spatial distribution).
If spiral galaxies only pass once through pericenter, there is a selection
against radial orbits at a given small distance to the cluster center,
explaining their quasi-isotropic inner orbits.

Finally, although only marginally significant in the {\tt sigv} stack and not
in the others, we conjecture that the possible
     tangential anisotropy of the ellipticals may be caused by tidal selection
     where those on small pericenters are tidally stripped and fall below the
     sample mass threshold. 

\begin{acknowledgements} 
  GAM thanks Radek Wojtak and Stefano Ettori for useful comments and
  Clifford M. Hurvich for statistical advice.
  We also thank the anonymous referee for her/his constructive and
  enlightening comments.
  AB is grateful to the IAP for its hospitality.
The work of AC was supported by the STARFORM Sinergia Project funded by the
Swiss National Science Foundation.
GAM's version of the MAMPOSSt code (the version used in this work)  is publicly available at
https://gitlab.com/gmamon/MAMPOSSt.
\end{acknowledgements}

\bibliography{master}

\onecolumn
\begin{appendix}
\section{Equivalence of the Solanes \& Salvador-Sol\'e (1990) and Aguerri et
  al. (2017) anisotropy inversions}
\label{sec:anisSS90A17}

In their sect.~2.6, \citeauthor{Aguerri+17} (\citeyear{Aguerri+17}, hereafter, A+17) present a
formula (but do not derive it) for anisotropy inversion that appears similar to that of
\citeauthor{Solanes&Salvador-Sole90}
(\citeyear{Solanes&Salvador-Sole90}, hereafter,
SS90), partially using the notations of SS90, without referring to these authors
where they present their 
equations. In this appendix, we confirm that the two formulae
are indeed equivalent and provide a more robust version of it.

Following the notations of mass inversion study of \cite{Mamon&Boue10},
we express the radial dynamical pressure as $p=\nu\left\langle v_r^2\right\rangle$, and introduce a
projected pressure $P=\Sigma \sigma_{\rm los}^2$ (which does not have the
dimension of pressure, but that of pressure times size).
We denote $q = \nu v_c^2$, which has the dimension of pressure.
The $H$ function introduced by both SS90 and A+17 is equal to $P/2$, while
$\Psi$ of A+17 ($\Psi_1$ of SS90) satisfies
$\Psi(r) = -q/r$.

The kinematic projection equation \citep{Binney&Mamon82}
\begin{equation}
  P(R) = 2\,\int_R^\infty {r\over
    \sqrt{r^2-R^2}}\,\left[1-\beta(r)\,\left({R\over r}\right)^2\right]\,p(r)\,\d r
  \ ,
  \label{PofR}
\end{equation}
can be inverted for general anisotropy \citep{Mamon&Boue10,Wolf+10}, and for isotropy
($\beta=0$) this simply involves the standard Abel inversion applied to
determine $p$ from $P$ (instead of $\nu$ from $\Sigma$) to yield
\begin{equation}
    p(r) \equiv p_{\rm iso}(r) = -{1\over \pi}\,\int_r^\infty {P'(R)\over
      \sqrt{R^2-r^2}}\,\d R 
    = -{1\over \pi\,r}\,{\d I\over \d r} \ ,
    \label{piso}
\end{equation}
where (using the notation of SS90)
\begin{equation}
  I(r) = \int_r^\infty {R\over \sqrt{R^2-r^2}}\,P(R)\,\d R
  \label{Iofr}
\end{equation} 
(see eq.~[22] of \citeauthor{Mamon&Boue10}, who denoted $I$ by $J$, while A+17 used $K$ in place of $I$).

Both SS90 and A+17 provide two equations, one for $\beta\,p$ and one for
$(3-2\beta)p$. They can then deduce  $\beta(r)$ from the ratio of the two
equations (thus eliminating $p$).
The first of the two final equations of SS90 (their eq. 23) and A+17 (their
eq. 10) are clearly the same
apart from notation. 
In our notation, this writes
\begin{equation}
  \left[3-2\beta(r)\right] \,p(r) = \int_r^\infty {q(s)\over s}\,\d s
  + 2\,p_{\rm iso}(r) \ .
\label{SS90_eq1}
\end{equation} 

The second of the final equations of SS90 and A+17 are also clearly the same,
and,  transformed to our notation, are
\begin{equation}
  \beta(r) \,p(r) = -{1\over r^3}\,\int_0^r s^2\,q(s)\,\d s - p_{\rm iso}(r) -
           {3\over \pi r^2}\,I(r) + {3\over \pi\,r^3}\,\int_0^r I(R)\,\d R
           \ .
\label{SS90_eq2}
\end{equation}

However, the last term of equation~(\ref{SS90_eq2}) is an  integral of an
integral, which can be transformed to
\[
   \int_0^r I(s)\,\d s
 = \int_0^r \d s\,\int_s^\infty {R\over \sqrt{R^2-s^2}}\,P(R)\,\d R
 = {\pi\over 2} \,\int_0^r R\,P(R)\,\d R + \int_r^\infty \sin^{-1} \left({r\over
   R}\right)\,R\,P(R)\,\d R \ ,
\]
where the last integral is obtained after reversing the order of integration
in the double integral.
It thus seems preferable to re-write equation~(\ref{SS90_eq2}) as
\begin{equation}
  (\beta p)_{\rm new}(r) = -{1\over r^3}\,\int_0^r s^2\,q(s)\,\d s
  - p_{\rm iso}(r) + {3\over 2r^3}\int_0^r R\,P(R)\,\d R + {3\over \pi
        r^3}\int_r^\infty \left[\sin^{-1}\left({r\over R}\right)-{r\over
        \sqrt{R^2-r^2}}\right]\,R\,P(R)\,\d R \ .
\end{equation}

\section{Equivalence of the Bicknell et al. (1989) and Dejonghe \& Merritt (1992) anisotropy inversions}
\label{sec:anisB89DM92}

\citeauthor{Bicknell+89} (\citeyear{Bicknell+89}, hereafter B+89) and
\citeauthor{Dejonghe&Merritt92} (\citeyear{Dejonghe&Merritt92}, hereafter
DM92) both first compute the radial
dynamical pressure to then solve the Jeans equation for the anisotropy profile. 
Using the same notations as in Appendix~\ref{sec:anisSS90A17}, we can express
the pressure of B+89 (their eqs. [3.5] to [3.8]) as
\begin{equation}
  p_{\rm B+89}(r) = {1\over 3}\int_r^\infty {q(s)\over s}\,\d s -{2\over 3\,r^3}\,\int_0^r
  s^2 q(s)\,\d s + {1\over r^3}\,\int_0^r R\,P(R)\,\d R - {2\over \pi r^3}\,
  \int_r^\infty \left[{r\over \sqrt{R^2-r^2}}-\sin^{-1} \left ({r\over
      R}\right)\right]\,R\,P(R)\,\d R  \ ,
  \label{pB89}
\end{equation}
where we assumed that their $A = (3/2)\,\int_0^\infty R\,P(R)\,\d R /
\int_0^\infty q r^2\,\d r$ is unity, as expected from virial equilibrium (see
the discussion by DM92).\footnote{The beautiful anisotropy inversion of DM92
  suffers from several typos: in their equations~(43) and (44), $\psi$ should
be replaced by $\Phi = -\psi$ and in their eq.~(45) $D_z$ should be $D_x$.}
With our notation, the radial pressure of DM92 (their eq. [57a]) is
\begin{equation}
  p_{\rm DM92}(r) = {1\over 3}\int_r^\infty {q(s)\over s}\,\d s +{2\over
    3\,r^3}\,\int_r^\infty  s^2 q(s)\,\d s
  - {2\over \pi r^3}\,
  \int_r^\infty \left[{r\over \sqrt{R^2-r^2}}+\cos^{-1} \left ({r\over
        R}\right)\right]\,R\,P(R)\,\d R \ ,
\label{pDM92}
\end{equation}
where we assumed again the relation $A=1$
(which translates to $G(\infty) = 0$ in DM92's notation).
Replacing $\sin^{-1}x$ by $\pi/2-\cos^{-1}x$, one can re-express
the pressure of B+89 of equation~(\ref{pB89}) as
\[
  p_{\rm B+89}(r) = {1\over 3}\!\int_r^\infty {q(s)\over s}\,\d s
  -{2\over 3r^3}\!\int_0^\infty \!\!s^2 q(s)\d s
  +{2\over 3r^3}\!\int_r^\infty \!\!s^2 q(s)\d s
  + {1\over r^3}\!\int_0^\infty \!\!RP(R)\d R
  - {2\over \pi r^3}\!\int_r^\infty \left[{r\over \sqrt{R^2\!-\!r^2}}+\cos^{-1} \left ({r\over
      R}\right)\right]RP(R)\d R
\]
\begin{equation}
  \qquad\quad\ \ = p_{\rm DM92}(r)
  + {1\over6\,\pi\, r^3}\,\left[
\int_0^\infty  6\pi\,R\,P(R)\,\d R
    -
   \int_0^\infty 4\pi\,r^2\, q(r)\,\d r
\right] \ .
    \label{pB89bis}
\end{equation}
Now, the term in brackets in equation~(\ref{pB89bis}) is zero, from the
virial theorem, as the first integral represents the kinetic energy,
while the second integral represents the absolute value of the
potential energy (see eq. [51] of DM92).
This can be checked by inserting $P(R)$ from the equation of kinematic projection~(\ref{PofR}) into the
first integral, yielding (after inversion of the order of the integrals)
\[
\int_0^\infty R\,P(R)\,\d R = {2\over 3}\,\int_0^\infty \left[3-2\,\beta(r)\right]\,p(r)\,r^2\,\d r
\]
and by inserting $q(r)$ from 
the Jeans equation, which in our simplified
notation is
\begin{equation}
  {\d p\over \d r} + 2\,\beta {p(r)\over r}=-{q(r)\over r} \ ,
\label{jeanssimp}
\end{equation}
yielding (again after inversion of the
order of integration)
\[
\int_0^\infty r^2\,q(r)\,\d r = \int_0^\infty
  \left[3-2\,\beta(r)\right]\,p(r)\,r^2\,\d r \ .
  \]
Therefore, the B+89 and DM92 anisotropy inversions are equivalent, as
expected.
The anisotropy is obtained by solving the Jeans equation~(\ref{jeanssimp})
for the anisotropy, yielding
\begin{equation}
  \beta(r) = - {r\,p'(r) + q(r)\over 2\,p(r)}
  = -{1\over 2}\,\left[{q(r)\over  p(r)}+{\d \ln p\over \d \ln r} \right] \ .
  \label{beta2}
\end{equation}
Thus both B+89 and DM92 algorithms involve differentiation of the data, as
seen from the derivative $p'(r)$ in equation~(\ref{beta2}) and the data
terms involving $P(R)$ in equations~(\ref{pB89}) and (\ref{pDM92}). 
The DM92 algorithm seems preferable, as it is simpler and avoids
inwards extrapolation of the data to $R=0$.
\footnote{\cite{Dejonghe&Merritt92} discuss at length the possibility of an
  extra $C/r^3$ term to the radial pressure, where the constant $C$ is an integral
over physical radius  from zero to infinity, meaning that inwards integration is required as well as
  the outwards integrals of equation~(\ref{pDM92}). On one hand, they argue that
  $C=0$ to ensure that 1) both the radial and tangential
  velocity variances remain positive at large radii, and 2) the radial velocity
  dispersion does not reach unphysically large values in cases where the
  tracer density falls much faster than $1/r^3$.
  On the other hand, they argue that some choices for the gravitational
  potential (or equivalently the mass profile) may lead to situations where
  the virial theorem is inconsistent with the 2nd moment equations, in which
  case the term $C/r^3$ should be incorporated into the solution of the
  radial pressure.
}

The derivative of the radial pressure appearing in equation~(\ref{beta2}) can be
written as an integral involving the derivative of an analytical fit to the
observed projected pressure, as follows.
Differentiating equation~(\ref{pDM92}) leads to 
\begin{equation}
  p'(r) = -{q(r)\over r} - {2\over \pi}\, {\d (J/r^3)\over \d r}
  = -{q(r)\over r} + {6\over \pi r^4}\,J(r) - {2\over \pi r^3}\,{\d J\over \d r}
  \ ,
\label{pprime}
\end{equation}
where
\begin{equation}
  J(r) = \int_r^\infty \left[\cos^{-1} \left ({r\over R}\right)+{r\over
      \sqrt{R^2-r^2}}\right]\,
  R\,P(R)\,\d R \ .
\label{Jofr}
\end{equation}
With the substitution $R = r/\cos\theta$, we can write
\begin{equation}
  J(r) = r^2\,\int_0^{\pi/2} \left( \theta + \cot\theta \right)
  \,
P\left ({r\over \cos\theta}\right) {\sin\theta\over \cos^3\theta}\,\d \theta
\ ,
\label{Jofr2}
\end{equation}
which leads to
\begin{eqnarray} 
  {\d J\over \d r} &=&
  2\,r \,\int_0^{\pi/2}
  \left( \theta + \cot\theta \right)
  \,
  P\left ({r\over \cos\theta}\right) {\sin\theta\over \cos^3\theta}\,\d \theta
  + r^2
  \int_0^{\pi/2}
  \left( \theta + \cot\theta \right)
  \,
  P'\left ({r\over \cos\theta}\right) {\sin\theta\over \cos^4\theta}\,\d
  \theta
  \\
  &=& {2\over r}\,\int_r^\infty \left[\cos^{-1}\left({r\over
      R}\right)+{r\over \sqrt{R^2-r^2}}\right]\,R\,P(R)\,\d R
  + {1\over r}\,\int_r^\infty \left[\cos^{-1}\left({r\over
      R}\right)+{r\over \sqrt{R^2-r^2}}\right]\,R^2\,P'(R)\,\d R \ .
  \label{dJdr}
\end{eqnarray} 
Inserting the expression of $\d J/\d r$ of equation~(\ref{dJdr}) into
the 2nd equality of equation~(\ref{pprime}) yields
\begin{equation}
  r\,p'(r)+q(r) = {2\over \pi r^3}\,
  \int_r^\infty \left[\cos^{-1}\left({r\over R}\right)+{r\over
      \sqrt{R^2-r^2}}\right]\,R\,\left[P(R)-R P'(R)\right]\,\d R \ .
\label{rppplusq}
\end{equation}
Inserting equation~(\ref{rppplusq}) into the 1st equality of equation~(\ref{beta2}) then yields
the anisotropy
\begin{equation}
  \beta(r) = 
  {\int_r^\infty \left[\cos^{-1}\left({r/ R}\right)+{r/\sqrt{R^2-r^2}}\right]\,
  R\,\left[P(R)-R\,P'(R)\right]\,\d R
  \over
  2\,\int_r^\infty \left[\cos^{-1}\left({r/ R}\right)+{r/\sqrt{R^2-r^2}}\right]\,
  R\,P(R)\,\d R - \pi r^2/3 \int_r^\infty q(s)\,\left(r/s+2s^2/r^2\right)\,\d
  s} \ .
\end{equation}

\section{Additional figures}
\label{sec:figures}

\begin{figure*}[ht]
  \centering
  \includegraphics[width=0.9\hsize]{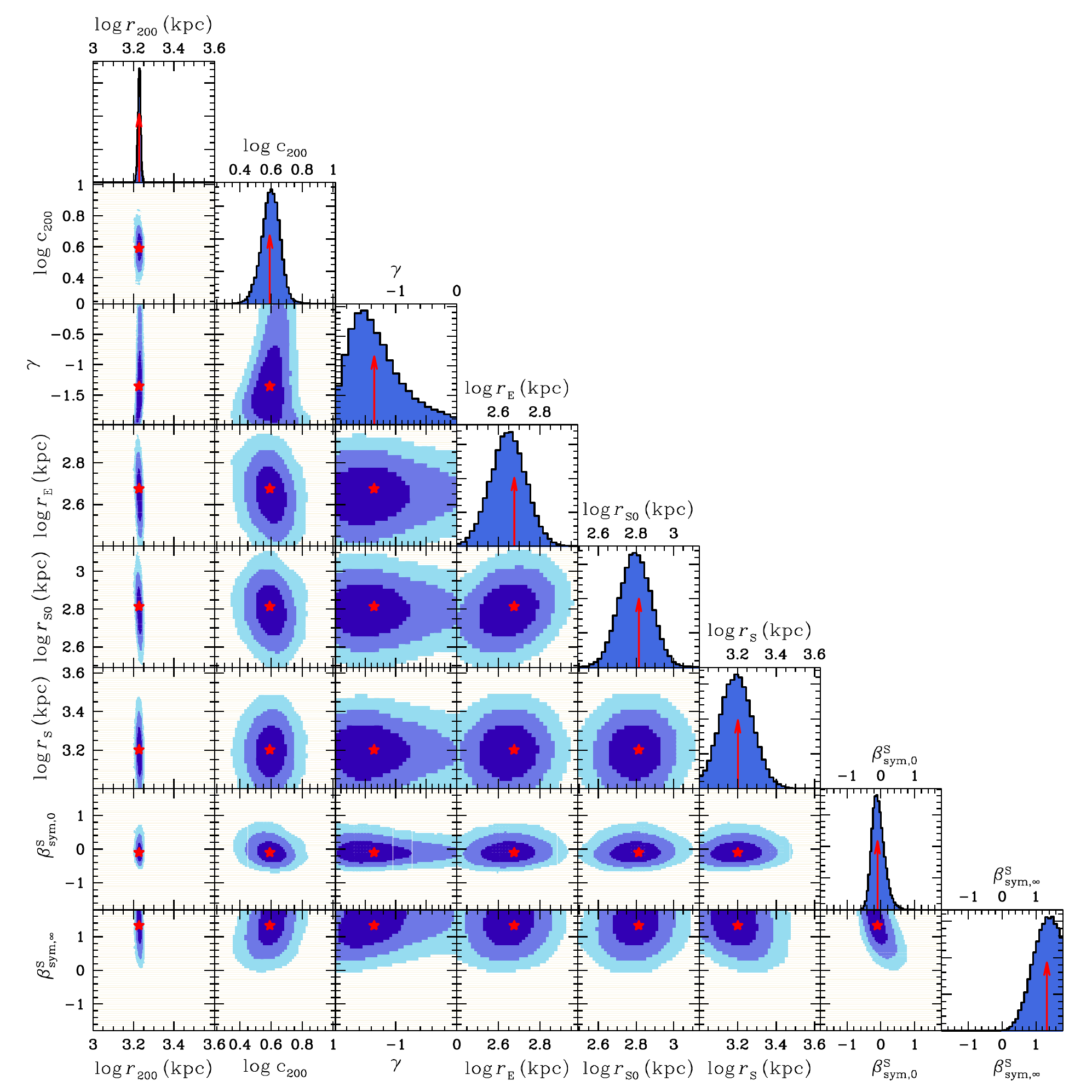}
  \caption{Same as Fig.~\ref{fig:mosaicBIC}, but for model~\gNFWisoELgTTAL.}
\label{fig:mosaicgNFW}
\end{figure*}

\begin{figure*}[ht]
  \centering
  \includegraphics[width=\hsize]{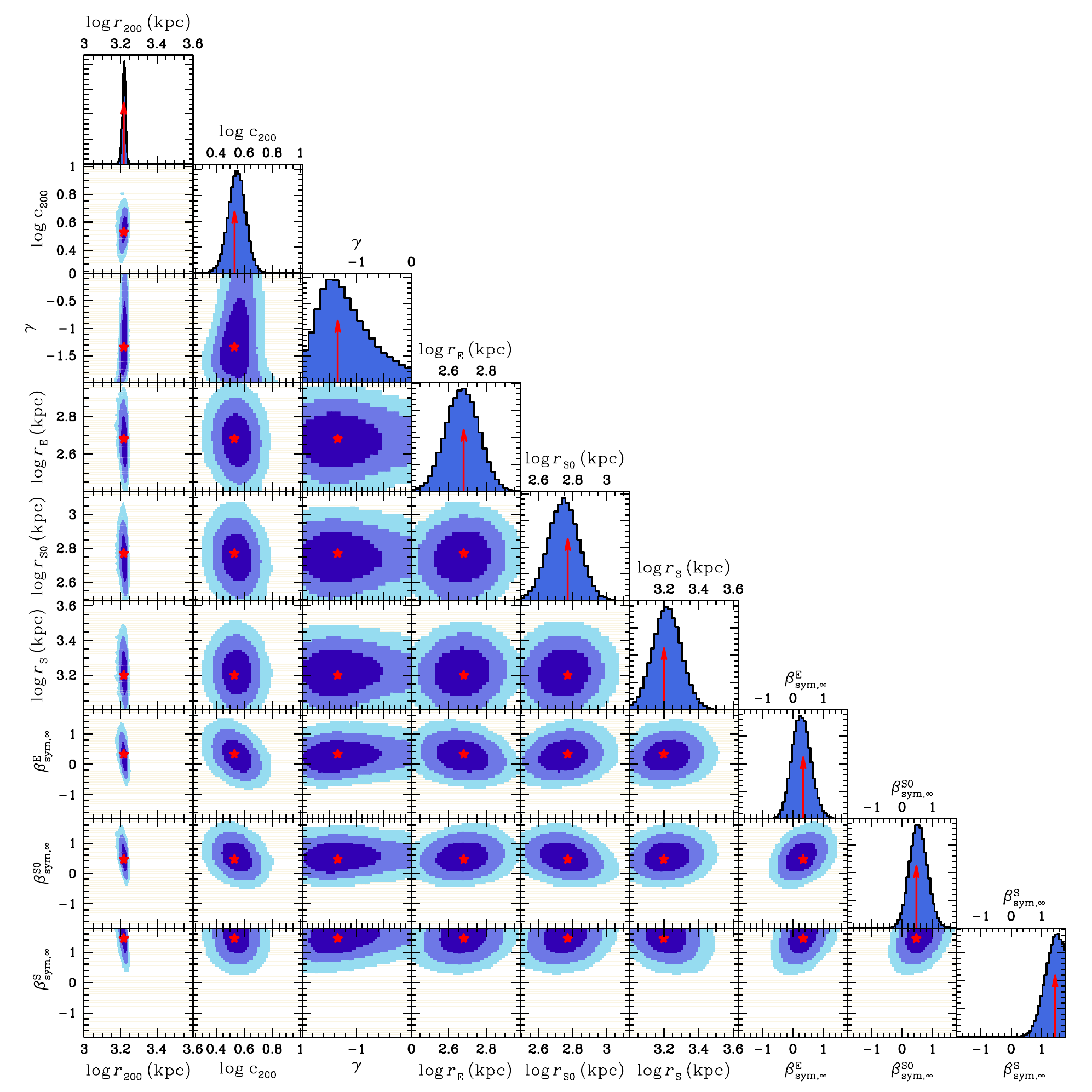}
  \caption{Same as Fig.~\ref{fig:mosaicBIC}, but for model~\gNFWTTAL.}
\label{fig:mosaicgNFW2}
\end{figure*}

This appendix provides additional MCMC mosaic plots for
models~\gNFWisoELgTTAL\ (Fig.~\ref{fig:mosaicgNFW})
and \gNFWTTAL\ (Fig.~\ref{fig:mosaicgNFW2}).

\section{Two-body relaxation time for galaxies falling into clusters}
\label{sec:relax}
\subsection{Tranverse velocity formalism}

We consider equal-mass galaxies of total (i.e. subhalo)
mass $m$ infalling into a pre-existing spherical NFW cluster of scale radius $a$. For
simplicity, we will (incorrectly) assume that the cluster is stationary.
We also assume that the galaxy number density profile is proportional to the
cluster mass density profile, and that the galaxy masses are constant, unaffected by
tidal stripping.

The change of velocity of a galaxy in a single encounter can be written as
(eq. 4-2 of \citealp{Binney&Tremaine87})
\begin{equation}
  |\delta v_\perp| = 2\,{G m \over p\, v_{\rm rel}} \ ,
\label{dv}
\end{equation}
where $v_{\rm rel}$ is the relative velocity of the two galaxies, $p$ is the
impact parameter, and $G$ is Newton's gravitational constant.
During a time $\d t$, the galaxy will suffer a number of encounters with
other galaxies with impact parameter between $p$ and $p+\d p$ of
\begin{equation}
  \d^2 {\cal N} = (\nu \,v_{\rm rel}\, \d t) \times (2 \pi p \,\d p) \ .
  \label{numcoll}
\end{equation}
Equations~(\ref{dv}) and (\ref{numcoll}) lead to a change in squared
transverse velocity over an orbit of 
\begin{equation}
  \Delta v_\perp^2 = \int_{\rm orbit} \d t \int_{p_{\rm min}}^{p_{\rm max}} {\d^2
    {\cal N}
    \over \d p \,\d t}\,\,\delta v_\perp^2\,\d p\
  = 8\pi\,G^2 m^2\int_{\rm orbit} {\nu(t)\over v_{\rm
      rel}(t)}\,\ln\Lambda(t)\,\d t  \ ,
\end{equation}
where $\Lambda = p_{\rm max}/p_{\rm min}$.
One generally assumes $p_{\rm min} = G m/v_{\rm rel}^2$ and $p_{\rm max} = r$,
yielding $\Lambda = r v_{\rm rel}^2 / (G m)$.

We now need to write the equation of motion of infalling galaxies to change
the outer integration variable from time to radius.

\subsection{Equation of motion of orbit of known pericenter and apocenter}

We begin by expressing the
conservation of energy $E$ and angular momentum $L$:
\begin{eqnarray} 
  E &=& {1\over 2} v^2 + \Phi(r)
  = {1\over 2}\,v_{\rm peri}^2 + \Phi(r_{\rm peri}) =
  {1\over 2}\,v_{\rm apo}^2 + \Phi(r_{\rm apo})
  \ ,
  \label{Econs}\\
  L^2 &=& r^2\,v_{\rm t}^2 = r_{\rm peri}^2 v_{\rm peri}^2 =
  r_{\rm apo}^2 v_{\rm apo}^2 \ ,
  \label{Lcons}
\end{eqnarray}
where `peri' and `apo' respectively denote the pericenter and apocenter of
the orbit, while $v_{\rm t}^2$ is the squared tangential velocity at any
time.
Applying $v^2 = \dot r^2 + v_{\rm t}^2$ to the first two equalities
of eqs.~(\ref{Econs}) and (\ref{Lcons}), we deduce the equation of motion
\begin{equation}
  \left({\d r\over \d t}\right)^2 - \left[1- \left({r_{\rm peri} \over
      r}\right)^2\right]\,v_{\rm peri}^2 - 2\,\left[\Phi(r_{\rm
      peri})-\Phi(r)\right] = 0 \ .
  \label{eqmom1}
\end{equation}
Eliminating $v_{\rm apo}$ in the 3rd equalities of eqs.~(\ref{Econs}) and (\ref{Lcons}) yields
\begin{equation}
  v_{\rm peri}^2 = {2\,\over
    1-1/\lambda^2}\,
  \left[\Phi(\lambda r_{\rm peri} )- \Phi(r_{\rm peri})\right]  \ ,
\label{vp}
\end{equation}
where $\lambda = r_{\rm apo}/r_{\rm peri}$.
Combining equations~(\ref{eqmom1}) and (\ref{vp}) leads to an equation of
  motion expressed in terms of the pericenter and the ratio of apo- to
  pericenter:
  \begin{equation}
    \left({\d r\over \d t}\right)^2
    -{2\,\over 1-1/\lambda^2}\,\left[1- \left({r_{\rm peri} \over r}\right)^2\right]
    \,\Phi(\lambda r_{\rm peri})
    -2\,\left\{1-{1\over 1-1/\lambda^2}\,\left[1-\left({r_{\rm peri}\over r}\right)^2\right]\right\}
      \,\Phi(r_{\rm peri}) 
    + 2\,\Phi(r) = 0 \ .
  \label{eqmom2}
   \end{equation}

  \subsection{Change of transverse velocity over an orbit}
  Expressing $\d t = \epsilon\,dr/\left|{\d r / \d t}\right|$,
  where $\epsilon = -1$
  and 1 for galaxies falling in and bouncing out, respectively,
  we can express the squared change in transverse
  velocity over an orbit as
  \begin{equation}
    \Delta v_\perp^2 = 16\pi\,G^2m^2\int_{r_{\rm peri}}^{r_{\rm apo}}
      {\nu(r)\over v_{\rm rel}(r)}\,
      \ln\left({r v_{\rm rel}^2(r)\over G\,m}\right)
      \,{\d r\over \left|{\d r / \d t}\right|} \ ,
\label{Dvperp2}
  \end{equation}
  where  $\left|{\d r / \d t}\right|$ is deduced from
  equation~(\ref{eqmom2}).
  Equation~(\ref{Dvperp2}) can be re-written in dimensionless form by
  dividing the squared velocities by the circular velocity  $\sqrt{G
    M(a)/a)}$ at the scale
  radius $a$ and using equation~(\ref{nuofr}), which leads to
  \begin{equation}
    {\Delta v_\perp^2 \over G M(a)/a} = 4\,N(a)\,\left({m\over M(a)}\right)^2
    \int_{r_{\rm peri}/a}^{r_{\rm apo}/a} {\widetilde \nu(x) \over y_{\rm rel}(x)}\,\ln
    \left ({x \,y_{\rm rel}^2\over m/M(a)}\right)\,{\d x \over |y_r|} \ ,
    \label{Dvperp2tilde}
  \end{equation}
  where $N(a)$ is the number of galaxies in the sphere of radius $a$,
  $x=r/a$, $y_{\rm rel} = v_{\rm rel}/\sqrt{GM(a)/a}$, and
  $y_r = {\d r / \d t} / \sqrt{GM(a)/a}$,  noting that the NFW gravitational
  potential can be written as
\begin{eqnarray} 
  \Phi(r) &=& - {GM(a)\over a}\,\widetilde \Phi \left ({r\over a}\right)
  \label{Phiofr}\\
  \widetilde \Phi(x) &=& {1\over \ln 2-1/2}\,{\ln(1+x)\over x} \ .
\label{phitildeNFW}
  \end{eqnarray} 
Since the typical velocities of galaxies in clusters are of the order of the
circular velocity at the scale radius, equation~(\ref{Dvperp2tilde}) directly
provides the inverse of the typical number of orbits for a galaxy to isotropize by two-body
relaxation.

Infalling galaxies encounter outgoing galaxies as well as
virialized galaxies and other infalling galaxies.
Thus, the relative velocities will, on average, be greater than
the velocities of the test galaxies, i.e. $|v_{\rm rel}| > |v|$. Hence,
assuming $v_{\rm rel} = v$, i.e. that the galaxies that the test galaxies
encounter are static, produces an upper limit to the amount of two-body
relaxation in pumping angular momentum into the infalling galaxies,
i.e. building up $\Delta v_\perp^2$.
We thus write
\begin{equation}
  y_{\rm rel} = \sqrt{y_r ^2 + {2\over 1-1/\lambda^2}\,\left ({x_{\rm
        peri}\over x}\right)^2\,
    \left[\widetilde\Phi(x_{\rm peri})-\widetilde\Phi(\lambda\,x_{\rm
        peri})\right]} \ ,
  \label{yrel}
\end{equation}
where we used $x_{\rm peri} = r_{\rm peri}/a$, the 2nd equality in
eq.~(\ref{Lcons}) and equation~(\ref{Phiofr}). 

According to our MAMPOSSt fits of the distribution of E, S0 and S galaxies in
PPS, we have, for model~\NFWisoELTTAL,
$r_{200} = 1.7\,\rm Mpc$, $c_{200} = 3.9$, leading to
$M_{200} = 10^{14.7}\rm \,M_\odot$ and $a = 437\,\rm kpc$.
Solving equation~(\ref{Ntot2}) for $N(a) = N(r_\nu)$ yields $N(a) = 13.8$.
Given the mean stellar mass of $10^{10}\,\rm M_\odot$, we deduce the subhalo
mass $m$ by solving the stellar mass as a function of halo mass from the
abundance matching formula of \cite{Behroozi+13}. For our mean stellar mass
of $10^{10}\,\rm M_\odot$ (Sect.~\ref{sec:sample}), this yields a subhalo mass of
$m = 4.6\times 10^{11} \rm \,M_\odot$ at $z=0$ and
$5.4\times 10^{11}\rm \,M_\odot$ at $z=1$. We take the higher subhalo  mass to be
conservative.
This in turns gives a mass ratio of $m/M(a) = 0.004$. 

\begin{figure}[ht]
  \centering
  \includegraphics[width=0.5\hsize]{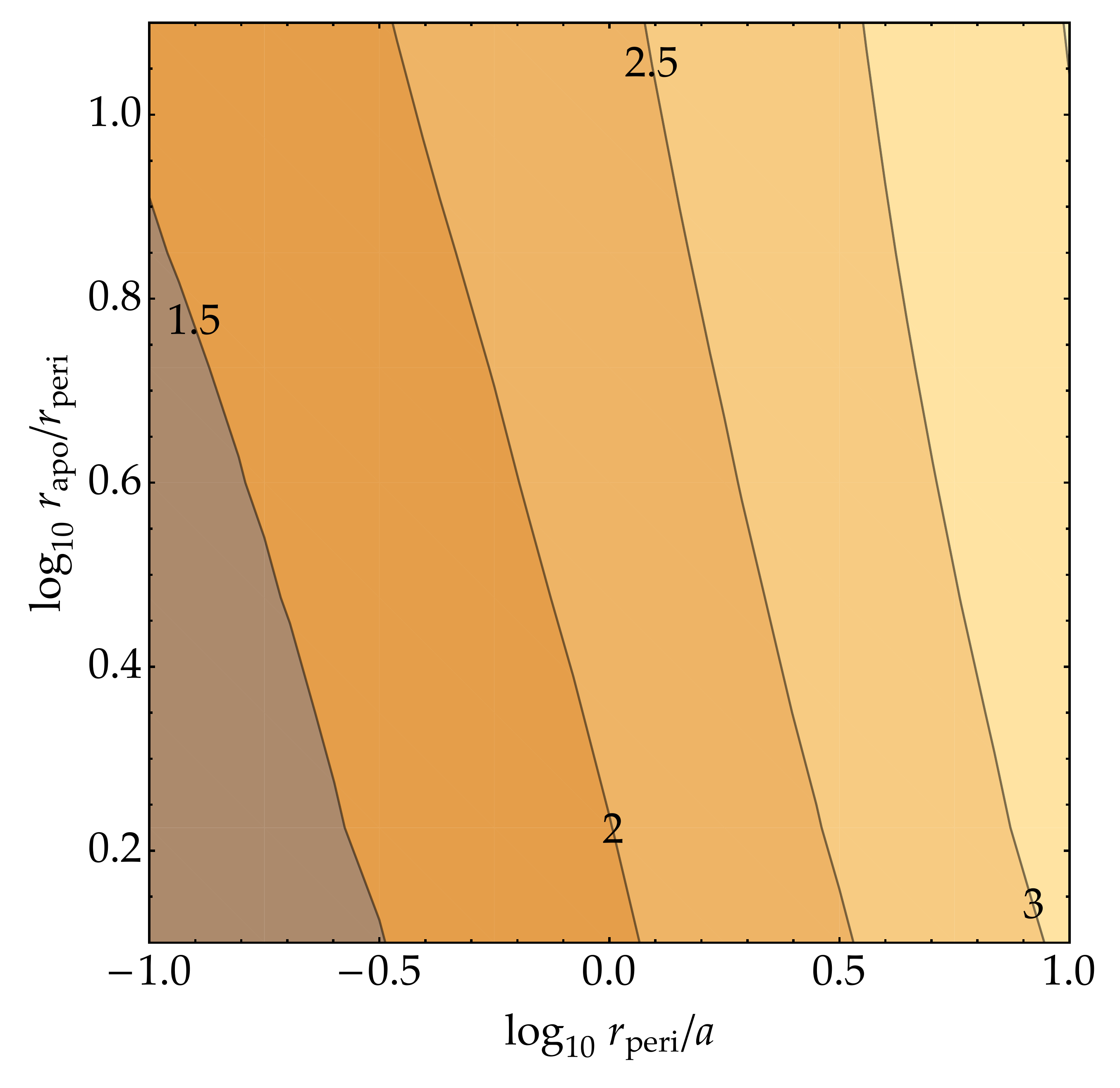}
  \caption{Contours of the inverse of $\log_{10} \{\Delta v_\perp^2 /[ G M(a)/a]\}$, representing
    the decimal logarithm of the
    number of orbits for an infalling galaxy to isotropize by two-body
    relaxation, using equation~(\ref{Dvperp2tilde}) with equations~(\ref{nutildeNFW}),
    (\ref{phitildeNFW}) and (\ref{yrel}), and assuming $m/M(a) = 0.004$ and $N(a)=14$.}
  \label{fig:relax}
\end{figure}
Figure~\ref{fig:relax} indicates the log number of orbits to isotropize, i.e.
$[G\,M(a)/a]\ / \Delta v_\perp^2$,
whose inverse is given in equation~(\ref{Dvperp2tilde}), given our
estimates of $m/M(a)$ and $N(a)$ for our cluster sample and its associated
galaxies.
The figure indicates that two-body relaxation is slow for the bulk of our
galaxies, with $10^{1.5} \simeq 30$ orbits required for deeply penetrating
orbits with $\lambda=5$ (the typical value of apo- to pericenter ratio found
in $\Lambda$CDM halos, \citealp{Ghigna+98}), and even more for orbits with
greater pericenters.
In comparison, a na\"{\i}ve application of the $N/(8 \ln N)$ formula
(eq. [4-9] of \citealp{Binney&Tremaine87}), would yield only 5.7 orbits at
$r=a$ for
$N = M(a)/m = 1/0.004 = 250$, highlighting the need for orbit averaging, as done here.

More massive galaxies do not relax faster with equation~(\ref{Dvperp2tilde}),
because they encounter too few galaxies of comparable mass to isotropize.
Recall that this calculation is conservative as many of the encounters of
infalling galaxies involve galaxies moving in the other directions with
relative velocities that are double the velocity of the test galaxy.

\section{Apocenter-to-pericenter ratio for isotropic Hernquist models}
\label{sec:apoperi}
Given a system of particles orbiting in a fixed gravitational potential, one
can determine the pericentric and apocentric radii by expressing the
conservation of energy
and  angular momentum:
The first equalities of eqs.~(\ref{Econs}) and (\ref{Lcons}) imply that the
pericenter and apocenter are the roots of 
\begin{equation}
  {1\over 2}\,{L^2\over r^2} + \Phi(r) - E = 0 \ .
  \label{apoperieq}
\end{equation}

We built an isotropic \cite{Hernquist90} model following the method of
\cite*{Kazantzidis+04_6d}, where we first draw random radii, compute the
gravitational potential and then we draw velocities from
$f(v|r) = v^2 f(v^2/2+\Phi(r))$, where $f \equiv f(E)$ is the 6D distribution
function of the isotropic Hernquist model, given by \cite{Hernquist90}.
We tested that the velocity anisotropy profile was near zero at all radii
(median value of $\beta= -0.01$).
Once we drew $100\,000$ 6D coordinates, we solved
equation~(\ref{apoperieq}), where the two roots correspond to the pericenter
and apocenter.

The extraction of the roots of equation~(\ref{apoperieq}) for each of the
$10^5$ particles
was performed in vectorial fashion: for a list of
6001 geometrically spaced radii $r_i$ between 0.001 and 1000 (in units of the
Hernquist scale radius), we estimate
the \emph{left-hand-side} (LHS) of equation~(\ref{apoperieq}) with $r=r_i$.
We first
set $r_{\rm peri} = 0$ and $r_{\rm apo} = 10^6 a$ (where $a$ is the scale
radius of the Hernquist model) for all the particles.
Noting that the LHS of equation~(\ref{apoperieq}) must be less than or
equal to 0, since it represents $-{1\over 2} (\d r / \d t)^2$,
we then vectorially adjusted
$r_{\rm peri}$ and $r_{\rm apo}$ with the conditions
\[
\begin{array}{l}
  \tt if \ |LHS| < |oldLHS| \ \& \ LHS < oldLHS  \\
  \qquad \tt update \ {\it r}_{peri} \\
  \tt if \  |LHS| < |oldLHS| \ \& \ LHS > oldLHS \\
  \qquad \tt update \ {\it r}_{apo} \\
  \tt save \ LHS \ to \ oldLHS \\
\end{array}
\]

It took 15 (1.5) seconds to process 100\,000 particles with 0.001 (0.01) dex
precision in this manner with a 
script language (SM, aka SuperMongo) on a single processor. We found a
median $r_{\rm apo}/r_{\rm peri}$ of 3.79 with an uncertainty of 0.02 (from
10 trials).
\end{appendix}
\end{document}